\documentclass[a4paper,11pt,twoside,leqno,openright]{scrreprt}
\usepackage[utf8]{inputenc}
\usepackage[english]{babel}
\usepackage{amsmath,amssymb,amsfonts}
\usepackage{pdfpages}
\usepackage{hyperref,url,cite}
\usepackage{graphicx}
\usepackage{chngcntr}
\usepackage{enumitem}
\usepackage{color}
\newcommand{\bbone}{{\text{\usefont{U}{dsss}{m}{n}\char49}}}
\counterwithout{table}{chapter}
\counterwithout{equation}{chapter}
\begin{document}
\renewcommand{\bibname}{}
\pagenumbering{Roman}
\includepdf{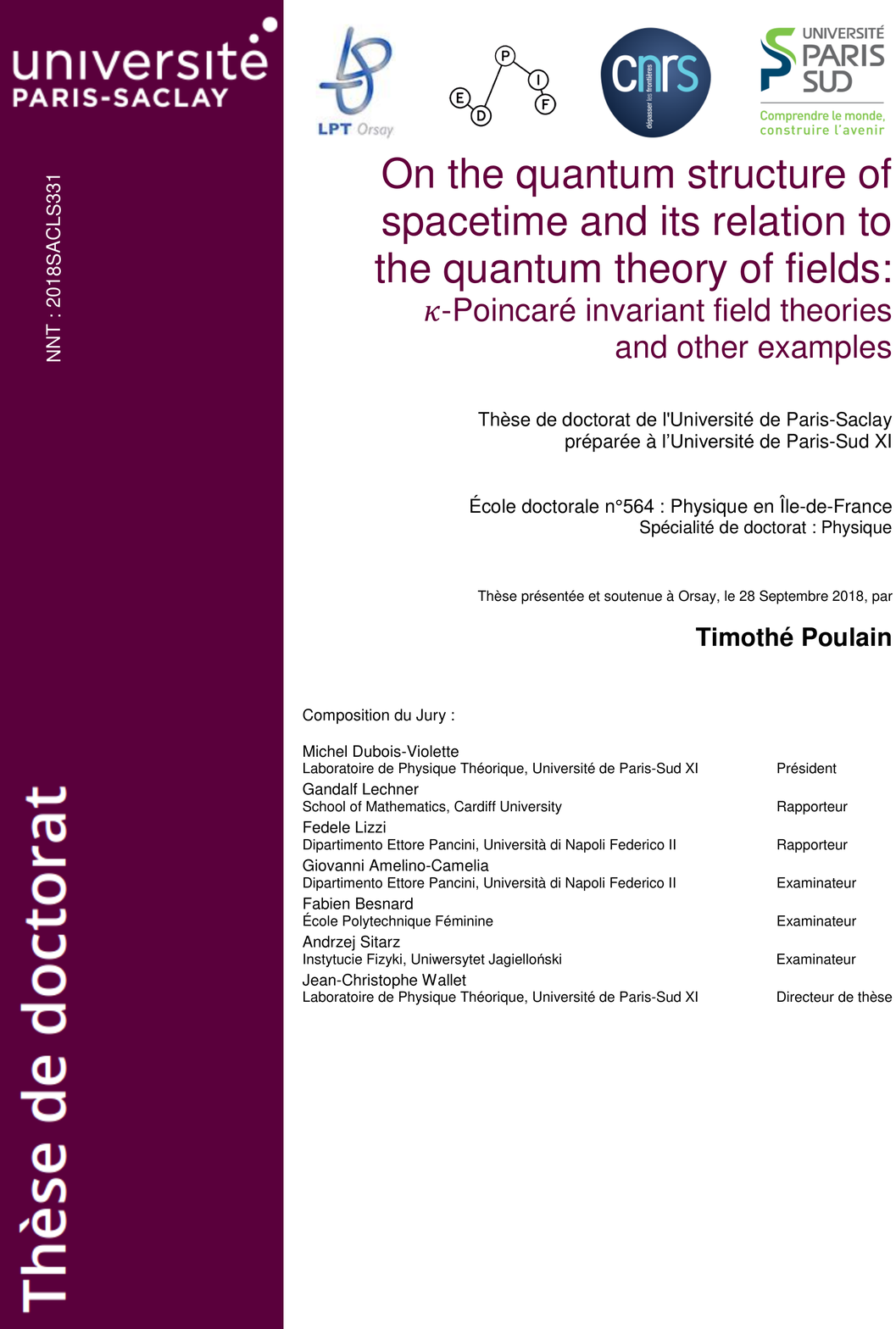}

\chapter*{}
\textit{To my parents.}



**Title:** On the quantum structure of spacetime and its relation to the quantum theory of fields

**Keywords:** Noncommutative geometry, quantum field theory, quantum gravity

**Abstract:** As many theoretical studies point out, the classical description of spacetime, at the basis of general relativity and the standard model of particles physics, might be no longer adequate to reconcile gravity with quantum mechanics at Planck scale. Instead, a more appropriate description could consist in trading the usual local coordinates on smooth manifold with noncommuting operators, therefore assuming a quantum structure to the spacetime. Accordingly, it is to expect the behaviour of field theories on noncommutative background to differ from the one of field theories on classical background. This is the aim of noncommutative field theory (NCFT) to study these new properties, toward the description of a theory of quantum gravity.

The first part of this dissertation is devoted to the construction of such quantum spacetimes, using the star product formulation of noncommutative geometry. (Recall that, within this approach, quantum spaces *are defined* as noncommutative algebras of functions endowed with deformed (star) product.) More precisely, we provide an operational method for deriving expressions for star products to be used in the construction of NCFT. Whenever the algebra of coordinates is of Lie algebra type, one canonical way to construct star products amounts to make use of the Weyl quantisation scheme, together with tools from harmonic analysis on locally compact group $\mathcal{G}$. Here, we focus on two families of quantum spaces.

The first family is characterised by semisimple Lie algebras of coordinates. In this case, we show that the quantisation map can be conveniently characterised by its evaluation on plane waves, and that families of inequivalent star products are classified by the second cohomology group $H^2(\mathcal{G}, \mathcal{A})$ of $\mathcal{G}$ with value in an Abelian group $\mathcal{A}$. In view of practical use of these star products in the context of NCFT, we show that it is further convenient to select a representative in one of the classes of equivalence belonging to $H^2(\mathcal{G}, \mathcal{A})$, and represent the abstract algebra of coordinates as an algebra of differential operators the above representative depends on. We then restrict our attention to the case $\mathcal{G} = SU(2)$. The physical motivations for studying the corresponding quantum space, which is known as $\mathbb{R}^3_\theta$, lie in 2+1 quantum gravity, group field theory, and brane models. In this case, we derive a star product equivalent to the Kontsevich product for the Poisson manifold dual to $\mathfrak{su}(2)$.

The second family of quantum spaces we consider is related to the celebrated $\kappa$-deformations. In this case, the algebra of coordinates is given by a solvable Lie algebra. The importance of this quantum spacetime, known as $\kappa$-Minkowski, lies into the fact that its symmetries are provided by the (quantum) $\kappa$-Poincaré algebra (a deformation of the classical Poincaré Hopf algebra), together with the fact that the dimensionful deformation parameter $\kappa$ provides us with a natural energy scale at which the quantum gravity effects may become relevant. Even though the algebraic properties of $\kappa$-Minkowski and $\kappa$-Poincaré, as well as the classical properties of NCFT built on $\kappa$-Minkowski, have been widely studied, the quantum properties of these NCFT have (almost) not been explored. This is probably due to the complicated expressions of the star products which were used to undertake these studies, together with the fact that requiring the $\kappa$-Poincaré invariance of the action functional (encoding the dynamics of interacting scalar fields) induces the loss of cyclicity of the Lebesgue integral, *i.e.* $\int f \star g \neq \int g \star f$. Here, we construct a natural Weyl-like star product associated with $\kappa$-Minkowski which admits a relatively simple expression. On the other hand, we interpret the loss of cyclicity as reflecting the occurrence of a KMS condition at the level of the algebra of fields modelling $\kappa$-Minkowski. In particular, the Lebesgue integral now defines a twisted trace. The possibility for this KMS condition to induce a KMS condition at the level of the algebra of observables, as well as its possible relation to the interesting proposal of A. Connes and C. Rovelli about the thermal origin of time, are discussed.

The second part of the dissertation is devoted to the construction and the study of quantum field theories built on quantum spacetimes. The main result of this part is the comprehensive characterization of the one-loop quantum properties for various models of $\kappa$-Poincaré invariant interacting scalar field theory. Both the one-loop 2-point and 4-point functions are computed for various choices of kinetic operators (chosen as being square of Dirac operators), and quartic interaction potentials. We consider three different kinetic operators and two families of inequivalent interactions, for a total of six inequivalent models. We find that only in the case the kinetic operator is given by the square of an $\mathcal{U}_\kappa(iso(4))$-equivariant Dirac operator the one-loop 2-point function has milder UV divergences than in the commutative case. We then restrict our attention to this model to compute the one-loop 4-point function which is found to be finite. Our computation provides a real technical progress in NCFT and quantum gravity models based on $\kappa$-deformations since no such derivation had been performed before. Furthermore, this opens the way toward a more systematic study of the quantum properties of noncommutative field theory and noncommutative gauge theory built on $\kappa$-Minkowski. One-loop renormalisation and possible extensions of our models are briefly discussed.

Finally, the one-loop quantum properties for various models of NCFT with quartic interactions built on $\mathbb{R}^3_\theta$ from the above-mentioned Kontsevich product are studied. In this case, we find that the one-loop 2-point function is finite, the deformation parameter $\theta$ playing the role of a natural UV and IR cutoff for the NCFT. This result is discussed in light of what we found in the case of $\kappa$-Minkowski.

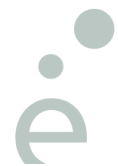



**Résumé :** Comme le suggèrent de nombreuses études théoriques, la description de l'espace-temps à la base de la relativité générale et du modèle standard de la physique des particules pourrait ne pas fournir le cadre adéquat à la réconciliation de la gravité avec la mécanique quantique à l'échelle de Planck. Une meilleure description pourrait consister à munir l'espace-temps d'une structure quantique en remplaçant, par exemple, les coordonnées locales sur la variété par un ensemble d'opérateurs ne commutant pas deux-à-deux. Il s'ensuit que le comportement des théories de champs construites sur de tels espaces diffère en général de celui des théories de champs ordinaires. L'étude de ces possibles nouvelles propriétés est l'objet de la *théorie non-commutative des champs* (TNCC) dont nous étudions certains des aspects dans ce mémoire.

La première partie de cette thèse est consacrée à la construction d'espaces(-temps) quantiques. L'approche adoptée se fonde sur un résultat bien connu de géométrie algébrique, à savoir le théorème de Gelfand-Naimark, qui permet d'interpréter dans un langage algébrique le contenu topologique d'un espace (ou d'une variété). Dans ce contexte, un espace quantique *est défini* comme étant une algèbre non-commutative de fonctions, et caractériser une telle algèbre revient en outre à caractériser sa loi de composition.
Ainsi, nous proposons une méthode pour construire de manière systématique de tels produits non-commutatifs (ou ⋆-produits) dans le cas où l'algèbre d'opérateurs positions est de type algèbre de Lie. À l'instar de la formulation statistique de la mécanique quantique, notre construction se base sur le schéma de quantification de Weyl et s'appuie sur des résultats d'analyse harmonique sur les groupes localement-compacts $\mathcal{G}$. Deux familles d'espaces quantiques sont considérées.

La première famille étudiée est caractérisée par des algèbres de Lie semi-simples. Dans ce cas, nous montrons que l'opérateur de quantification $Q$ – à partir duquel le ⋆-produit est défini via la relation $Q(f \star g) \coloneqq Q(f)Q(g)$ – est caractérisé par son action sur les ondes planes, *i.e.* $Q(e^{ipx})$. En identifiant les $Q(e^{ipx})$ à des représentations projectives de $\mathcal{G}$, les familles de ⋆-produits inéquivalents sont classifiées par le second groupe de cohomologie $H^2(\mathcal{G}, \mathcal{A})$ de $\mathcal{G}$ à valeur dans un groupe abélien $\mathcal{A}$. Dans la pratique, et plus particulièrement en vue d'une utilisation d'un ⋆-produit en TNCC, il est commode de sélectionner un représentant d'une des classes d'équivalence appartenant à $H^2(\mathcal{G}, \mathcal{A})$ et d'en déterminer une expression explicite. Nous montrons que cela peut être réalisé en représentant les opérateurs abstraits comme des opérateurs différentiels. Nous appliquons ensuite cette approche au cas $\mathcal{G} = SU(2)$ et en déduisons l'expression pour un ⋆-produit équivalent au produit de Kontsevich. L'intérêt physique de cet espace quantique, connu sous le nom de $\mathbb{R}^3_\theta$, réside dans son lien avec la gravité quantique à 2+1 dimensions, la « group field theory » et certains modèles de « branes ».

La seconde famille d'espaces considérée est connue sous le nom de $\kappa$-Minkowski et est caractérisée par une algèbre de Lie résoluble. L'intérêt de cet espace-temps quantique réside dans le fait qu'il est défini comme l'espace homogène associé à l'algèbre de Hopf de $\kappa$-Poincaré. Cette dernière définit une déformation, à l'échelle de Planck, de l'algèbre de Poincaré et s'avère être étroitement liée à certains modèles de gravité quantique. Bien que les propriétés algébriques de $\kappa$-Minkowski et de $\kappa$-Poincaré, ainsi que les propriétés classiques des TNCC associées, ait été très largement étudiées, il n'en est pas de même de leurs propriétés quantiques. Ceci est dû à l'expression compliquée des ⋆-produits utilisés dans ces études, ainsi qu'à la perte de cyclicité de l'intégrale de Lebesgue, *i.e.* $\int f \star g \neq \int g \star f$, apparaissant dans l'expression de l'action fonctionnelle, dès lors que celle-ci est requise invariante sous l'action du groupe de $\kappa$-Poincaré. Ici, nous construisons un ⋆-produit à l'aide de la méthode de quantification de Weyl. Ce produit admet une expression relativement simple qui nous permet de calculer efficacement les corrections à une boucle des fonctions 2-points et 4-points pour six modèles de TNCC scalaires dont la limite commutative correspond au modèle $\phi^4$. Chaque modèle est caractérisé par un choix d'opérateur cinétique correspondant au carré d'un opérateur de Dirac, et par un choix de potentiel d'interaction quartique. La perte de cyclicité est, quant à elle, interprétée comme signalant l'apparition d'une condition KMS au niveau de l'algèbre des champs modélisant $\kappa$-Minkowski. La possibilité pour cette condition KMS d'être étendue à l'algèbre des observables est discutée, ceci à la lumière de la proposition de A. Connes et C. Rovelli sur l'origine thermique du temps.

La seconde partie de la thèse est consacrée à l'étude à proprement dite des propriétés quantiques (à une boucle) de divers modèles de TNCC construits à partir des ⋆-produits précédemment mentionnés. Dans le cas de $\mathbb{R}^3_\theta$, muni du produit de Kontsevich, nous trouvons que la fonction 2-point est finie à une boucle. Il est intéressant de noter que, dans ce cas, le paramètre de déformation $\theta$ joue le rôle d'une coupure ultraviolette et infrarouge. Ceci confère un statut unique à ce genre de théorie, puisque ni les théories commutatives, ni les TNCC construites sur Moyal ou $\kappa$-Minkowski ne possèdent ce genre de propriété (*i.e.* le paramètre de déformation ne régularise pas, en général, le comportement UV des théories de champs). Dans le cas de $\kappa$-Minkowski, nous considérons six modèles inéquivalents de TNCC scalaires invariantes sous l'action de $\kappa$-Poincaré. Notre calcul des propriétés une boucle des fonctions 2-points et 4-points constitue le premier calcul de ce genre et fournit une véritable avancée technique dans le domaine des TNCC et de la gravité quantique. Nous trouvons que seul le modèle dont l'opérateur cinétique correspond au carré d'un opérateur de Dirac $\mathcal{U}_\kappa(iso(4))$-équivariant diverge moins que dans le cas commutatif (*i.e.* linéairement). Dans ce cas, la fonction 4-point est trouvée finie à une boucle. Les propriétés de renormalisation, ainsi que les possibles extensions de ces modèles, sont finalement discutées dans la conclusion.

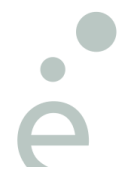


\chapter*{Preface.\addcontentsline{toc}{part}{Preface.}}
{This manuscript summarises the research conducted during my Ph.D. studies, from October 2015 to September 2018, under the supervision of Jean-Christophe Wallet at the Laboratoire de Physique Th\'{e}orique of the Universit\'{e} de Paris-Sud XI in Orsay, France. The material presented here is based on the following papers:
\begin{itemize}
\item{T. Juri\'{c}, T. Poulain, and J.-C. Wallet, \href{https://doi.org/10.1007/JHEP05(2016)146}{\textit{``Closed star product on noncommutative $\mathbb{R}^3$ and scalar field dynamics,"}} J. High Energy Phys. \textbf{2016}(05):146, 2016;}
\item{T. Juri\'{c}, T. Poulain, and J.-C. Wallet, \href{https://doi.org/10.1007/JHEP07(2017)116}{\textit{``Involutive representations of coordinate algebras and quantum spaces,"}} J. High Energy Phys. \textbf{2017}(07):116, 2017;}
\item{T. Poulain, and J.-C. Wallet, \href{https://doi.org/10.1088/1742-6596/965/1/012032}{\textit{``Quantum spaces, central extensions of Lie groups and related quantum field theories,"}} J. Phys.: Conf. Ser. \textbf{965}:012032, 2017;}
\item{T. Poulain, and J.-C. Wallet, \href{https://doi.org/10.1103/PhysRevD.98.025002}{\textit{``$\kappa$-Poincar\'{e} invariant quantum field theories with Kubo-Martin-Schwinger weight,"}} Phys. Rev. D\textbf{98}:025002, 2018;}
\item{T. Juri\'{c}, T. Poulain, and J.-C. Wallet, \href{http://arxiv.org/abs/arXiv:1805.09027}{\textit{``Vacuum energy and the cosmological constant problem in $\kappa$-Poincar\'{e} invariant field theories,"}} Preprint arXiv:1805.09027, 2018;}
\item{T. Poulain, and J.-C. Wallet, \href{http://arxiv.org/abs/arXiv:1808.00350}{\textit{``$\kappa$-Poincar\'{e} invariant orientable field theories at one-loop: scale-invariant couplings,"}} Preprint arXiv:1808.00350, 2018.}
\end{itemize}\bigskip

\paragraph{Note pour la version fran{\c{c}}aise :}
\noindent{Un r{\'e}sum{\'e}, en fran{\c{c}}ais, de mon travail de th{\`e}se est donn{\'e} en appendice ; \textit{cf.} Appendix \ref{app-resumefr}.}
}
\chapter*{Acknowledgements.}
{I am deeply grateful to Jean-Christophe Wallet for his patience and guidance during the three years of my Ph.D. I am also grateful to him for having introduced me to, and sharing his expertise with me on, noncommutative geometry and quantum field theory both from the point of view of physics and mathematics. \bigskip

I warmly thank M. Dubois-Violette, G. Lechner, F. Lizzi, G. Amelino-Camelia, F. Besnard, and A. Sitarz, for refereeing my work, and for having accepted to be part of my defense committee.\bigskip

I would like to acknowledge COST Action MP1405 for financially supporting my visits at the Ru{\dj}er Bo{\v{s}}kovi{\'c} Institute in Zagreb (Croatia), in September 2016, and at the Ettore Pancini Department of Physics of the Universit{\`a} Federico II in Naples (Italy), in December 2017. I warmly thank S. Meljanac, A. Samsarov, and T. Juri{\'c}, as well as P. Vitale, and F. Lizzi, for their hospitality during my visits, and for the helpful and stimulating discussions we had at these occasions.\bigskip

Finally, I would like to acknowledge the \textit{{\'E}cole Doctorale n$^\circ$564}, and the \textit{Laboratoire de Physique Th{\'e}orique} -- UMR 8627 Universit{\'e} de Paris-Sud and CNRS -- for funding my research activities during the past three years.}

\tableofcontents
\cleardoublepage
\phantomsection
\addcontentsline{toc}{part}{\listtablename.}
\listoftables
\part*{Introduction.\addcontentsline{toc}{part}{Introduction.}\pagenumbering{arabic}}
The fail of classical mechanics to describe physical phenomena which involve atomic systems ultimately led to the discovery of quantum mechanics in the 1920s. One of the most characteristic conceptual feature of quantum mechanics, which contrasts sharply with -- at the time well-established -- classical determinism, is the fundamental limit to the precision with which the (classical) properties of a physical system can be known. This fact, known as the Heisenberg uncertainty principle, is reflected at the mathematical level by the introduction of noncommuting variables acting on some Hilbert space which replace the classical dynamical quantities at the quantum level. An important example of this type is provided by the classical phase space whose point coordinates $(x,p)$ are replaced, at the quantum level, by noncommuting operators $(\hat{x},\hat{p})$, \textit{i.e.} such that $\hat{x}\hat{p}-\hat{p}\hat{x}\neq0$, therefore precluding the use of the classical notion of trajectory. Consequently, the classical idea of spacetime, as the place of the events, is, in some sense, already lost in quantum mechanics, as the spacetime cannot be reconstructed as a continuum from the measurement of successive positions of a point particle; even though the coordinate operators still commute among themselves within this picture.\smallskip

The idea to extend the concept of noncommutative (phase) space to the spacetime itself dates back to the early days of the relativistic quantum theory of fields and is probably due to the founders of quantum mechanics themselves; see, e.g., 
\cite{Letter1,Letter2}.\footnote{See also, e.g., in \cite{Peierls:1933,Dunne:1993} where noncommutativity of spacelike coordinates of a quantum mechanical system emerges as a consequence of the presence of a (strong, constant,) magnetic field background.} One of the initial motivation
\cite{Snyder:1947,Snyder:1947b,Yang:1947} was to cure (not all, but at least some of,) the ultraviolet (UV) divergences plaguing the theory of quantum electrodynamics by means of the introduction of a fundamental length scale $\ell>0$,\footnote{In this context, the parameter $\ell$ plays a role similar to that of the Planck constant $\hbar$ in quantum mechanics.} such that a natural UV cutoff would be provided by $\Lambda\sim\ell^{-1}$. (Un)fortunately, the pioneering works, which led ultimately to the powerful and fruitful renormalisation procedure, arose about the same time (1946-49), overshadowing the idea of noncommutative spacetime for a while. Nevertheless, some of the central ideas behind the reintroduction of the concept of noncommutative spacetime to reconcile gravity with quantum mechanics was already present in the original Snyder's paper.\footnote{In the following, we shall use the terminology ``quantum gravity," in a possibly broader sense than the usual, to designate any approach aiming to reconcile, or unify, gravity with quantum mechanics.} In particular, it was mentioned that ``the roots of the trouble in field theory [could] lie in the assumption of point interactions between matter and fields," or, equivalently, that the classical description of spacetime, as a continuum, might ``not provide a suitable framework within which interacting matter and fields can be described." And indeed, noncommutative field theory (NCFT), namely field theory on noncommutative background, can usually be regarded as ordinary field theory but with nonlocal interactions; see Part \ref{part-ncft}. As we are going to see, it is by now known that, the introduction of a fundamental length scale is not sufficient, in general, to fully regularise the quantum field theory, however. Instead, it is conjectured that the ``ultraviolet behaviour of a field theory on noncommutative spaces is sensitive to the topology of the spacetime considered, namely to its compactness"; see, e.g.,
\cite{Chaichian:2000}. Typical examples of this idea are provided by NCFT built on the Moyal space
\cite{Filk:1996}, as well as on the $\kappa$-Minkowski space
\cite{moi:2018a}, for which the one-loop 2-point functions for the $\phi^4$ scalar field theory are still UV divergent, while NCFT built on $\mathbb{R}^3_\theta$, a class of quantum spaces with $\mathfrak{su}(2)$ noncommutativity, are found to be finite, the deformation parameter $\theta\sim\ell$ playing the role of a natural UV cutoff
\cite{JCW:2013b,moi:2016}; see Table \ref{tableau}.

\begin{table}[h!]
\centering
\includegraphics[scale=0.46]{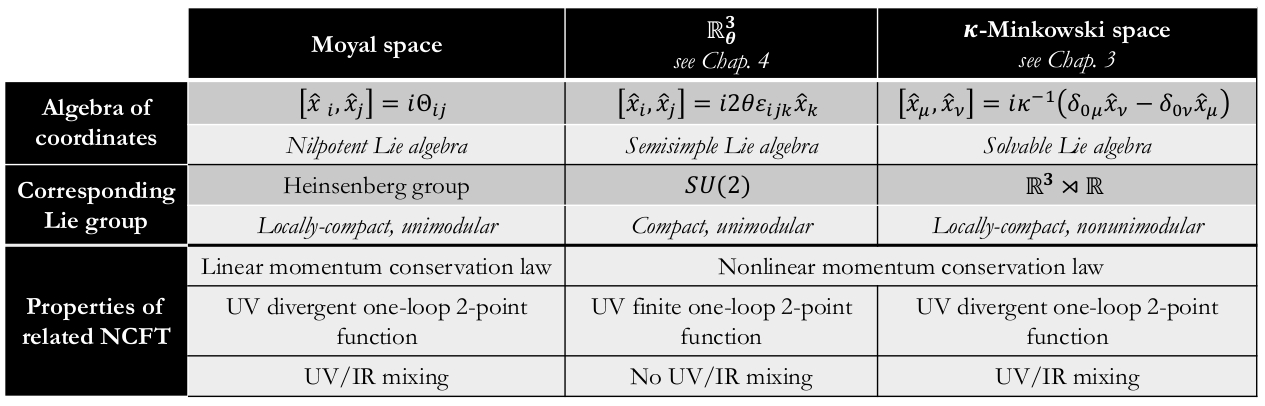}
{\caption{\label{tableau} \small \noindent Examples of quantum spacetimes, and some properties of related models of noncommutative $\vert\phi\vert^4$ scalar field theories. Note that, the expressions for the momentum conservation laws are given by the Baker-Campbell-Hausdorff formula associated with the corresponding Lie algebras of coordinate operators, while the UV behaviour of the NCFT depends strongly on the compactness of the Lie group underlying the quantum spacetime. Here, $\Theta_{ij}$ is a skew symmetric constant tensor, while $\theta,\kappa>0$ are real.}}
\end{table}

From another perspective, it has been suggested, as early as the birth of Einstein's theory of general relativity, that quantum effects should lead, in one way or another, to modifications in the description of the gravitational interaction, that could be reflected in the abandonment of the classical concept of (Riemannian) geometry. For a comprehensive historical review on the development of quantum gravity see, e.g.,
\cite{Stachel:1999,Rovelli:2002,Prugovecki:1996}, and references therein. Besides the many attempts to quantise the gravitational field, the minimal length scale hypothesis, as a way to reconcile gravity with quantum mechanics, slowly strengthened. This was initially justified by heuristic arguments based on some kind of gravitational Heisenberg microscope argument, namely the energy used to probe a sufficiently small region of spacetime might -- according to basic arguments of general relativity and quantum mechanics -- give rise to the formation of a black hole horizon, hence making the measurement process obsolete unless a fundamental length scale under which spacetime cannot be probed exists;\footnote{In the context of quantum gravity, this length scale is often interpreted as the Planck length, \textit{i.e.}
\begin{equation}
\ell_p:=\sqrt{\frac{\hbar G}{c^3}}\approx 1.6\times10^{-35}\text{m},
\end{equation}
the (theoretical) scale at which both quantum and gravitational effects become (equally) important.} see, e.g.,
\cite{Mead:1964,Maggiore:1993a,Maggiore:1993b,Doplicher:1994,Doplicher:1995}. More significantly, a renewed interest in the possible noncommutative structure of spacetime arose in the 1990s from the theoretical evidences for the existence of a minimal length in both the context of string theory
\cite{Amati:1989} and that of loop quantum gravity
\cite{Ashtekar:1992}; note that such theoretical evidence was already pointed out, e.g., in
\cite{Born:1938}, as early as 1938.
More recently, a proposal
\cite{Amelino:2000,Amelino:2001} to incorporate a fundamental (observer independent) length scale into a theory compatible with the Einstein relativity principle has emerged, providing this way a good starting point to investigate testable (phenomenological) scenarios aiming to confront quantum gravity models with observations, for instance by measuring deviations from usual dispersion relations
\cite{Amelino:1998}; for a review on doubly special relativity, see, e.g.,
\cite{Kowalski:2005,Amelino:2010}. Although we will not, strictly speaking, consider models which include gravity in the present study, we do believe that the above examples provide reasonable motivations for studying noncommutative spacetimes and noncommutative field theory.\smallskip

Although the way forward to reconcile classical gravity with quantum mechanics is far from being unique, the noncommutative structure of spacetime at some (possibly Planckian) scale appears to be a common feature shared by most of the current approaches to quantum gravity. In this respect, noncommutative geometry
\cite{Connes:1990,Landi:1997,Madore:1999,Gracia:2001} provides a suitable mathematical framework for undertaking the study of the quantum structure of spacetime and its consequences on the description of physical phenomena. At the commutative level, algebraic geometry provides us with a full dictionary between geometric objects and algebraic ones. For instance, the Gelfand-Naimark theorem
\cite{Gelfand:1943} states that there is a one-to-one correspondence between commutative C*-algebras $\mathcal{A}$ and locally compact Hausdorff spaces $X$. In particular, the points of $X$ are identified with the characters on the commutative C*-algebra of continuous functions $C_0(X)$ vanishing at infinity. On the other hand, any abstract commutative C*-algebra $\mathcal{A}$ can be regarded as the algebra of continuous functions over some topological space. Hence, the topology of a space can be encoded into an algebraic language. To capture the geometry of the space more ingredients are needed. It has been proven by A. Connes that certain commutative algebras supplemented with some additional structures, the socalled spectral triples $(\mathcal{A},\mathcal{H},\mathcal{D})$ which are made of an algebra $\mathcal{A}$ acting on some Hilbert space $\mathcal{H}$, together with a Dirac operator $\mathcal{D}$, are in one-to-one correspondence with Riemannian spin manifolds. The idea of noncommutative geometry is then to replace commutative algebras by noncommutative ones, and interpret the result as encoding some of the characteristics of what could be a noncommutative spacetime, or quantum geometry. Among the celebrated example of physical applications of this (noncommutative) spectral triple approach to the geometry, it is to mention the Connes-Chamseddine description of the standard model of particle physics and gravity
\cite{Chamseddine:1997,Chamseddine:2007}. However, it is another approach to noncommutative geometry we adopt in the present dissertation, and no use of spectral triples is made. We will be mainly concerned with noncommutative quantum field theory, using the star product formulation of noncommutative geometry. We will investigate one-loop quantum properties for various models of scalar field theory with quartic interactions, whose dynamics is characterised by various choices of kinetic operators chosen to be square of Dirac operators. For this purpose, the most important picture is that of Gelfand-Naimark: namely, to interpret noncommutative algebras of functions (endowed with star product) as hypothetical quantum spacetimes.\smallskip
 
The (renew of) interest in noncommutative field theory slowly appeared in the physics literature from the middle of the 1980s
\cite{Witten:1986,Dubois:1990a,Dubois:1990b,Madore:1991,Grosse:1992}. This interest was further increased by the observation that NCFT might emerge from some regime of string theory, and matrix theory, in external (magnetic) backgrounds
\cite{Seiberg:1999,Schomerus:1999,Connes:1998}. From the beginning of the 2000s, unusual renormalisation properties of the NCFT  built on the Moyal space $\mathbb{R}_\theta^4$
\cite{Minwalla:2000,Chepelev:2000} triggered a growing interest, in particular to cope with UV/IR mixing. Schematically, this phenomenon results from the existence of nonplanar diagrams which, albeit UV finite, become singular at exceptional low external momenta. This generates UV divergences in higher order diagrams in which they are involved as subdiagrams, signalling that UV and IR scales are nontrivially related. Recall that, one presentation of $\mathbb{R}_\theta^4$ is provided by $\mathbb{C}[\hat{x}_i]/\mathcal{R}$, the quotient of the free algebra generated by four Hermitian coordinates $\hat{x}_i$ by the relation $\mathcal{R}$ defined by $[\hat{x}_i,\hat{x}_j]=i\Theta_{ij}$, where $\Theta_{ij}$ is a skew symmetric constant tensor. This deformation of $\mathbb{R}^4$ can be described as a (suitable) algebra of functions on $\mathbb{R}^4$ equipped with the celebrated Groenewold-Moyal product
\cite{Groenewold:1946,Moyal:1949} obtained from the Weyl quantization scheme. For a review on the various presentations of the Moyal space, as well as related NCFT, see, e.g.,
\cite{Douglas:2001,Szabo:2003,Wallet:2008}. The investigation of the various properties of NCFT built on $\mathbb{R}^4_\theta$ has generated many works leading to the first all order renormalisable scalar field theory with quartic interactions
\cite{Grosse:2003,Grosse:2005,Grosse:2005b} where the UV/IR mixing was rendered innocuous through the introduction of a harmonic term; see also, e.g.,
\cite{Grosse:2004,Disertori:2007,Langmann:2003,Langmann:2004,degoursac:2008,degoursac:2011}. Other examples of quantum spaces can be obtained from algebras of coordinates which are of Lie algebra type; see, e.g.,
\cite{Gracia:2002,Kupriyanov:2008,moi:2017a,moi:2018b}. One such example is provided by quantum spaces with $\mathfrak{su}(2)$ noncommutativity. This type of noncommutativity was proposed, for example, in the context of spin network theory
\cite{Penrose:1971}, that of $2+1$ quantum gravity
\cite{Freidel:2006,Freidel:2008,Guedes:2013}, and that of brane models
\cite{Alekseev:1999}. This space, known as $\mathbb{R}^3_\theta$
\cite{Hammou:2002}, can be related to the universal enveloping algebra of $\mathfrak{su}(2)$, \textit{i.e.} $\mathcal{U}\big(\mathfrak{su}(2)\big):=\mathbb{C}[\hat{x}_i]/\mathcal{R}'$, where the relation $\mathcal{R}'$ is defined by $[\hat{x}_i,\hat{x}_j]=i\theta\varepsilon_{ijk}\hat{x}_k$, $\theta>0$. Scalar field theories built on $\mathbb{R}^3_\theta$ have been studied, e.g., in
\cite{JCW:2013b,Vitale:2014,moi:2016}, and will be discussed in more details in Chap. \ref{sap-ncftsu2}. Some of these NCFT have been shown to be free of perturbative UV/IR mixing and are characterised by the occurrence of a natural UV cutoff, both stemming from the group algebraic structure underlying $\mathbb{R}^3_\theta$
\cite{JCW:2016}.\smallskip

Another famous example of Lie algebra type noncommutative spacetime is provided by the $\kappa$-Minkowski space. This latter appears in the physics literature to be one of the most studied noncommutative spaces with Lie algebra type noncommutativity, and is sometimes regarded as a good candidate for a quantum spacetime to be involved in a description of quantum gravity, at least in some limit \cite{Amelino:2004,Freidel:2004,Cianfrani:2016}. Informally, the $d+1$ dimensional $\kappa$-Minkowski space may be viewed as the enveloping algebra of the Lie algebra generated by the $d+1$ operators $\hat{x}_\mu$ satisfying $[\hat{x}_0,\hat{x}_i]=i\kappa^{-1} \hat{x}_i,\ [\hat{x}_i,\hat{x}_j]=0,\ i,j=1,\cdots, d$, where the deformation parameter $\kappa$ has dimension of a mass. This latter algebra of coordinates has been characterised a long time ago in
\cite{Majid:1994} by exhibiting the Hopf algebra bicrossproduct structure of the $\kappa$-Poincar\'{e} quantum algebra
\cite{Lukierski:1991,Lukierski:1992} which (co)acts covariantly on $\kappa$-Minkowski and may be viewed as describing its quantum symmetries. A considerable amount of literature has been devoted to the exploration of algebraic aspects related to $\kappa$-Minkowski and $\kappa$-Poincar\'e (see, e.g.,
\cite{Pachol:2011}, and references therein), in particular dealing with concepts inherited from quantum groups, and (twist) deformations. For a comprehensive recent review on these algebraic developments see, e.g.,
\cite{Lukierski:2017}, and references therein. Besides, the possibility to have testable/observable consequences from related phenomenological models has raised a growing interest and has resulted in many works dealing, for example, with doubly special relativity, modified dispersion relations, and relative locality
\cite{Amelino:2000,Amelino:2001,Amelino:2002,Amelino:2009,Gubitosi:2013,Amelino:2011}, as well as in the study of the classical properties of noncommutative field theories built on $\kappa$-Minkowski
\cite{Agostini:2004,Agostini:2004b,Agostini:2007,Dimitrijevic:2003,Dimitrijevic:2004,Borowiec:2009,Meljanac:2011,Meljanac:2011b}. In contrast, their quantum properties have amazingly not been so widely explored compared to the present status of the above mentioned NCFT built on $\mathbb{R}^3_\theta$ and $\mathbb{R}^4_\theta$. This is probably due to the nonunimodularity of the Lie group underlying the construction of $\kappa$-Minkowski, together with the fact that requiring the $\kappa$-Poincar{\'e} invariance of the action functional (encoding the dynamics of interacting scalar fields) induces a loss of cyclicity of the Lebesgue integral, \textit{i.e.} $\int f\star g\neq\int g\star f$. These technical difficulties have been further amplified by the complicated expressions of the star products used in these studies. Nevertheless, the UV/IR mixing within some scalar field theories on $\kappa$-Minkowski has been examined in
\cite{Grosse:2006} and found to possibly occur. The corresponding analysis was based on a star product for the $\kappa$-deformation derived in
\cite{Dimitrijevic:2004c} from a general relationship between the Kontsevich formula and the Baker-Campbell-Hausdorff formula that can be conveniently used when the noncommutativity is of Lie algebra type
\cite{Kathotia:1998}. However, the cumbersome expression of this star product leads to very involved formulas which drastically restrain the study and the analysis of the quantum properties of NCFT built on $\kappa$-Minkowski. In the present dissertation, we will use another star product to investigate the quantum properties of $\kappa$-Poincar{\'e} invariant interacting scalar field theories. This product -- which is equivalent to the star product derived in \cite{Durhuus:2013} -- is based on a generalisation of the Weyl quantisation scheme; see Chap. \ref{sec-Minkowski}. The relatively simple expression of this star product (see eq. \eqref{star-4d}) enabled us to provide the first complete analysis of the one-loop quantum behaviour of various models of $\kappa$-Poincar{\'e} invariant scalar field theory with quartic interactions bypassing this way the above mentioned difficulties; see Chap. \ref{ch-ncft}.\smallskip

The NCFT we consider in Chap. \ref{ch-ncft} are $\kappa$-Poincar\'e invariant, which is a physically reasonable requirement keeping in mind the important role played by the Poincar\'e invariance in ordinary field theories together with the fact that the $\kappa$-Poincar\'e algebra can be viewed as describing the quantum symmetries of the $\kappa$-Minkowski spacetime. Indeed, an important question to address when considering NCFT is the fate of the symmetries of a noncommutative spacetime. This has triggered a lot of works using various approaches which basically depend if one insists on preserving (almost all) the classical symmetries or if one considers deformed ones. For example, in
\cite{Doplicher:1994}, the attention was focused on preserving the classical (undeformed) Lorentz or Poincar\'e symmetries for the Moyal space, as well as in
\cite{Dkabrowski:2010,Dkabrowski:2011b} for $\kappa$-Minkowski space. In this latter work, the authors ensures classical covariance of $\kappa$-Minkowski space starting from a generalised version of it introduced in
\cite{Lukierski:2002}, \textit{i.e.} $[\hat{x}_\mu,\hat{x}_\nu]=i\kappa^{-1}(v_\mu \hat{x}_\nu - v_\nu \hat{x}_\mu)$. The authors of \cite{Lukierski:2002} show that, under some assumptions, deformed (quantum) symmetries are not the only viable and consistent solution for treating such models. Note however that the original $\kappa$-Minkowski space, $[\hat{x}_0,\hat{x}_i]=i\kappa^{-1} \hat{x}_i,\ [\hat{x}_i,\hat{x}_j]=0$, which we consider in the present study, does not fit in that description and breaks the classical relativity principle. This leads us to the other approach widely studied in the literature, namely the extension of the usual notion of Lie algebra symmetries to the one of (deformed) Hopf algebra symmetries aiming to encode the new (canonical) symmetries for the quantum spacetimes. This point of view is motivated by the fact that, in the commutative case, the Minkowski spacetime can be regarded as the homogeneous space the Poincar\'e symmetry group acts on transitively. Hence, a deformation of the former should (in principle) implies a deformation of the latter and vice versa. This idea underlies the original derivation of $\kappa$-Minkowski as the homogeneous space associated to $\kappa$-Poincar\'e
\cite{Majid:1994}. Another interesting example (to put in perspective with
\cite{Doplicher:1994}) is given in
\cite{Chaichian:2004,Chaichian:2005}, where it is shown that the symmetries for the Moyal space can be obtained through formal (Drinfeld) twist deformation of the Lorentz sector of the Poincar\'e algebra while translations remain undeformed. Finally, similar considerations applied to $\mathbb{R}^3_\theta$ have been considered, e.g., in
\cite{Majid:1988,Majid:1991}. General discussions on the fate of the Poincar\'e symmetries within the context of noncommutative spacetimes can be found, e.g., in
\cite{Amelino:2002c}, and references therein.\smallskip

\paragraph{Outline.} The first part of this dissertation is devoted to the study of various classes of Lie algebra type noncommutative spaces. We construct families of star products associated with such spaces we may eventually use in the study of noncommutative field theory. The guideline underlying our constructions lies in abstract harmonic analysis on, and representation theory of, Lie groups. In Chapter \ref{sec-Minkowski}, we construct a star product associated with $\kappa$-Minkowski using standard tools from harmonic analysis and group C*-algebras. The derivation is performed in the spirit of the Weyl quantisation scheme leading to the celebrated Groenewold-Moyal product in quantum mechanics. Here, the Heisenberg group is replaced (in the 2-dimensional case) by the $ax+b$ group. The group and its algebra are characterised. In particular, we point out that the group is nonunimodular. The star product is defined as usual through the introduction of an invertible quantisation map which, in the Weyl quantisation scheme, is given (up to Fourier transform) by a bounded *-representation of the convolution algebra of the $ax+b$ group. The construction is then extended to any (spatial) dimensions, in which case the group is given by the semidirect product of two Abelian groups, namely $\mathcal{G}_{d+1}:=\mathbb{R}\ltimes\mathbb{R}^d$. The C*-algebra of fields modelling $\kappa$-Minkowski is presented. In particular, due to the nonunimodularity of $\mathcal{G}_{d+1}$, the natural involution on $\kappa$-Minkowski (\textit{i.e.} compatible with the structure of group C*-algebra) does not coincide with the ordinary complex conjugation. This plays an important role in the construction of action functionals aiming to describe the dynamics of $\mathbb{C}$-valued interacting scalar fields on $\kappa$-Minkowski background. Finally, the relation between $\kappa$-Minkowski and the $\kappa$-Poincar\'{e} algebra is recalled, emphasising the canonical action of the latter on the former, which can be interpreted as the action of a symmetry group on its homogeneous space. Basics on abstract harmonic analysis, and $\kappa$-Poincar\'{e}, are collected in Appendix \ref{sap-harmonic}, and Appendix \ref{sap-poincare}, respectively.\smallskip

In Chapter \ref{sec-products}, we consider various families of quantum spaces whose algebras of coordinate operators, say $\mathfrak{g}$, are semisimple Lie algebras. Note that, as a solvable Lie algebra, the $\kappa$-Minkowski algebra of coordinates does not pertain to this category. In particular, the groups associated with $\mathfrak{g}$, say $\mathcal{G}$, are unimodular. Therefore, a natural involution to be used in the construction of noncommutative field theories on these spaces is provided by the complex conjugation. The approach adopted to construct the star products associated with these spaces is slightly different from the one adopted in Chap. \ref{sec-Minkowski}. The construction is done in two steps. First, we show that the abstract Lie algebra $\mathfrak{g}$ can be conveniently represented as an algebra of differential operators acting on some Hilbert space of functions. In view of the important role played by the involutive structures in both the construction of the algebra of fields and the study of noncommutative field theory, we require the differential representations to define morphisms of *-algebras. It is worth mentioning that this requirement is not always taken into account in the literature. We show that under these assumptions (\textit{i.e.} the differential representations to be morphisms of Lie algebras preserving the involutions) the admissible representations are classified by a set of four differential equations we call ``master equations." Next, independently of the above set of master equations, we turn to the construction of star products which we define as usual by $f\star g:=Q^{-1}\big(Q(f)Q(g)\big)$. We show that the quantisation maps $Q$ -- which are regarded as differential operators when applied to functions -- are fully determined by their evaluation on the plane waves. Namely, $Q\big(e^{ip\cdot x}\big)$ which we call deformed plane waves. Then, we show that, keeping in mind the Weyl quantisation scheme, the deformed plane waves can be interpreted as projective representations of $\mathcal{G}$. Thus, we highlight the fact that, under this assumption, families of inequivalent star products (which merely results from the multiplication of deformed plane waves) are classified by the second cohomology group, $H^2(\mathcal{G},\mathcal{A})$, of $\mathcal{G}$ with value in an Abelian group $\mathcal{A}$. However, the study of quantum properties of noncommutative field theory usually necessitates an explicit expression for the star product. This amounts to choose a representative in one of the classes of equivalence belonging to $H^2(\mathcal{G},\mathcal{A})$, which is further facilitated by using the above mentioned differential representation of the algebra of coordinate operators.\smallskip

In Section \ref{sec-su2}, we apply this procedure to the case of $SU(2)$. We show that explicit expressions for the deformed plane waves can be obtained upon using $SO(3)$-equivariant differential *-representations of $\mathfrak{su}(2)$ together with polar decomposition of the deformed plane waves and the Wigner theorem for $SU(2)$. Making use of the master equations, we show that $SO(3)$-equivariant differential *-representations are labelled by three real functional of the Laplacian on $\mathbb{R}^3$. Finally, we find that the deformed planes waves are characterised by two (representation dependent) functions of the momenta defined themselves by two Volterra integrals. As a consequence, we show that the Wigner-Weyl quantisation map associated to the symmetric ordering, \textit{i.e.} such that $W(e^{ip\cdot x})=e^{ip\cdot \hat{x}}$, cannot be obtained within such approach. More precisely, it has to be modulated by some weight $W(e^{ip\cdot x})\to \omega(p)e^{ip\cdot \hat{x}}$. In Section \ref{sukont}, we select a specific representation among the family of $SO(3)$-equivariant differential *-representations, and show that the corresponding star product is equivalent to the Kontsevich product for the Poisson manifold dual to $\mathfrak{su}(2)$,  namely closed for the trace functional defined by the usual Lebesgue integral $\int d^3x$. We use this product in Chap. \ref{sap-ncftsu2} for studying noncommutative field theory on $\mathbb{R}^3_\theta$, a deformation of $\mathbb{R}^3$ of $\mathfrak{su}(2)$ noncommutativity.\smallskip

The second part of the dissertation is devoted to the study of the quantum properties of various model of noncommutative field theories built from the star products constructed in Part \ref{ch-ncst}. In Chapter \ref{ch-ncft}, we give the first comprehensive derivation of the one-loop order corrections to both the 2-point and 4-point functions for various models of $\kappa$-Poincar\'{e} invariant $\mathbb{C}$-valued scalar field theory with quartic interactions. In Section \ref{sec-action}, we discuss and analyse the properties a physically reasonable action functional aiming to describe the dynamics of scalar fields on $\kappa$-Minkowski background should satisfy. First of all, owing to the natural action of the $\kappa$-deformed Poincar\'{e} (Hopf) algebra on the $\kappa$-Minkowski space -- the $\kappa$-Poincar\'{e} algebra playing the role of the algebra of symmetries for the quantum space -- together with the important role played by the Poincar\'{e} algebra in ordinary quantum field theory, it is physically relevant to require the $\kappa$-Poincar\'{e} invariance of any physically reasonable action functional. This is supported by the fact that both the $\kappa$-Minkowski space and $\kappa$-Poincar\'{e} algebra tend, respectively, to the ordinary Minkowski spacetime and Poincar\'{e} algebra in the commutative (low energy) limit, $\kappa\to\infty$. It is known that the Lebesgue integral is $\kappa$-Poincar\'{e} invariant. That latter is not cyclic with respect to the star product constructed in Chap. \ref{sec-Minkowski}, however. We emphasise that the Lebesgue integral defines a twisted trace instead. But whenever there is a twisted trace, there is a related KMS condition. We show that the positive linear functional given by $\zeta(f):=\int d^4x f(x)$ actually defined a KMS weight on the C*-algebra of fields modelling $\kappa$-Minkowski, which is equivalent to have a KMS condition. The related modular group and Tomita modular operator are characterised. To summarise, to enforce the $\kappa$-Poincar\'{e} invariance of the action functional trades the cyclicity of the Lebesgue integral for a KMS condition. This new interpretation we give to the loss of cyclicity sheds new light on the possible role played by $\kappa$-Minkowski, $\kappa$-Poincar\'{e}, and related noncommutative field theories, in the description of physics at Planck scale. After recalling the initial motives for introducing the KMS condition in quantum statistical mechanics, we discuss the possible applications and implications of this KMS condition in the context of quantum gravity and Planckian physics. Next, decomposing the action functional into a kinetic term and interaction term, we discuss their admissible expressions. In particular, we choose the kinetic operator to be related to the square of some Dirac operators. We restrict our attention on polynomial interactions which tend to the usual $\vert\phi\vert^4$ model in the commutative limit. We find that there is essentially four inequivalent interactions.\smallskip

In Section \ref{sec-2point}, we compute the one-loop 2-point functions for two kinetic operators one being given by the first Casimir of the $\kappa$-Poincar\'{e} algebra, the other being given by the square of an $\mathcal{U}_\kappa(\text{iso}(4))$-equivariant Dirac operator. (A third case, for which the kinetic operator is given by the square of a modular Dirac operator, is also briefly discussed.) We show that, thanks to the relatively simple (integral) expression of the star product, we can identify noncommutative field theories with ordinary, albeit nonlocal, field theories. This enables us to use standard techniques from path integral quantisation and perturbation theory. The related material is recalled in Appendix \ref{sap-perturbation} for completeness. We find that one model has milder UV divergences than its commutative counterpart. The other model is found to diverge slightly worst. In both case, whenever the interaction considered is nonorientable, we find UV/IR mixing. In Section \ref{sec-4point}, we restrict our attention to the model with equivariant kinetic operator and orientable interactions. We compute the corresponding 4-point function at one-loop order and show that all the contributions are UV finite with no IR singularity. One-loop renormalisation is briefly discussed.\smallskip

In Chapter \ref{sap-ncftsu2}, we study the quantum behaviour of both real and complex scalar field (both massive and massless) theories, with quartic interactions, built on $\mathbb{R}^3_\theta$. Using the Kontsevich product derived in $\S$\ref{sec-kont}, we exhibit two type of contributions to the one-loop 2-point function. In all of the cases, we find that the contributions are UV and IR finite, the parameter $\theta$ encoding the deformation playing the role of a natural UV and IR cutoff for the models. No UV/IR mixing is found within these models.\smallskip

We finally summarise and comment our result in the last part of the dissertation. 
\part{Noncommutative spacetimes and star products.}\label{ch-ncst}
In the spirit of the Gelfand-Naimark theorem, it is common to define quantum spaces as noncommutative, associative, C*-algebras of functions. One natural way to construct such algebra amounts to deform, in some \textit{smooth way}, (the algebraic structures of) the commutative algebra of $\mathbb{C}$-valued smooth functions associated with the classical spacetime,\footnote{Here, by ``classical spacetime" we mean smooth manifold $\mathcal{M}$.} say $\mathcal{A}:=\big(\mathfrak{F}(\mathcal{M}),\cdot\big)$, into a noncommutative algebra of $\mathbb{C}$-valued smooth functions, $\mathcal{A}_\theta:=\big(\mathfrak{F}(\mathcal{M}),\star_\theta\big)$.\footnote{We insist on the fact that, for a real deformation parameter $\theta>0$, only the algebraic structure of $\mathcal{A}$ is deformed, \textit{i.e.} $\cdot\mapsto\star_\theta$, while the structure of linear space is not. In other words, the fields remain classical (in both the sense of $\hbar$ and $\theta$), only the way they compose is modified. Moreover, note that the space of functions, $\mathfrak{F}(\mathcal{M})$, we start from may differ from one case to the other, depending on both the nature of $\mathcal{M}$ and the nature of the deformation, \textit{i.e.} which type of noncommutative algebra of coordinate operators we consider. In practice, a good starting point is to consider the space of Schwartz functions, then to enlarge this space by successive completions; see below.} Formally, the associative (noncommutative) star product, resulting from the deformation of $\mathcal{A}$, may be defined in such a way that it differs from the usual pointwise product by terms of order, at least, $\theta$; namely
\begin{equation}\label{star-expansion}
(f\star_\theta g)(x)=(f\cdot g)(x)+\mathcal{O}(\theta),
\end{equation}
with $(f\cdot g)(x)=f(x)g(x)$, for any $f,g\in\mathfrak{F}(\mathcal{M})$. Furthermore, it is to expect the involutive structure, as well as the C*-norm, on $\mathcal{A}$ to be deformed in accordance with the star product so that the whole structure of C*-algebra is transferred to $\mathcal{A}_\theta$. In any case, the noncommutative algebra $\mathcal{A}_\theta$ is required to reduce to the commutative algebra $\mathcal{A}$ we start from when taking the commutative limit, $\theta\to0$. In particular, this implies $(f\star_\theta g)(x)\to f(x)g(x)$ when $\theta\to0$, a condition which is formally satisfied when considering star product of the form of \eqref{star-expansion}.\smallskip

This latter requirement simply reflects the desire to recover a known physical theory in some limit of the noncommutative model, for example when the typical length scale of the physical system under consideration becomes large compared with $\theta$. In a quantum gravity prospect -- identifying the deformation parameter $\theta$ with the Planck length $\ell_p:=\sqrt{\hbar G/c^3}$ -- this requirement would be to recover either a known quantum field theory in the limit $G\to0$ or Einstein's theory of gravitation in the limit $\hbar\to0$, both limits corresponding to $\ell_p\to0$. From a phenomenological point of view, this approach enables us to balance the lack of experimental/observational data to guide the construction of new physical models. Moreover, the semicommutative limit, namely (small) departures from the commutative theory, can be easily studied (in a controlled manner) upon expending the star product in power of the deformation parameter. \smallskip

Of course, the way to deform $\mathcal{A}$ is not unique, and to one commutative algebra may correspond several, possibly inequivalent, noncommutative algebras $\mathcal{A}_\theta$, each algebra being characterised by a specific star product. The characterisation and classification of such deformation procedures have been the subject of many mathematical studies and led to a huge amount of literature. Once more, the initial motivation finds its origin in quantum mechanics, more precisely the quantisation of classical systems, and dates back to the early works of H. Weyl
\cite{Weyl:1927}, E. Wigner
\cite{Wigner:1932}, and J. von Neumann
\cite{vNeumann:1931}, which ultimately led to the celebrated Groenewold-Moyal star product
\cite{Groenewold:1946,Moyal:1949}. It is instructive to recall the main steps leading to the Groenewold-Moyal product as we are going to use a similar approach to construct star products associated with $\kappa$-Minkowski.\medskip

Recall that one important feature of this scheme is the notion of ``twisted convolution" of two functions on (the 2-dimensional) phase space, that we denote by $f\hat{\circ}g$, whose explicit expression was first given in
\cite{vNeumann:1931}; for a more recent treatment see, e.g., \cite{Hennings:2010}.\\
This product is defined by
\begin{subequations}\label{Weyl-Heisenberg}
\begin{equation}
W(f\hat{\circ}g):=W(f)W(g),\ \forall f,g\in L^1(\mathbb{R}^2),
\end{equation}
where the Weyl operator is given by $W(f):=\int d\xi_1d\xi_2 f(\xi_1,\xi_2) e^{i(\xi_1\hat{p}+\xi_2\hat{q})}$, in which the unitary operator in the integrand can be viewed as (a unitary representation of) an element of the unimodular simply connected Heisenberg group $\mathcal{G}$ obtained by exponentiating the Heisenberg algebra, \textit{i.e.} a central extension of the Abelian Lie algebra, $[\hat{q},\hat{p}]=0$, say $[\hat{q},\hat{p}]=i\hbar$, where $\hbar$ is a central element.\\
From this, it follows the expression of the Groenewold-Moyal product defining the deformation of $\mathbb{R}^2$. It is defined by
\begin{equation}
f\star_\hbar g:=\mathcal{F}^{-1}\big(\mathcal{F}f\hat{\circ}\mathcal{F}g\big),
\end{equation}
where the invertible Wigner-Weyl quantisation map is given by
\begin{equation}
Q:L^2(\mathbb{R}^2)\to\mathcal{L}\big(L^2(\mathbb{R})\big),\ Q(f):=W(\mathcal{F}f),
\end{equation}
\end{subequations}
and $\mathcal{F}$ is the ordinary Fourier transform on $\mathbb{R}^2$.\footnote{Note that, within the above derivation, we have implicitly identified functions on the Heisenberg group with functions on $\mathbb{R}^2$. This simply reflects the fact that we have parametrised, in term of real parameters, the group elements of the Heisenberg group.}\\
In fact, this picture fits perfectly within the theory of harmonic analysis on locally compact groups, the relevant group being, in the present case, provided by the Heisenberg group. Moreover, the Wigner-Weyl quantisation map provides an isometric isomorphism between the group C*-algebra of the Heisenberg group and the C*-algebra of Hilbert-Schmidt operator on $L^2(\mathbb{R})$, thus ensuring the equivalence between the probabilistic (star product) interpretation of quantum mechanics and the (more conventional) Hilbert space approach. Basics on harmonic analysis are recalled in Appendix \ref{sap-harmonic} for completeness.\medskip

The generalisation of the above quantisation scheme to spaces of functions equipped with more general Poisson structure than the classical symplectic phase space led to the theory of deformation quantisation; for a review see, e.g.,
\cite{Sternheimer:1998,Dito:2002}, and references therein. Note that, in this case, the star product \eqref{star-expansion} takes the form
\begin{equation}\label{star-poisson}
(f\star_\theta g)(x)=f(x)g(x)+\frac{i\theta}{2}\lbrace f,g\rbrace+\mathcal{O}(\theta^2),
\end{equation}
where $\lbrace f,g\rbrace$ is the Poisson bracket characterising the Poisson manifold to be deformed. It is worth noting that, as pointed out in
\cite{Rieffel:1990a}, to arbitrary star products of the form \eqref{star-poisson} does not always correspond a closed (integral) expression, but rather a formal (not necessarily convergent) power series in the deformation parameter $\theta$. Therefore, it may exist deformations for which $f\star_\theta g\notin\mathfrak{F}(\mathcal{M})$, \textit{i.e.} the algebra $\mathcal{A}_\theta$ is not closed under star multiplication, which is troublesome for defining reasonable notion of quantum spacetime. In addition, star products which have integral expression are easier to manipulate in (noncommutative) quantum field theory than formal power series. In view of eq. \eqref{Weyl-Heisenberg}, see also Appendix \ref{sap-harmonic}, this problem can in principle be bypassed, at least in the case the algebra $\mathfrak{g}$ of coordinate operators is of Lie algebra type, by adapting (up to technicalities) the above Weyl quantisation scheme to the harmonic analysis on the Lie group $\mathcal{G}$, related to $\mathfrak{g}$, in order to derive a suitable expression for the star product. A suitable definition for the C*-algebra of fields modelling the quantum spacetime, which is characterised by $\mathfrak{g}$, is then provided by the group C*-algebra $C^*(\mathcal{G})$ of $\mathcal{G}$. Early considerations supporting this interpretation can be found in
\cite{Rieffel:1990b}, see also
\cite{Majid:1988,Majid:1990,Majid:1994}.\bigskip

In Chapter \ref{sec-Minkowski}, we extend the above Weyl quantisation scheme to the construction of a star product associated with $\kappa$-Minkowski. This is achieved by replacing the Heisenberg group by the nonunimodular $ax+b$ group. Note that, this approach has already been used in
\cite{Durhuus:2013} to construct a star product for $\kappa$-Minkowski. This product will be used in Part \ref{part-ncft} to construct action functionals aiming to describe the dynamics of various families of 4-dimensional $\kappa$-Poincar\'{e} invariant interacting scalar field theory. We will see that, the relatively simple expression of this star product leads to very tractable expressions for the propagator and interaction potential characterising the NCFT. This will enable us to compute radiative corrections without resorting to expansions in $\kappa$. In another (independent) chapter, Chap. \ref{sec-products}, we present the construction of various families of star products in the case the Lie algebra of coordinate operators is semisimple. We show that the construction can conveniently be carried out by representing the abstract coordinate operators as differential operators acting on some Hilbert space. Families of quantisation maps, giving rise to the star products, are characterised by their action on the plane waves together with polar decomposition of operator. We finally emphasise that inequivalent families of star products can be obtain from considerations stemming from group cohomology. Although differential representations have been widely used in the literature (much more than the group algebraic approach we adopt in Chap. \ref{sec-Minkowski}), it is worth mentioning that many of these studies do not care about the preservation of the involutive structures while constructing star products. In contrast, in our derivation particular attention is drawn to the preservation of the various involutive structures underlying the quantum spaces, this all along the various steps leading to the expressions of the star products. It seems to us that this requirement is of primary importance, for both mathematical and physical purposes, in order to prevent any inconsistency in the derivation of the star products and the algebras of fields. This will be discussed in more details in Chap. \ref{sec-products}. 
\chapter{\texorpdfstring{$\kappa$}{k}-Minkowski as a group algebra.}\label{sec-Minkowski}
In this chapter, we focus on the specific example of $\kappa$-Minkowski, whose algebra $\mathfrak{m}_\kappa$ of coordinate operators is given, in the $(d+1)$-dimensional case, by
\begin{equation}\label{kappa-Lie}
[\hat{x}_0,\hat{x}_i]=\frac{i}{\kappa}\hat{x}_i,\ [\hat{x}_i,\hat{x}_j]=0,\ i,j=1, ...,d,
\end{equation}
where $\kappa>0$ is a real, dimensionful, parameter labelling the deformation, and the coordinates $\hat{x}_\mu$ are assumed to be selfadjoint operators acting on some Hilbert space $\mathcal{H}$.\smallskip

The real interest of $\kappa$-Minkowski, among all of the other (currently known) possible choices of quantum spacetimes, lies in its relation to the $\kappa$-Poincar\'{e} algebra
\begin{equation}\label{bicrossproduct}
\mathfrak{P}_\kappa=\mathcal{U}\big(\mathfrak{so}(1,3)\big)\triangleright\!\!\!\blacktriangleleft \mathfrak{T}_\kappa.
\end{equation}
The $\kappa$-Poincar\'{e} algebra has been originally obtained in
\cite{Lukierski:1991,Lukierski:1992} by In\"{o}n\"{u}-Wigner contraction
\cite{Inonu:1953} of $SO_q(3,2)$.\footnote{With $SO_q(3,2)$ a q-deformation of the classical anti-de Sitter space.} Another presentation of $\mathfrak{P}_\kappa$ amounts to exhibit its bicrossproduct structure
\cite{Majid:1994}, eq. \eqref{bicrossproduct}, which merely reflects the fact that deforming the action of the Lorentz sector on the translations induces a backreaction of these latter on the former. Within this picture, $\kappa$-Minkowski naturally arises as the Hopf dual of the translation Hopf subalgebra $\mathfrak{T}_\kappa\subseteq\mathfrak{P}_\kappa$ which acts covariantly on it, and, in fact, because of the bicrossproduct structure of $\mathfrak{P}_\kappa$, the whole $\kappa$-Poincar\'{e} algebra can be shown to act covariantly on $\mathfrak{m}_\kappa$. For more details on this derivation, as well as algebraic properties of $\mathfrak{P}_\kappa$, see Appendix \ref{sap-poincare}, and references therein. Therefore, the $\kappa$-Poincar\'{e} algebra can be interpreted as describing the (quantum) symmetries of $\kappa$-Minkowski. This implies that not only the Minkowski spacetime is transposed at the noncommutative level, but the whole (dual) picture $\lbrace\text{spacetime}+\text{symmetries}\rbrace$, with the nice property that $\kappa$-Minkowski (resp. $\kappa$-Poincar\'{e}) tends to the ordinary Minkowski spacetime (resp. Poincar\'{e} algebra) in the commutative (low energy) limit $\kappa\to\infty$.\smallskip

In addition to this, it is worth mentioning that, the deformation parameter $\kappa$ is of mass dimension. Hence, this deformation, eq. \eqref{kappa-Lie} and \eqref{bicrossproduct}, provides a natural energy scale at which the effects thereof, \textit{i.e.} departure from the commutative, should become significant. This scale has been interpreted in the literature as being the Planck scale or, at least, some intermediate quantum gravity scale. One of the first (successful) attempt to (provide a framework to) consistently incorporate such dimensionful (observer independent) parameter into a physical theory, in a way compatible with the relativity principle, is provided by the theory of doubly special relativity
\cite{Amelino:2000,Amelino:2001}. Of course, due to relativistic effects such as length contractions, it is not possible to (naively) incorporate a fundamental scale of mass or length dimension within the framework of Einstein's theory of special relativity. One way to proceed, then, consists in deforming the relativity symmetry group of special relativity, hence the ordinary dispersion relation, in such a way that both the speed of light $c$ and the new (quantum gravity) scale are now observer independent. It turns out
\cite{Kowalski:2001,Bruno:2001} that this can be achieved in the framework of $\kappa$-Poincar\'{e} Hopf algebra, with the benefit that the ordinary symmetry group, as well as dispersion relations, and relativity principle, can be recovered in a smooth way by simply taking the limit $\kappa\to\infty$ thanks to the construction of quantum groups. Finally, note that, geometrically, the $\kappa$-deformation of the Poincar\'{e} algebra implies that the corresponding (group) manifold of energy-momentum is curved, the curvature being proportional to $\kappa$, hence reflecting the non-Abelian structure of the group whose vector fields are (locally) provided by the $\kappa$-Minkowski algebra of coordinate operators
\cite{Majid:1994,Kowalski:2013}. For a recent comprehensive review on the development of $\kappa$-Poincar\'{e}, $\kappa$-Minkowski and their possible physical implications see, e.g.,
\cite{Borowiec:2010,Kowalski:2017}, and references therein. We will come back to these points later on when constructing $\kappa$-Poincar\'{e} invariant action functional aiming to describe the dynamics of interacting scalar fields on $\kappa$-Minkowski background; see Chap. \ref{ch-ncft}. For the moment, let us restrict our attention on eq. \eqref{kappa-Lie} and construct the corresponding C*-algebra of fields, endowed with star product, modelling the $\kappa$-Minkowski space.\smallskip

As already mentioned, a convenient presentation of the $\kappa$-Minkowski space is obtained by exploiting standard objects from the framework of harmonic analysis and group C*-algebras. This approach, which has been used in
\cite{Durhuus:2013} to derive a star product for the 2-dimensional $\kappa$-Minkowski space, is the one we mainly follow in this section. It can be mentioned that, besides the use of this approach to derive the celebrated Groenewold-Moyal product, eq. \eqref{Weyl-Heisenberg}, associated with the Moyal plan, this framework has also been used in recent studies on $\mathbb{R}^3_\theta$, a deformation of $\mathbb{R}^3$, related to the convolution algebra of the compact Lie group $SU(2)$; see, e.g., in
\cite{JCW:2015,JCW:2016}, and in $\S$\ref{sec-su2}, for more details. The material presented in this chapter is published in
\cite{moi:2018a}. Additional details on harmonic analysis on locally compact group can be found in Appendix \ref{sap-harmonic}.
\section{Characterisation of the convolution algebra.}\label{sec-convolution}
Let us define, for any $n\in\mathbb{N}$, the Lie subalgebra $\mathfrak{m}_\kappa^{(n+1)}:=[\mathfrak{m}_\kappa^{(n)},\mathfrak{m}_\kappa^{(n)}]\subseteq\mathfrak{m}_\kappa$, with initial condition $\mathfrak{m}_\kappa^{(0)}:=\mathfrak{m}_\kappa$. From eq. \eqref{kappa-Lie}, we easily infer that the derived Lie algebra $\mathfrak{m}_\kappa^{(1)}=\text{span}(\hat{x}_i)_{i=1,...,d}$, as vector space, is a nilpotent ideal of $\mathfrak{m}_\kappa$, while $\mathfrak{m}_\kappa^{(n)}=0$, $\forall n\geq 2$. Hence, the derived series
\begin{subequations}\label{kappa-alg}
\begin{equation}
\mathfrak{m}_\kappa\supseteq\mathfrak{m}_\kappa^{(1)}\supseteq0,
\end{equation}
forms an elementary sequence, indicating that the Lie algebra $\mathfrak{m}_\kappa$ is (split) solvable.\\
In addition, $\mathfrak{m}_\kappa$ decomposes into a semidirect product of Lie algebras, \textit{i.e.}
\begin{equation}\label{alg-decomposition}
\mathfrak{m}_\kappa=\mathbb{R}\hat{x}_0\oplus_{d\tau}\mathfrak{m}_\kappa^{(1)},
\end{equation}
\end{subequations}
where the Lie algebra homomorphism $d\tau:\mathbb{R}\hat{x}_0\to\text{Der}\ \mathfrak{m}_\kappa^{(1)}$ characterises the (adjoint) action of $\hat{x}_0$ on the ``spacelike directions" $\hat{x}_i$.\footnote{See, e.g, propositions 1.22-3 in ref.
\cite{Knapp:2002}.}\smallskip

It turns out that the group $\mathcal{G}_{d+1}$, obtained by exponentiating $\mathfrak{m}_\kappa$, is a solvable, simply connected, Lie group, diffeomorphic to an Euclidean space.\footnote{See, e.g., theorem 5.9 in ref.
\cite{Onishchik:1993}.} Hence, $\mathcal{G}$ is amenable.\footnote{See, e.g., theorem 2.3.3 in ref.
\cite{Greenleaf:1969}.} Moreover, because of the very structure of $\mathfrak{m}_\kappa$, eq. \eqref{kappa-alg}, $\mathcal{G}_{d+1}$ satisfies the sequence
\begin{subequations}
\begin{equation}
\mathcal{G}_{d+1}\supseteq\mathcal{G}_{d+1}^{(1)}\supseteq\lbrace1\rbrace,
\end{equation}
entailing the semidirect product structure for $\mathcal{G}_{d+1}$, \textit{i.e.}
\begin{equation}\label{grp-decomposition}
\mathcal{G}_{d+1}=\mathbb{R}\ltimes_\tau\mathcal{G}_{d+1}^{(1)},
\end{equation}
\end{subequations}
where $\mathcal{G}^{(1)}_{d+1}=\mathbb{R}^d$ is normal in $\mathcal{G}_{d+1}$.\footnote{See, e.g., corollary 1.126 in ref.
\cite{Knapp:2002}.} Recall that, a group, say $A$, is said to act on another group, say $B$, by automorphism if there exists a smooth map $\tau:A\times B\to B$ such that $g\mapsto\tau(g,\cdot)$ is a group homomorphism between $A$ and $\text{Aut}B$, the group of automorphisms of $B$. Hence, we call semidirect product of $A$ on $B$, and we write $C:=A\ltimes_\tau B$, the Lie group, with Cartesian product topology, whose composition law and inverse are given, for any $a_i\in A$ and $b_i\in B$, by
\begin{equation}\label{group-law2}
(a_1,b_1)(a_2,b_2)=\big(a_1a_2,\tau(a_1,b_2)b_1\big),\ (a_3,b_3)^{-1}=\big(a_3^{-1},\tau(a_3^{-1},b_3^{-1})\big).
\end{equation}
If $C=BA$ with $A,B\subseteq C$ as subgroups, $B$ normal in $C$, and $A\cap B=\lbrace1\rbrace$, then $\tau$ is nothing but the adjoint action of $A$ on $B$, namely $\tau(a,b)=aba^{-1}$; conditions that are fulfilled by the group, eq. \eqref{grp-decomposition}, underlying the construction of $\kappa$-Minkowski.\smallskip

Note that the decompositions \eqref{alg-decomposition} and \eqref{grp-decomposition} hold independently of the dimension $d$ of $\mathfrak{m}_\kappa^{(1)}$. Therefore, it is convenient to begin with the construction of the star product for $d=1$, then extend the construction to the desired dimension.\smallskip

Let us set $d=1$. In two dimensions, there exists, up to isomorphism, only one non-Abelian Lie algebra of which eq. \eqref{Lie-bracket} provides a specific choice of basis. The corresponding group is given by $\mathcal{G}_{2}=\mathbb{R}\ltimes_\tau\mathbb{R}$, isomorphic to the $ax+b$ group widely studied in the mathematical literature. For basic mathematical details see, e.g.,
\cite{Khalil:1974,Williams:2007}, and references therein. In view of the decomposition \eqref{grp-decomposition}, this group can be conveniently characterised by 
\begin{equation}\label{group-elem}
W(p^0,p^1):=e^{ip^1\hat{x}_1}e^{ip^0\hat{x}_0},
\end{equation} 
where the parameters $p^0,p^1\in\mathbb{R}$ will be interpreted as (Fourier) momenta in due time. Note that, the above group elements, eq. \eqref{group-elem}, can be related to the more conventional exponential form of the Lie algebra \eqref{kappa-Lie} through mere redefinition of $p^1$. To see this, use the simplified Baker-Campbell-Hausdorff formula $e^Xe^Y=e^{\lambda(u)X+Y}$, valid whenever $[X,Y]=uX$, to obtain
\begin{equation}\label{group-elem2}
W(p^0,p^1)=e^{i(p^0\hat{x}_0+\lambda(p^0/\kappa)p^1\hat{x}_1)},
\end{equation}
where $\lambda(u)=ue^u(e^u-1)^{-1}$; see, e.g.,
\cite{Vanbrunt:2015}. For the ensuing computations, the group elements are easier to manipulate with the parametrisation \eqref{group-elem} than \eqref{group-elem2}, however.\smallskip

Now, using the identity $e^Xe^Y=e^{Y}e^{e^uX}$, which holds true whenever $[X,Y]=uX$, we easily obtain the composition law for $\mathcal{G}_2$, which is given by
\begin{subequations}
\begin{equation}\label{group-law}
W(p^0,p^1)W(q^0,q^1)=W(p^0+q^0,p^1+e^{-p^0/\kappa}q^1).
\end{equation}
It follows that the unit element and inverse are given by
\begin{equation}\label{group-unit-inverse}
1=W(0,0),\ W^{-1}(p^0,p^1)=W(-p^0,-e^{p^0/\kappa}p^1).
\end{equation}
\end{subequations}
Equation \eqref{group-law} provides us with what will be identified with the energy-momentum composition law when we will consider NCFT on $\kappa$-Minkowski background in Chap. \ref{ch-ncft}. This essentially reflects the nontrivial coproduct structure of the $\kappa$-Poincar\'{e} algebra; see eq. \eqref{hopf1}. The more usual composition law for the $ax+b$ group is easily recovered by representing the group elements \eqref{group-elem} as upper triangular matrices
 \begin{equation}\label{grp-matrix}
W(p^0,p^1)\mapsto (a,b):=\begin{pmatrix}a&b\\0&1\end{pmatrix},
\end{equation}
such that $a:=e^{-p^0/\kappa}$ and $b:=p^1$. Then, from the decomposition
\begin{equation}
\begin{pmatrix}a&b\\0&1\end{pmatrix}=\begin{pmatrix}1&b\\0&1\end{pmatrix}\begin{pmatrix}a&0\\0&1\end{pmatrix},
\end{equation}
we easily infer the semidirect product structure of the $ax+b$ group, namely $\mathcal{G}_2\cong BA$, with $A:=\big\lbrace (a,0),\ a>0\big\rbrace$ and $B:=\big\lbrace (1,b),\ b\in\mathbb{R}\big\rbrace$, together with group law \eqref{group-law2} and group action $\tau:A\times B\to B$ given by $\big((a,0),(1,b)\big)\mapsto (a,0)(1,b)(a^{-1},0)$. In view of \eqref{group-law}, this latter (adjoint) action is reflected at the level of the parameters $p^\mu$ in
\begin{equation}\label{group-action}
\tau:\mathbb{R}\times\mathcal{G}^{(1)}_{2}\to\mathcal{G}^{(1)}_{2},\ \tau(p^0,p^1)=e^{-p^0/\kappa}p^1,
\end{equation}
with $\mathcal{G}^{(1)}_{2}=\mathbb{R}$.\smallskip

According to the results of Appendix \ref{sap-harmonic}, the convolution algebra of $\mathcal{G}_2$ is characterised, for any functions $f,g\in L^1(\mathbb{R}^2)$, by the product
\begin{subequations}\label{set-convolution-2d}
\begin{equation}
(f\hat{\circ}g)(p_1^0,p_1^1)=\int_{\mathbb{R}^2} f\big(p_1^0-p^0_2,p^1_1-p^1_2e^{-(p_1^0-p^0_2)/\kappa}\big)g\big(p^0_2,p^1_2\big)\ dp_2^0dp_2^1,
\end{equation}
together with the involution\footnote{Note that the definition of the involution, eq. \eqref{involution2-ap}, has to be slightly adjusted when working with right invariant Haar measure, see eq. \eqref{broll1}. In this case we have $f^*(x):=\Delta_{\mathcal{G}}(x)\overline{f^\flat(x)}$.}
\begin{equation}\label{invol-Fourier}
f^*(p^0,p^1)=e^{p^0/\kappa}\bar{f}(-p^0,-e^{p^0/\kappa}p^1),
\end{equation}
and the modular function
\begin{equation}\label{function-modulaire-2d}
\Delta_{\mathcal{G}_2}(p^0,p^1)=e^{p^0/\kappa},
\end{equation}
\end{subequations}
where we have identified functions on $\mathcal{G}_2$ with functions on $\mathbb{R}^2$ in view of \eqref{grp-decomposition} and the parametrisation \eqref{group-elem}. In particular, the right  invariant Haar measure $d\nu(p^0,p^1)$ coincides with the usual Lebesgue measure on $\mathbb{R}^2$, while the left invariant Haar measure is given by $d\mu(p^0,p^1)=e^{p^0/\kappa}dp^0dp^1$. From now on, except otherwise stated, we shall work with the right  invariant measure.
\section{Weyl quantisation map and related star product.}
Let $\pi_U:\mathcal{G}_2\to\mathcal{B}(\mathcal{H}_\pi)$ be a (strongly continuous) unitary representation of $\mathcal{G}_2$ on some Hilbert space $\mathcal{H}_\pi$, and $\mathcal{B}(\mathcal{H}_\pi)$ be the C*-algebra of bounded operators on $\mathcal{H}_\pi$.\footnote{We can think, for example, to the right  regular representation $\pi_U:\mathcal{G}_2\to L^2(\mathbb{R}^2)$ defined by $(\pi_U(s)f)(t)=f(ts)$.} Accordingly, any representation of the convolution algebra defined, for any $f\in L^1(\mathbb{R}^2)$, by
\begin{equation}\label{rep-minkowski}
\pi:L^1(\mathbb{R}^2)\to \mathcal{B}(\mathcal{H}_\pi),\ \pi(f):=\int_{\mathbb{R}_2} f(p^0,p^1)\pi_U(p^0,p^1) dp^0dp^1,
\end{equation}
is a nondegenerate bounded *-representation.\footnote{In the case of $\pi_U$ is the right  regular representation, we have $\pi(f)g=f\hat{\circ}g$.} Indeed, let $\langle\cdot,\cdot\rangle$ denote the Hilbert product on $\mathcal{H}_\pi$, such that
\begin{equation}
\langle u,\pi(f)v \rangle=\int_{\mathcal{G}_2} f(s) \langle u,\pi_U(s)v \rangle d\nu(s),\ u,v\in\mathcal{H}_\pi, f\in L^1(\mathcal{G}_2).
\end{equation}
On the one hand, we have
\begin{subequations}\label{broll1}
\begin{equation}
\langle u,\pi(f^*)v \rangle=\int_{\mathcal{G}_2} \Delta_{\mathcal{G}_2}(s)\bar{f}(s^{-1}) \langle u,\pi_U(s)v \rangle d\nu(s),
\end{equation}
while, on the other hand, $\langle u,\pi(f)^\dag v \rangle:=\langle \pi(f)u,v \rangle=\overline{\langle v,\pi(f)u \rangle}$ such that
\begin{equation}
\langle u,\pi(f)^\dag v \rangle=\int_{\mathcal{G}_2}\bar{f}(s) \langle \pi_U(s) u,v \rangle d\nu(s)=\int_{\mathcal{G}_2}\bar{f}(s) \langle u,\pi_U(s^{-1})v \rangle d\nu(s),
\end{equation}
\end{subequations}
from which we conclude that $\pi(f)^\dag=\pi(f^*)$.\smallskip

We now turn to the construction of the quantisation map from which a star product associated with $\kappa$-Minkowski can formally be defined by $Q(f\star g):=Q(f)Q(g)$.\\
In the following, we shall denote by
\begin{equation}
\mathcal{F}f(p^0,p^1):=\int_{\mathbb{R}^2}f(x_0,x_1)e^{-i(p^0x_0+p^1x_1)}dx_0dx_1,
\end{equation}
the Fourier transform of $f\in L^1(\mathbb{R}^2)$, and by $S_c(\mathbb{R}^2)$ the spaces of Schwartz functions on $\mathbb{R}^2=\mathbb{R}\times\mathbb{R}$ with compact support in the first (\textit{i.e.} timelike) variable.\\
Following the line of the Weyl quantisation scheme, eq. \eqref{Weyl-Heisenberg}, we define $Q$ by 
\begin{equation}\label{Weyl-Minkowski}
Q(f):=\pi(\mathcal{F}f),\ \forall f\in L^1(\mathbb{R}^2)\cap \mathcal{F}^{-1}\big(L^1(\mathbb{R}^2)\big),
\end{equation}
where $\pi$ is a representation given by eq. \eqref{rep-minkowski}; see, e.g., in ref.
\cite{Durhuus:2013,moi:2018a}. Notice that, in view of eq. \eqref{group-elem}, the functions appearing in eq. \eqref{set-convolution-2d}, and \eqref{rep-minkowski}, are interpreted as Fourier transforms of functions of spacetime coordinates. This interpretation is supported by the fact that, taking the formal commutative limit, $\kappa\to\infty$, in eq. \eqref{group-elem}, and \eqref{set-convolution-2d}, we recover all of the usual notions of plane waves, convolution and involution. Hence the occurrence of $\mathcal{F}f$ in the right-hand-side of eq. \eqref{Weyl-Minkowski}.\\
Requiring $Q$ to define a morphism of *-algebras, we can write
\begin{subequations}
\begin{align}
&Q(f\star g)=Q(f)Q(g)=\pi(\mathcal{F}f)\pi(\mathcal{F}g)=\pi\big(\mathcal{F}f\hat{\circ}\mathcal{F}g\big),\label{q-star}\\
&Q(f^\ddagger)=\pi(\mathcal{F}f^*)
\end{align}
\end{subequations}
identifying eq. \eqref{q-star} with $Q(f\star g)=\pi\big(\mathcal{F}(f\star g)\big)$, we finally obtained the expressions for the star product and the involution
\begin{subequations}\label{star-general}
\begin{align}
&f\star g=\mathcal{F}^{-1}\big(\mathcal{F}f\hat{\circ}\mathcal{F}g\big),\\
&f^\ddagger=\mathcal{F}^{-1}\big(\mathcal{F}f^*\big),\label{invol-general}
\end{align}
\end{subequations}
where $\mathcal{F}^{-1}$ is the inverse Fourier transform on $\mathbb{R}^2$. Observe that both the star product and the involution are representation independent despite the fact that $Q$ depends on $\pi$.\smallskip

Now, upon combining the explicit expressions for the convolution and involution, eq. \eqref{set-convolution-2d}, with eq. \eqref{star-general}, we find that, for any $f,g\in\mathcal{F}^{-1}\big(S_c(\mathbb{R}^2)\big)$,
\begin{subequations}\label{involstar-2d}
\begin{align}
&(f\star g)(x_0,x_1)=\int \frac{dp^0}{2\pi} dy_0\ e^{-iy_0p^0}f(x_0+y_0,x_1)g(x_0,e^{-p^0/\kappa}x_1),\label{star-2d}\\
&f^\ddagger(x_0,x_1)= \int \frac{dp^0}{2\pi} dy_0\ e^{-iy_0p^0}{\bar{f}}(x_0+y_0,e^{-p^0/\kappa}x_1),\label{invol-2d}
\end{align}
\end{subequations}
with $f\star g, f^\ddagger\in\mathcal{F}(\mathcal{S}_c)$, which coincide with the star product and involution of
\cite{Durhuus:2013}.\smallskip

Before proceeding to the extension of the above results to $d=3$, it is worth mentioning that it as been shown in ref.
\cite{Durhuus:2013} that eq. \eqref{involstar-2d} extend to (a subalgebra of) the multiplier algebra $\mathcal{N}_c(\mathbb{R}^2)$ of $\mathcal{F}^{-1}\big(S_c(\mathbb{R}^2)\big)$ involving the smooth functions on $\mathbb{R}^2$, with compact support in the first variable, which satisfy standard polynomial bounds, together with all their derivatives.\footnote{More precisely, any $f\in \mathcal{N}_c(\mathbb{R}^2)$ satisfies polynomial bounds of the form
\begin{equation}
\vert\partial_0^n\partial_1^m f(p^0,p^1)\vert\leq c_{n,m}(1+\vert p^0\vert)^{N_n}(1+\vert p^1\vert)^{M_{n,m}},\ n,m\in\mathbb{N},
\end{equation}
where $N_n$, $M_{n,m}$, and $c_{n,m}$ are some constants, with $c_{n,m}:=0$; for more details see ref.
\cite{Durhuus:2013}.
} In particular, $x_0$, $x_1$ and the unit function belong to $\mathcal{N}_c(\mathbb{R}^2)$. Therefore, from \eqref{star-2d} and \eqref{invol-2d}, we easily obtain
\begin{equation}
x_0\star x_1=x_0x_1+\frac{i}{\kappa}x_1,\ x_1\star x_0=x_0x_1,\ x_\mu^\ddagger=x_\mu,\ \mu=0,1,
\end{equation}
consistent with the defining relation \eqref{kappa-Lie} for $d=1$.\smallskip

In view of eq. \eqref{alg-decomposition} and \eqref{grp-decomposition}, the extension of the above construction to the 4-dimensional ($d=3$) case is straightforward and merely amounts to substitute $p^1$ with $\vec{p}$ in the various above expressions. Explicitly, we have $\mathcal{G}_4=\mathbb{R}\ltimes_\tau\mathbb{R}^3$ with $\tau(p^0,\vec{p})=e^{-p^0/\kappa}\vec{p}$ and
\begin{subequations}\label{group4d-parametrization}
\begin{equation}
W(p^0,\vec{p}):=e^{i\vec{p}\cdot \vec{x}}e^{ip^0x_0}.
\end{equation}
The group law \eqref{group-law} becomes
\begin{equation}
W(p^0,\vec{p})W(q^0,\vec{q})=W(p^0+q^0,\vec{p}+e^{-p^0/\kappa}\vec{q}),
\end{equation}
while unit and inverse \eqref{group-unit-inverse} are now given by 
\begin{equation}
1=W(0,\vec{0}),\ W^{-1}(p^0,\vec{p})=W(-p^0,-e^{p^0/\kappa}\vec{p}).
\end{equation}
\end{subequations}
Then, the construction leading to \eqref{involstar-2d} can be thoroughly reproduced, replacing $\mathbb{R}^2$ by $\mathbb{R}^4$ and \eqref{function-modulaire-2d} by
\begin{equation}\label{function-modulaire-4d}
\Delta_{\mathcal{G}_4}(p^0,\vec{p})=e^{3p^0/\kappa}.
\end{equation}
Setting for short $x:=(x_0,\vec{x})$, we obtain, for any functions $f,g\in\mathcal{F}^{-1}\big(\mathcal{S}_c(\mathbb{R}^4)\big)$,
\begin{subequations}\label{starinvol4d}
\begin{align}
&(f\star g)(x)=\int \frac{dp^0}{2\pi} dy_0\ e^{-iy_0p^0}f(x_0+y_0,\vec{x})g(x_0,e^{-p^0/\kappa}\vec{x}),\label{star-4d}\\
&f^\ddagger(x)= \int \frac{dp^0}{2\pi} dy_0\ e^{-iy_0p^0}{\bar{f}}(x_0+y_0,e^{-p^0/\kappa}\vec{x}),\label{invol-4d}
\end{align}
\end{subequations}
with $f\star g\in\mathcal{F}^{-1}\big(\mathcal{S}_c(\mathbb{R}^4)\big)$ and $f^\ddagger\in\mathcal{F}^{-1}\big(\mathcal{S}_c(\mathbb{R}^4)\big)$. Moreover, we can show that
\begin{equation}\label{antivolution}
(f\star g)^\ddagger=g^\ddagger\star f^\ddagger.
\end{equation}
For latter convenience, note that, thanks to the Paley-Wiener theorem, functions in $\mathcal{F}^{-1}\big(\mathcal{S}_c(\mathbb{R}^4)\big)$ are by construction analytic in the (timelike) variable $x_0$, being Fourier transforms of functions with compact support in the (timelike) variable $p^0$.\smallskip

Finally, it is instructive to get more insight on $C^*(\mathcal{G}_4)$, the group C*-algebra which models the (4-dimensional) $\kappa$-Minkowski space. In view of the discussion given at the end of Appendix \ref{sap-harmonic}, the completion of $L^1(\mathcal{G}_4)$ with respect to the norm related to the right regular representation on $L^2(\mathcal{G}_4)$ yields the reduced group C*-algebra $C_r^*(\mathcal{G}_4)$. Furthermore, since $\mathcal{G}_4$ is amenable, as any solvable Lie group, we have $C_r^*(\mathcal{G}_4)\cong C^*(\mathcal{G}_4)$, involving as dense *-subalgebra the set of Schwartz functions with compact support equipped with the above convolution product.\smallskip

For the ensuing construction of (4-dimensional) $\kappa$-Poincar\'{e} invariant NCFT, in Chap. \ref{ch-ncft}, it will be sufficient to consider the algebra $\mathcal{F}^{-1}\big(\mathcal{S}_c(\mathbb{R}^4)\big)$ which we denote by $\mathcal{M}_\kappa$.\smallskip

Until then, we now turn to the presentation of the construction of various families of star products associated with quantum spaces whose algebras of coordinate operators are given by semisimple Lie algebras.
\chapter{Other examples of quantum spaces.}\label{sec-products}
In the previous chapter, we have derived a star product for $\kappa$-Minkowski by adapting the Weyl quantisation scheme to the $ax+b$ group. This approach -- which easily extends to the construction of star products associated with any other quantum spacetime of Lie algebra type noncommutativity -- appears to be well suited to the study of NCFT as it provides us with (almost) all the needed ingredients for constructing an action functional aiming to describe the dynamics thereof. More precisely, let $\mathfrak{g}$ denote the noncommutative Lie algebra of coordinate operators characterising the quantum space and $\mathcal{G}$ the corresponding Lie group. We have seen that the C*-algebra of fields modelling the quantum space can conveniently be identified with the group C*-algebra $C^*(\mathcal{G})$ of $\mathcal{G}$. Natural candidates for a star product, an involution, and a measure of integration, are then canonically provided by the harmonic analysis on $\mathcal{G}$; see Chap. \ref{sec-Minkowski}.\smallskip

Another approach for constructing star products, based on the use of differential representations, has been (much more) extensively studied in the physics literature, often forgetting about the above group algebraic structure underlying the quantum space.\\
Although the star product is still defined through the introduction of an invertible linear morphism of algebras $Q$, \textit{i.e.} $f\star g:=Q^{-1}\big(Q(f)Q(g)\big)$, it is now assumed to associate with any function $f\in\mathfrak{F}(\mathbb{R}^n)$ a differential operator $Q(f)$ such that $(f\star g)(x):=Q(f)\triangleright g(x)$, where $\triangleright$ denotes the left action of operator. In addition, we require that $f\star1=1\star f=f$.\\
In practice, quantisation maps fulfilling the above requirements can be determined by first representing the abstract involutive algebra $\mathfrak{g}$ of coordinate operators as an involutive algebra $\mathfrak{h}$ of differential operators acting on some Hilbert space $\mathcal{H}$, then formally constructing the differential operators $Q(f)$ as functions of the generators of $\mathfrak{h}$. Thus, $Q$ is characterised by a choice of differential representation for $\mathfrak{g}$, this choice being itself strongly constrained by the requirement $Q(f)\triangleright 1=f$. In fact, because of the linearity of $Q$, it is in principle sufficient to determine the action of the quantisation map on the plane waves to fully determine the expression of the star product.\smallskip

The purpose of this section is to present a method for deriving expressions for the deformed plane waves, $E_k(\hat{x}):=Q(e^{ik\cdot x})$, appearing in the expression of the star product. Our derivation follows broadly the lines sketched above insisting on the preservation of the various involutive structures, however. Moreover, we informally make the link between the objects appearing within this approach and those of harmonic analysis, hence keeping track of the group algebraic structures underlying the quantum space. This leads us to the conclusion that families of inequivalent star products can be obtained, in principle, from group cohomological considerations. The material presented in this section is published in
\cite{moi:2017a}.
\section{Deformed plane waves, star products, and group cohomology.}\label{sec-dpw}
Let $\mathfrak{g}$ be a semisimple Lie algebra with Lie bracket
\begin{equation}\label{Lie-bracket}
[\hat{x}_\mu,\hat{x}_\nu]=i\theta C_{\mu \nu}^{\hspace{11pt} \rho}\hat{x}_\rho,
\end{equation}
where $\theta>0$ is a dimensionful, $[\theta]=L$, real parameter and $C_{\mu\nu\rho}\in\mathbb{R}$ are the structure constants determining $\mathfrak{g}$.\footnote{Note that, the algebra $\mathfrak{m}_\kappa$, eq. \eqref{kappa-Lie}, characterising $\kappa$-Minkowski is not semisimple. Indeed, as a solvable Lie algebra, $\text{rad}(\mathfrak{m}_\kappa)\neq0$. Recall that the radical $\text{rad}(\mathfrak{g})$ of a Lie algebra $\mathfrak{g}$ is the unique solvable ideal of $\mathfrak{g}$ containing every solvable ideals of $\mathfrak{g}$, while a (nonzero) Lie algebra $\mathfrak{g}$ is said to be semisimple if $\text{rad}(\mathfrak{g})=0$.} It follows that, the (connected component of the) corresponding Lie group $\mathcal{G}$ is semisimple, hence unimodular.\footnote{See, e.g., proposition 2.29 of ref.
\cite{Folland:1995}.} In view of eq. \eqref{involution2-ap} this means that a natural involution to be involved in the construction of an action functional for NCFT on such quantum space (\textit{i.e.} of semisimple Lie algebra type) is provided by the complex conjugation $f\mapsto \bar{f}$. In particular, the reality of the action functional can be conveniently controlled by introducing the Hilbert product on $\mathcal{H}=L^2(\mathbb{R}^n)$, $\langle f,g \rangle_2:=\int f(x) \bar{g}(x)d^n x$.
\subsection{Differential *-representations.}
Explicit expressions for the deformed plane waves can be obtained by representing the abstract coordinate operators as differential operators acting on some suitable Hilbert space $\mathcal{H}$. Let
\begin{equation}\label{diff-rep}
\pi:\mathfrak{g}\to\mathfrak{h}:=\pi(\mathfrak{g}),\ \ \pi(ab)=\pi(a)\pi(b),\ \forall a,b\in\mathfrak{g},
\end{equation}
be such differential representation. It follows that the Lie algebraic structure \eqref{Lie-bracket} of $\mathfrak{g}$ is automatically transferred to $\mathfrak{h}$, namely $[\pi(\hat{x}_\mu),\pi(\hat{x}_\nu)]=i\theta C_{\mu\nu\rho}\pi(\hat{x}^\rho)$. From now on, we shall write $\hat{x}_\mu$ for designating both the abstract operators and their representations $\pi(\hat{x}_\mu)$. An as natural as important requirement for this representation is to define a *-algebras' morphism, \textit{i.e.} to satisfy $\pi(a^{*})=\pi(a)^\dag$, $\forall a\in\mathfrak{g}$, where $^*$ ($^\dag$) denotes an involution of $\mathfrak{g}$ ($\mathfrak{h}$). This requirement ensures the involutive structures of the algebra of fields modelling the noncommutative space to be preserved under representation and, in particular, that selfadjoint (resp. unitary) operators made of $\hat{x}_\mu$ are represented as selfadjoint (resp. unitary) differential operators. Besides the purely algebraic motive, this requirement makes possible the implementation of reasonable reality conditions in the construction of an expression for the action functional when considering NCFT. \\
It is worth mentioning that this condition is not always satisfied in the literature.\smallskip

It is further convenient to consider representations of the form
\begin{equation}\label{diff-rep2}
\hat{x}_\mu=x^\nu\varphi_{\nu\mu}(\partial)+\chi_\mu(\partial),
\end{equation}
where the functionals $\varphi_{\nu\mu}$ and $\chi_\mu$ are viewed as formal expansions in the deformation parameter $\theta$.\footnote{In the Poincar\'{e}-Birkhoff-Witt basis, the functionals $\varphi_{\nu\mu}$ and $\chi_\mu$ take formally the form
\begin{equation}
\varphi_{\nu\mu}(\partial)=\sum (\tilde{\varphi}_{k_1\cdots k_n})_{\nu\mu}\theta^{\sum_i k_i}\partial_1^{k_1}\cdots\partial_n^{k_n},\ \ \chi_\mu(\partial)=\sum (\tilde{\chi}_{k_1\cdots k_n})_{\mu}\theta^{\sum_i k_i}\partial_1^{k_1}\cdots\partial_n^{k_n}.\nonumber
\end{equation}
} Requiring $\hat{x}_\mu\to x_\mu$ when $\theta\to0$, implies
\begin{equation}
\varphi_{\nu\mu}(\partial)=\delta_{\nu\mu}+\mathcal{O}(\theta),\ \chi_\mu(\partial)=\mathcal{O}(\theta).
\end{equation}
Combining \eqref{diff-rep2} with \eqref{Lie-bracket}, and using the algebraic relation $[\hat{x}_\lambda,h(x,\partial)]=-\partial h/\partial(\partial^\lambda)$ valid for any functional $h$ depending on $x$ and $\partial$, we find that the Lie algebraic structure of $\mathfrak{g}$ is preserved under representation provided the functionals $\varphi_{\nu\mu}$ and $\chi_\mu$ satisfy
\begin{subequations}\label{diff-rep3}
\begin{align} 
&\frac{\partial \varphi_{\lambda\mu}}{\partial (\partial_\rho)} \varphi_{\rho\nu} - \frac{\partial \varphi_{\lambda\nu}}{\partial (\partial_\rho)} \varphi_{\rho\mu} = i\theta C_{\mu \nu}^{\hspace{11pt} \rho} \varphi_{\lambda\rho},\label{diff-rep3a}\\
&\frac{\partial\chi_\mu}{\partial(\partial_\rho)}\varphi_{\rho\nu}-\frac{\partial\chi_\nu}{\partial(\partial_\rho)}\varphi_{\rho\mu}= i\theta C_{\mu \nu}^{\hspace{11pt} \rho}\chi_\rho.
\end{align}
\end{subequations}
The above set of differential equations generates infinitely many solutions for the representation $\pi$ defined by eq. \eqref{diff-rep2}. Further constraints on $\varphi_{\nu\mu}$ and $\chi_\mu$ stem from the requirement the differential representation $\pi$ to be a *-representation. This is achieved by requiring that $\langle f,\hat{x}_\mu g \rangle=\langle \hat{x}_\mu f,g \rangle$ for any $f,g\in\mathcal{H}$, namely $\hat{x}_\mu^\dag=\hat{x}_\mu$ to be selfadjoint. Upon using $\partial^\dag_\mu=-\partial_\mu$ and $h^\dag(\partial)=\bar{h}(-\partial)$, we can compute
\begin{align}\label{diff-rep-calcul}
\langle f,\hat{x}^\dag_\mu g \rangle &= \langle\big(x^\alpha\varphi_{\alpha\mu}(\partial)+\chi_\mu(\partial)\big)f,g \rangle=\langle f,{\bar{\varphi}}_{\alpha\mu}(-\partial)x^\alpha g\rangle+\langle f,{\bar{\chi}}_\mu(-\partial)g\rangle\\
&=\langle f,x^\alpha{\bar{\varphi}}_{\alpha\mu}(-\partial) g\rangle_2+\langle f,\frac{\partial{\bar{\varphi}}_{\alpha\mu}(-\partial)}{\partial(\partial_\alpha)}g\rangle+\langle f,{\bar{\chi}}_\mu(-\partial)g\rangle.\nonumber
\end{align}
Comparing the last line of \eqref{diff-rep-calcul} with $\langle f,\hat{x}_\mu g \rangle=\langle f, \big(x^\alpha\varphi_{\alpha\mu}(\partial)+\chi_\mu(\partial)\big)g \rangle$, we conclude that the representation \eqref{diff-rep2} is selfadjoint if, and only if,
\begin{subequations}\label{diff-rep4}
\begin{align}
&{\bar{\varphi}}_{\alpha\mu}(-\partial)= \varphi_{\alpha\mu}(\partial),\label{diff-rep4a}\\
&\frac{\partial{\bar{\varphi}}_{\alpha\mu}(-\partial)}{\partial(\partial_\alpha)}= \chi_\mu(\partial)-{\bar{\chi}}_\mu(-\partial).
\end{align}
\end{subequations}
From \eqref{diff-rep4a}, we readily infer that $\varphi_{\alpha\mu}$ must have the following decomposition
\begin{equation}\label{diff-phi-decomposition}
\varphi_{\alpha \mu} (\partial) = \Phi_{\alpha \mu}(\partial) + i \Psi_{\alpha \mu}(\partial) \ ,
\end{equation}
with the real functional $\Phi_{\alpha \mu}$ (resp. $\Psi_{\alpha \mu}$) of even (resp. odd) degree in $\partial$.\smallskip

The four master equations, supplemented with eq. \eqref{diff-rep3} and \eqref{diff-rep4}, provides us with a set of differential equations from which admissible expressions for the differential *-representation $\pi$ can be derived. Obviously, the solution is not unique and depends on the Lie algebra of coordinate operators we start from. This will be exemplified in $\S$\ref{sec-su2} when considering the case of $\mathfrak{su}(2)$ Lie algebra. For the moment, we turn to the construction of the quantisation maps from which expressions for the star products can be derived.
\subsection{Quantisation maps and related star products.}\label{sec-qmapsu2}
As already mentioned at the beginning of this chapter, the *-algebras' morphism, called quantisation map, $Q$ can be characterised by its action on plane waves, \textit{i.e.} $Q(e^{ik\cdot x})$. Consequently, the corresponding star product which is defined, for any $f,g\in\mathfrak{F}(\mathbb{R}^n)$, by
\begin{equation}\label{sustar}
(f\star g)(x)=\int \frac{d^nk_1}{(2\pi)^n}\frac{d^nk_2}{(2\pi)^n}\ \tilde{f}(k_1)\tilde{g}(k_2)Q^{-1}\big(Q(e^{ik_1\cdot x})Q(e^{ik_2\cdot x})\big),
\end{equation}
where $\tilde{f}(k):=\int d^nxf(x)e^{-ik\cdot x}$, is fully characterised once the deformed plane waves
\begin{equation}\label{qpw}
E_k(\hat{x}):=Q(e^{ik\cdot x}),
\end{equation}
together with the inverse map $Q^{-1}$, are determined.\smallskip

First, we observe that, for a given differential *-representation, eq. \eqref{diff-rep}, the determination of $Q^{-1}$ can be conveniently carried out by enforcing the condition
\begin{equation}\label{q-act}
Q(f)\triangleright 1=f,
\end{equation}
since, for $Q$ invertible, we have by definition $Q^{-1}\big(Q(f)\big)=f=Q(f)\triangleright 1$.\smallskip

The next step consists in observing that the expression
\begin{equation}\label{q-fourier}
Q(f)(\hat{x})=\int \frac{d^nk}{(2\pi)^n}\ \tilde{f}(k) E_k(\hat{x})
\end{equation} 
looks like eq. \eqref{induced-rep} defining the induce representation of the convolution algebra of $\mathcal{G}$ we used in Chap. \ref{sec-Minkowski}. It follows that the deformed plane waves can informally be interpret as (stemming from) a representation of $\mathcal{G}$, namely
\begin{equation}\label{themap}
E:\mathcal{G}\to\mathcal{L}(\mathcal{H}),\ E:g\mapsto E(g):=E_k(\hat{x}),
\end{equation}
with $E(g^\dag)=E(g)^\dag$ and where $\mathcal{L}(\mathcal{H})$ is the set of linear operators acting on $\mathcal{H}$.\\
In eq. \eqref{diff-rep}, the group representation used was unitary. However, the operators $E(g)$ decompose in general into an angular part and a radial part, both stemming from the polar decomposition of operators. Explicitly, we can write
\begin{equation}\label{polar}
E(g)=U(g)|E(g)|,
\end{equation}
where $U:\mathcal{G}\to\mathcal{L}(\mathcal{H})$ is a unitary operator and $|E(g)|:=\sqrt{E(g)^\dag E(g)} \neq 0$.\\
Note that the previous situation, considered in Chap. \ref{sec-Minkowski}, is recovered if $|E(g)|=1$. Indeed, in this case, $E(g)=U(g)$ is unitary.\\
In view of the Stone's theorem, it is legitimate to parametrise the unitary operator as 
\begin{equation}\label{unitpw}
U(g)=e^{i\xi_g^\mu\hat{x}_\mu}, 
\end{equation}
where $\xi_g^\mu\in\mathbb{R}$ can be regarded as a kind of generalised Fourier parameter for the deformed plane waves appearing in eq. \eqref{q-fourier}. Hence, $U(g)$ define grouplike elements which compose as
\begin{equation}
e^{i\xi_{g_1}\cdot\hat{x}}e^{i\xi_{g_2}\cdot\hat{x}}=e^{iB(\xi_{g_1},\xi_{g_2})\cdot\hat{x}},
\end{equation}
where $B(\xi_{g_1},\xi_{g_2})\in\mathbb{R}$ stems from the Baker-Campbell-Hausdorff (BCH) formula for the Lie algebra $\mathfrak{g}=\text{Lie}(\mathcal{G})$ and satisfies
\begin{equation}\label{Bproperties}
B(\xi_{g_1},\xi_{g_2})=-B(-\xi_{g_2},-\xi_{g_1}) , \ B(\xi_g,0)=\xi_g .
\end{equation}
As we are going to see, $B(\xi_{g_1},\xi_{g_2})$ can actually be interpreted as the deformed composition law between momenta associated with the deformed plane waves. This explains why the use of the star product formalism is not always the most suitable for studying NCFT, even though integral expressions for the star product can formally be derived. Indeed, the BCH formula generally provides us with an expression for $B(\xi_{g_1},\xi_{g_2})$ which is an infinite sum of elements of $\mathfrak{g}$ and exact expressions for $B(\xi_{g_1},\xi_{g_2})$ do not always exist. This is for instance the case of the $\mathfrak{su}(2)$ Lie algebra. For an illustration of such difficulties see, e.g., in Chap. \ref{sap-ncftsu2}.\smallskip

The mapping $U:\mathcal{G}\to\mathcal{L}(\mathcal{H})$ clearly defines a unitary representation of $\mathcal{G}$, \textit{i.e.}
\begin{equation}\label{projectif-su2}
U(g_1)U(g_2)=U(g_1g_2),
\end{equation}
which holds up to unitary equivalence as mere application of the Wigner theorem.\\
On the other hand, we demand
$E:\mathcal{G}\to\mathcal{L}(\mathcal{H})$ to be a projective representation of $\mathcal{G}$, namely
\begin{equation}\label{projective}
E(g_1)E(g_2)=\Omega(g_1,g_2)E(g_1g_2),
\end{equation}
with $\Omega:\mathcal{G}\times\mathcal{G}\to{\mathbb{C}\hspace{-2pt}\setminus\hspace{-2pt}\lbrace0\rbrace}$ obeying a 2-cocycle condition, \textit{i.e.}
\begin{equation}
\Omega(g_1,g_2)\Omega(g_1g_2,g_3)=\Omega(g_1,g_2g_3) \Omega(g_2,g_3),
\end{equation}
such that the associativity of the related star product is ensured.\smallskip

Recall that projectively inequivalent representation of a group $\mathcal{G}$ are classified by $H^2\big(\mathcal{G},{\mathbb{C}\hspace{-2pt}\setminus\hspace{-2pt}\lbrace0\rbrace}\big)$, the second cohomology group of $\mathcal{G}$ with value in ${\mathbb{C}\hspace{-2pt}\setminus\hspace{-2pt}\lbrace0\rbrace}$; see, e.g.,
\cite{Gannon:2006}. Since eq. \eqref{projective} actually reflects the composition of plane waves which eventually determine the expression of the star product, we infer from the above remark that  inequivalent families of star products are classified by group cohomology. 
\section{Focus on \texorpdfstring{$\mathfrak{su}(2)$}{su(2)} noncommutative space.}\label{sec-su2}
We now apply the procedure developed in $\S$\ref{sec-dpw} to quantum spaces whose algebras of coordinate operators are characterised by the $\mathfrak{su}(2)$ Lie algebra,\footnote{Such quantum spaces are generally called $\mathbb{R}^3_\theta$, or $\mathbb{R}^3_\lambda$, in the literature and can be regarded as deformations of $\mathbb{R}^3$.} \textit{i.e.} satisfying 
\begin{equation}\label{su2per}
[\hat{x}_\mu,\hat{x}_\nu]=i2\theta\varepsilon_{\mu \nu}^{\hspace{11pt} \rho}\hat{x}_\rho.
\end{equation}
The corresponding set of master equations is easily obtained upon substituting $C_{\mu\nu\rho}$ with $2\varepsilon_{\mu\nu\rho}$ in  eq. \eqref{diff-rep3} and \eqref{diff-rep4}. Standard computations yield
\begin{subequations}\label{master}
\begin{align}
&i 2\theta \varphi_{\alpha \rho}= \varepsilon_{\rho}^{\hspace{4pt} \mu \nu} \frac{\partial \varphi_{\alpha \mu}}{\partial (\partial_\beta)} \varphi_{\beta \nu},\ \ \varphi^\dagger_{\alpha\rho}= \varphi_{\alpha \rho},\label{master1}\\
&i 2\theta \chi_\rho= \varepsilon_{\rho}^{\hspace{4pt} \mu \nu} \frac{\partial \chi_\mu}{\partial (\partial_\alpha)} \varphi_{\alpha \nu},\ \ \frac{\partial \varphi^\dagger_{\alpha \rho}}{\partial (\partial_\alpha)}= \chi_\rho - \chi_\rho^\dagger,\label{master2}
\end{align}
\end{subequations}
where we have used the algebraic relation $\delta_{\mu \gamma} \delta_\nu^{\hspace{4pt} \sigma} - \delta_\mu^{\hspace{4pt} \sigma} \delta_{\nu \gamma} = \varepsilon_{\mu \nu}^{\hspace{11pt} \rho} \varepsilon_{\rho \gamma}^{\hspace{11pt} \sigma}$.
\subsection{\texorpdfstring{$SO(3)$}{SO(3)}-equivariant *-representations.}\label{repsu}
More insight on the actual expressions of $\varphi_{\mu\nu}$ and $\chi_\mu$ can be obtained by observing that $\mathbb{R}^3_\theta\subseteq\mathcal{U}\big(\mathfrak{su}(2)\big)$ supports a natural action of $SU(2)/\mathbb{Z}_2\cong SO(3)$. Therefore, it seems natural to require the differential representation to be compatible with this structure, namely $\pi$ to be $SO(3)$-equivariant. We look for such representation in the sequel.\\
Mere application of the Schur-Weyl decomposition theorem for $SO(3)$ yields
\begin{equation}\label{phi-so3a}
\varphi_{\alpha \mu}(\partial) = \alpha(\Delta) \delta_{\alpha \mu} + \beta(\Delta) \left( \frac{1}{3} \delta_{\alpha \mu} \Delta - \partial_\alpha \partial_\mu \right) + \gamma(\Delta) \varepsilon_{\alpha\mu}^{\hspace{11pt} \rho} \partial_\rho,
\end{equation}
where $\alpha$, $\beta$ and $\gamma$ are $SO(3)$-invariant functionals depending on the Laplacian $\Delta$, to be determined in a while.\footnote{See, for instance, H. Weyl,  \textit{Classical groups}, Princeton University Press, 1946.} It will be further assumed that $\alpha$ and $\beta$ (resp. $\gamma$) are real (resp. purely imaginary) functionals so that \eqref{diff-phi-decomposition} is satisfied. This motivate the following factorisation 
\begin{subequations}\label{rep-so3}
\begin{equation}\label{phi-so3b}
\varphi_{\alpha \mu} (\partial) = f(\Delta) \delta_{\alpha \mu} + g(\Delta) \partial_\alpha \partial_\mu + i h(\Delta) \varepsilon_{\alpha\mu}^{\hspace{11pt} \rho} \partial_\rho,
\end{equation}
where the real $SO(3)$-invariant functionals $f$, $g$, $h$ can easily be related to the old quantities $\alpha, \beta, \gamma$ by mere comparison of \eqref{phi-so3a} with \eqref{phi-so3b}.
Similarly,
\begin{equation} 
\chi_\mu(\partial) = \ell (\Delta) \partial_\mu,
\end{equation}
\end{subequations}
where $\ell(\Delta)$ is a complex $SO(3)$-invariant functional to be determined. Hence, every $SO(3)$-equivariant differential *-representation $\pi$ is labelled by $f, g, h$ and $\ell$.\smallskip

Combining eq. \eqref{rep-so3} with eq. \eqref{master}, we find that the master equations reduce to two systems of differential equations for $f, g,h$ and $\ell$. On the one hand, the first system
\begin{subequations}\label{rep-sys1}
\begin{align}
&(f+g \Delta)h' - (h - \theta)g= 0,\label{rep-sys1a}\\
&(f+g \Delta)h' \Delta + (h - \theta)f= 0,\label{rep-sys1b}\\
&2(f+g \Delta)f' + (h - 2\theta)h - gf= 0,\label{rep-sys1c}
\end{align}
\end{subequations}
provides us with differential representations compatible with the $\mathfrak{su}(2)$ commutation relations. In particular, linear combination of the two first equations in eq. \eqref{rep-sys1} yields
\begin{equation}\label{rep-sys1d}
(f+g\Delta)(h - \theta) = 0.
\end{equation}  
On the other hand, the second system
\begin{subequations}\label{rep-sys2}
\begin{align}
&h= \theta,\label{rep-sys2a} \\
&2(f+g\Delta)' + 2g= \ell + \ell^\dag.\label{rep-sys2b}
\end{align}
\end{subequations}
selects, among the solutions of \eqref{rep-sys1}, those defining *-representations.\footnote{In eq. \eqref{rep-sys1} and \eqref{rep-sys2}, the prime $'$ denotes the derivative with respect to their arguments of the functions under consideration.}\smallskip

Before giving the general expression for admissible $SO(3)$-equivariant differential *-representations, one comment is in order. As long as $\chi_\mu = 0$, the first equation in \eqref{master2} is trivially satisfied and gives no constraints on neither $f$, $g$, nor $h$. In particular, \eqref{rep-sys2a} does not necessarily hold true. Assuming $h\neq\theta$, we find from \eqref{rep-sys1d} that $f+g\Delta=0$ and there exists only two solutions compatible with the remaining equations.\footnote{Whenever $\chi_\mu=0$, eq. \eqref{rep-sys1c} and \eqref{rep-sys2b} prevent $f+g\Delta=0$ and $h=\theta$ to be satisfied simultaneously.} Namely, either $\hat{x}_\mu = 0$ or $\hat{x}_\mu = i 2\theta x^\sigma \varepsilon_{\sigma \mu}^{\hspace{11pt} \rho}\partial_\rho$, solutions we shall disregard since in both cases $\hat{x}_\mu \triangleright 1 \neq x_\mu$ and $\hat{x}_\mu \triangleright f(x)\rightarrow 0$ when $\theta\rightarrow 0$ for any function $f\in\mathcal{H}$. Therefore, we now consider differential *-representation for which $h=\theta$ (with either $\chi_\mu\ne 0$ or $\chi_\mu=0$).\smallskip

The family of $SO(3)$-equivariant differential *-representations is finally found to be given by
\begin{subequations}\label{general_rep}
\begin{equation}
\hat{x}_\mu = x^\alpha \left[ f(\Delta) \delta_{\alpha \mu} + g(\Delta) \partial_\alpha \partial_\mu + i\theta \varepsilon_{\alpha \mu}^{\hspace{11pt} \rho} \partial_\rho \right] + \ell(\Delta) \partial_\mu,
\end{equation}
where the $SO(3)$-invariant functionals $f(\Delta)$, $g(\Delta)$ and $\ell(\Delta)$ satisfy, $f,g$ real,
\begin{align}
&2\left[(f+g\Delta)' + g \right]= \ell + \ell^\dagger, \label{1st_condition} \\
&2(f+g\Delta)f'= gf + \theta^2.\label{2nd_condition}
\end{align}
\end{subequations}
\subsection{Determination of the deformed plane waves.}
From eq. \eqref{projective}, we see that whenever the group under consideration is unitary, \textit{i.e.} $g^\dag g=1$ for any $g\in\mathcal{G}$, the radial part of the deformed plane waves is automatically determined by the 2-cocycle $\Omega$. Indeed, in this case, eq. \eqref{projective} becomes
\begin{equation}\label{calcul13}
E(g^\dag)E(g)=\Omega(g^\dag,g)E(1),\ E(1)=\bbone.
\end{equation}
Therefore, $\Omega(g^\dag,g)>0$ is real $\forall g\in\mathcal{G}$, and $\vert E(g)\vert=\sqrt{\Omega(g^\dag,g)}\ \bbone$ such that
\begin{equation}\label{cond-central}
[|E(g)|,U(g)]=0. 
\end{equation}
Combining \eqref{cond-central} with \eqref{polar} and \eqref{projectif-su2} we easily obtain
\begin{equation}\label{intermediaire}
E(g_1)E(g_2)=|E(g_1)||E(g_2)|U(g_1g_2)=|E(g_1)||E(g_2)||E(g_1g_2)|^{-1}E(g_1g_2).
\end{equation}
Setting for short $\omega_g:=\sqrt{\Omega(g^\dag,g)}$, we find that the plane waves compose as
\begin{subequations}\label{compopw}
\begin{align}
&E(g_1)E(g_2)=(\omega_{g_1}\omega_{g_2}\omega^{-1}_{g_1g_2})E(g_1g_2),\label{planew-multiplic}\\
&E(g_1g_2)=\omega_{g_1g_2}e^{iB(\xi_{g_1},\xi_{g_2})\cdot\hat{x}}.
\end{align}
\end{subequations}

We are now in position to fully determined the expression of the plane waves. To do so, it is convenient to reintroduce the explicit dependence in the momenta of the deformed plane waves. This is achieved by formally identifying $E(g)=\omega_g e^{\xi_g\cdot\hat{x}}$ with
\begin{equation}\label{generalform-ncexpo}
E_p(\hat{x})=\omega(p)e^{i\xi(p)\cdot\hat{x}}.
\end{equation}
Next, the two functions $\omega(p)$ and $\xi(p)$ can be obtained by taking full advantage of the family of differential representations \eqref{general_rep} derived in $\S$\ref{repsu} applied to eq. \eqref{q-act}.

\subsubsection{Derivation of the phase\texorpdfstring{ $\xi$.}{.}}
Let us first derive the expression for $\xi$. Combining eq. \eqref{qpw} with \eqref{q-act} we obtain
\begin{equation} \label{expo_appendix}
e^{i\xi(p)\cdot\hat{x}}\triangleright 1 = \frac{e^{ip\cdot x}}{\omega(p)}.
\end{equation}
Applying the same procedure to $e^{-i\xi(p)\cdot \hat{x}} \partial_\mu e^{i\xi(p) \cdot\hat{x}}$, we compute
\begin{align}
e^{-i\xi(p)\cdot\hat{x}} \partial_\mu e^{i\xi(p)\cdot \hat{x}} \triangleright 1 &= e^{-i\xi(p)\cdot \hat{x}} \partial_\mu \triangleright \frac{e^{ip\cdot x}}{\omega(p)} \\
&= e^{-i\xi(p)\cdot \hat{x}} \triangleright (ip_\mu) \frac{e^{ip\cdot x}}{\omega(p)}= (ip_\mu)  e^{-i\xi(p)\cdot \hat{x}} e^{i\xi(p) \cdot\hat{x}} \triangleright 1\nonumber
\end{align}
which combined with $e^{-i\xi(p)\cdot\hat{x}} e^{i\xi(p)\cdot\hat{x}} \equiv \bbone$ yield
\begin{equation} \label{operator_identity}
e^{-i\xi(p)\cdot\hat{x}} \partial_\mu e^{i\xi(p)\cdot\hat{x}} = (ip_\mu) \bbone,\ \forall p\in \mathbb{R}^3.
\end{equation}
In particular, if we rescale $p \mapsto \lambda p$, $\lambda \in \mathbb{R}$, in \eqref{operator_identity}, and take the derivative with respect to the parameter $\lambda$, we easily obtain for the right-hand-side of \eqref{operator_identity} $ip_\mu$, while the derivation of the left-hand-side leads to
\begin{equation}
\frac{d}{d\lambda} \left[ e^{-i\xi(\lambda p)\cdot\hat{x}} \partial_\mu e^{i\xi(\lambda p)\cdot\hat{x}} \right] = i \frac{d}{d\lambda} \left[ \xi^\nu(\lambda p) \right] \left( e^{-i\xi(\lambda p)\cdot\hat{x}} \varphi_{\mu \nu}(\partial) e^{i\xi(\lambda p)\cdot\hat{x}} \right),
\end{equation}
where we have explicitly used the expression of the representation \eqref{diff-rep2} through the relation $[\partial_\mu, \hat{x}_\nu] = [\partial_\mu,x^a\varphi_{a\nu}] = \varphi_{\mu \nu}$. In view of \eqref{operator_identity}, we readily infer that
\begin{equation}
e^{-i\xi(\lambda p)\cdot\hat{x}} \varphi_{\mu \nu}(\partial) e^{i\xi(\lambda p)\cdot\hat{x}} = \varphi_{\mu\nu}(i\lambda p) \bbone, 
\end{equation}
from which we conclude that
\begin{equation}\label{diff-xi}
\varphi_{\mu\nu}(i\lambda p) \frac{d}{d\lambda} \left[ \xi^\nu(\lambda p) \right] = p_\mu ,
\end{equation}
indicating that the function $\xi$ is merely determined by a first order differential equation. To solve this latter, it remains to invert $\varphi_{\mu\nu}$. In view of the Schur-Weyl decomposition for $SO(3)$, we are looking for solutions of the form
\begin{subequations}
\begin{equation} \label{phi-inverse-general}
(\varphi^{-1})_{\mu\nu}(\partial) = X(\Delta) \delta_{\mu \nu} + Y(\Delta) \partial_\mu \partial_\nu + Z(\Delta) \varepsilon_{\mu \nu}^{\hspace{11pt} \rho} \partial_\rho,
\end{equation}
such that $\varphi_{\mu\nu} (\varphi^{-1})^{\nu\sigma} = \delta_\mu^{\hspace{3pt}\sigma}$. Standard computations lead to the following system
\begin{equation}
fX - i\theta \Delta Z = 1,\ (f+\Delta g)Y + gX + i\theta Z = 0,\ fZ + i\theta X = 0,
\end{equation}
which admits the following unique solution, assuming $f^2\neq\theta^2\Delta$,\footnote{In the case $f^2-\theta^2\Delta=0$, $\varphi_{\mu\nu}$ is not invertible.}
\begin{equation}\label{XYZ-coord}
X(\Delta)=\frac{f(\Delta)}{f^2(\Delta)-\theta^2\Delta},\ Y(\Delta)= - \frac{2f'(\Delta)}{f^2(\Delta)-\theta^2\Delta},\ Z(\Delta)= - \frac{i \theta}{f^2(\Delta)-\theta^2\Delta},
\end{equation}
\end{subequations}
where we have used eq. \eqref{2nd_condition} to simplify the expression of $Y$. We conclude that
\begin{subequations}
\begin{equation}\label{phi-inverse}
(\varphi^{-1})^{\mu\nu}(ip) = \frac{1}{f^2+\theta^2 p^2} \left( f \delta^{\mu\nu} + 2f'p^\mu p^\nu + \theta \varepsilon^{\mu \nu \rho} p_\rho \right), 
\end{equation}
where $f$ and its derivative are (real) functions of $-p^2$.\smallskip

Finally, integrating $d\xi^\mu = (\varphi^{-1})^{\mu\nu}_{\vert_{i\lambda p}} p_\nu d\lambda$ between 0 and 1 on both side of the equality, we obtain
\begin{equation} \label{solution-xi}
\xi^\mu(p) = \int_0^1  (\varphi^{-1})^{\mu\nu}_{\vert_{i\lambda p}} p_\nu\ d\lambda,
\end{equation}
\end{subequations}
where the initial condition $\xi^\mu(0)=0$ stems from $E(1)=E_0(\hat{x})=\bbone$.\footnote{The notation $\varphi^{-1}_{\vert_y}$ means that the function $\varphi^{-1}$ is evaluated at $y$.}$^,$\footnote{Observe that $\xi$ depends only on one of the three functionals characterising the differential representation.}

\subsubsection{Derivation of the radius\texorpdfstring{ $\omega$.}{.}}
In order to fully characterise $E_p(\hat{x})$, it remains to determine its radial part, namely to derive the expression for $\omega(p)$. Again, the strategy amounts to rescale $p\mapsto \lambda p$ by some real parameter $\lambda$, then differentiating with respect to $\lambda$. On the one hand, we compute
\begin{equation}
\frac{d}{d \lambda} \left[ e^{i\xi(\lambda p)\cdot\hat{x}} \right] = i \frac{d}{d\lambda} \left[ \xi^\mu(\lambda p) \right] \hat{x}_\mu e^{i\xi(\lambda p)\cdot\hat{x}} = i (\varphi^{-1})^{\mu\nu}_{\vert_{i\lambda p}}  p_\nu \hat{x}_\mu e^{i\xi(\lambda p)\cdot\hat{x}},
\end{equation}
such that, evaluating this expression on 1, we have
\begin{align}\label{calcul11}
\frac{d}{d \lambda} \left[ e^{i\xi(\lambda p)\cdot\hat{x}} \right] \triangleright 1 &= i (\varphi^{-1})^{\mu\nu}_{\vert_{i\lambda p}}  p_\nu \left(x^\alpha \varphi_{\alpha \mu}(\partial) + \chi_\mu(\partial) \right) \triangleright \frac{e^{i\lambda p\cdot x}}{\omega(\lambda p)} \\
&= i (\varphi^{-1})^{\mu\nu}_{\vert_{i\lambda p}}  p_\nu \left(x^\alpha \varphi_{\alpha \mu}(i\lambda p) + \chi_\mu(i\lambda p) \right) \frac{e^{i\lambda p\cdot x}}{\omega(\lambda p)} \nonumber\\
&= i \left( x^\nu + \chi_\mu (\varphi^{-1})^{\mu\nu}_{\vert_{i\lambda p}} \right) p_\nu \frac{e^{i\lambda p\cdot x}}{\omega(\lambda p)}.\nonumber
\end{align}
On the other hand, we find 
\begin{equation}\label{calcul10}
\frac{d}{d \lambda} \left[ \frac{e^{i\lambda px}}{\omega(\lambda p)} \right] = \left( ix^\nu p_\nu - \frac{1}{\omega(\lambda p)} \frac{d}{d \lambda} \left[ \omega(\lambda p) \right] \right) \frac{e^{i\lambda px}}{\omega(\lambda p)}.
\end{equation}
Owing to the relation\footnote{Indeed, let $g(\lambda,x)=\hat{A}(\lambda)f(x)$, with $f\in\text{Dom}(\hat{A})$. Then,
\begin{equation}
\frac{dg}{d\lambda} =: \lim_{\epsilon \rightarrow 0} \frac{g(\lambda + \epsilon)-g(\lambda)}{\epsilon}= \lim_{\epsilon \rightarrow 0} \left( \frac{\hat{A}(\lambda + \epsilon)-\hat{A}(\lambda)}{\epsilon} \right) f(x),
\end{equation}
from which we conclude that
\begin{equation}
\frac{d}{d\lambda} \left[ \hat{A} f(x) \right] = \frac{d\hat{A}}{d\lambda} f(x).
\end{equation}}
\begin{equation}
\frac{d}{d\lambda} \left[ e^{i\xi(\lambda p)\cdot\hat{x}} \triangleright 1 \right] = \frac{d}{d\lambda} \left[ e^{i\xi(\lambda p)\cdot\hat{x}} \right] \triangleright 1,
\end{equation}
we can identify eq. \eqref{calcul10} with \eqref{calcul11}, to obtain
\begin{equation}
i \left( x^\nu + \chi_\mu (\varphi^{-1})^{\mu\nu}_{\vert_{i\lambda p}} \right) p_\nu = ix^\nu p_\nu - \frac{1}{\omega(\lambda p)} \frac{d}{d \lambda} \left[ \omega(\lambda p) \right] ,
\end{equation}
or equivalently
\begin{equation}
\frac{1}{\omega(\lambda p)} \frac{d}{d \lambda} \left[ \omega(\lambda p) \right] = - i \chi_\mu (\varphi^{-1})^{\mu\nu}_{\vert_{i\lambda p}} p_\nu.
\end{equation}
Integrating the above differential equation, we find the following solution for $\omega$
\begin{equation}\label{solution-omega}
\omega(p) = e^{-i \int_0^1 d\lambda \ \chi_\mu(i\lambda p) (\varphi^{-1})^{\mu\nu}_{\vert_{i\lambda p}} p_\nu}.
\end{equation}
This concludes the derivation of the deformed plane waves.

\subsubsection{Summary.}
Let us summarise and comment on our results.
\begin{enumerate}[label={$\textit{(\roman*)}$}]
\item{First, let us recap the main lines of our construction. We have seen, eq. \eqref{sustar}, that quantisation maps, hence star products, are fully characterised by the socalled deformed plane waves \eqref{qpw}. Having in mind the Weyl quantisation scheme, which essentially identifies quantisation maps with group representations,  we have then identified the deformed plane waves with projective representations of the group
\begin{equation}
E(g_1)E(g_2)=\Omega(g_1,g_2)E(g_1g_2),\ [\Omega]\in H^2\big(\mathcal{G},\mathbb{C}\!\setminus\!\!\lbrace0\rbrace\big),\ g_1,g_2\in\mathcal{G}.
\end{equation}
This led us to the conclusion that group cohomology could be used to classify the various inequivalent families of star products. However, to exhibit one representative of such deformed plane wave, only based on cohomological considerations, may not be an easy task in general. In practice, we have shown that an explicit expression for the deformed plane waves can be conveniently obtain by representing the abstract coordinate operators as differential ones. Recall that such representative is needed for performing actual computations in the context of NCFT;}
\item{In the case the group underlying the quantum space is given by $SU(2)$, we have shown that the star products can be indexed by three real functionals stemming from the requirement the differential star representation $\hat{x}_\mu$, eq. \eqref{general_rep}, defines an $SO(3)$-equivariant morphism of *-algebras. In this case, upon using the explicit expression of $\varphi^{-1}$, eq. \eqref{phi-inverse-general}, in both the expressions of $\xi$ and $\omega$, eq. \eqref{solution-xi} and \eqref{solution-omega}, we have found that the deformed plane waves
\begin{equation}
E(g)\mapsto E_p(\hat{x})=\omega(p)e^{i\xi(p)\cdot\hat{x}},
\end{equation}
are fully determined by a set of two Volterra integrals
\begin{subequations}\label{volterra}
\begin{align}
&\xi^\mu(p) = \int_{-p^2}^0 \frac{dt}{2\|\vec{p}\| \sqrt{-t}} \ \left[X(t) + t Y(t)\right] p^\mu,\\
&\omega(p) = e^{\int_{-p^2}^0 dt \left[X(t) + t Y(t)\right]\ell(t)},
\end{align}
\end{subequations}
in which $X$ and $Y$, eq. \eqref{XYZ-coord}, are representation dependent;}
\item{Next, from the positivity of $\omega$, eq. \eqref{calcul13}, together with the reality of $X$ and $Y$ (which depend only on the real functional $f$), it follows that $\ell(t)$ has to be real $\forall t\in\mathbb{R}$. This is achieved by requiring $\ell^\dag=\ell$. Hence, eq. \eqref{1st_condition}, entering the definition of the family of $^*$-representations \eqref{general_rep}, reduces to
\begin{equation}
\ell=(f+g\Delta)^\prime+g,
\end{equation}
therefore constraining the expression for $\ell$ once $f$ and $g$ satisfying \eqref{2nd_condition} are determined;}
\item{Finally, we conclude this chapter with the following important remark. Let us focus for a moment on the solution $\xi_\mu(p)=p_\mu$, $\omega(p)=1$, for all $p\in\mathbb{R}^3$. Thus, the deformed plane wave reduces to $E_p(\hat{x})=e^{ip\cdot\hat{x}}=:W(e^{ip\cdot{x}})$, which is nothing but the Wigner-Weyl quantisation map corresponding to the symmetric ordering of operators. Then, we can easily show that it is not possible to find any representation belonging to the family \eqref{general_rep} for which $W(e^{ip\cdot{x}})\triangleright1=e^{ip\cdot{x}}$. Indeed, let assume such representation exists. Then, $\xi$ and $\omega$ are such that
\begin{subequations}
\begin{align}
&\int_{-p^2}^0\ dt(X(t)+tY(t))\ell(t)=0,\label{integralone}\\
&\int_{-p^2}^0\ \frac{dt}{2\|\vec{p}\|\sqrt{-t}}(X(t)+tY(t))=1.\label{integraltwo}
\end{align}
\end{subequations}
Whenever $\chi(t)\ne0$, it can be easily checked that \eqref{integralone} and \eqref{integraltwo} cannot be simultaneously satisfied. Indeed, \eqref{integralone} implies $(X(t)+tY(t))\ell(t)=0$ and therefore $X(t)+tY(t)=0$, which clearly contradicts \eqref{integraltwo}. Despite this fact, it is worth mentioning that such Wigner-Weyl quantisation map has been used in the literature within approaches similar to the one presented here.}
\end{enumerate}
\section{Kontsevich product for \texorpdfstring{$\mathfrak{su}(2)$}{su(2)} noncommutative space.}\label{sukont}
In this section, we adapt the material of $\S$\ref{sec-su2} to the derivation of a suitable star product for $\mathbb{R}^3_\theta$. This product is obtained from a subfamily of $SO(3)$-equivariant differential *-representations indexed by only one real functional of $\Delta$, the ordinary Laplacian on $\mathbb{R}^3$. The expression of the deformed plane waves is given. Finally, making use of the Harish-Chandra map
\cite{Harish:1951}, the corresponding star product is shown to be equivalent to the Kontsevich product
\cite{Kontsevich:2003} for the Poisson manifold dual to the finite dimensional Lie algebra $\mathfrak{su}(2)$,  namely closed for the trace functional defined by the usual Lebesgue integral $\int d^3x$. This product will be used in Chap. \ref{sap-ncftsu2} to construct an action functional aiming to describe the dynamics of interacting scalar fields on $\mathbb{R}^3_\theta$.
\subsection{A suitable family of differential *-representations.}
Setting $f+g\Delta=:R(\Delta)$, with $R(\Delta)$ a real functional of $\Delta$, we see that eq. \eqref{2nd_condition} gives rise to a Riccati equation, \textit{i.e.}
\begin{equation}\label{riccati}
2g' = \left(\frac{2R'(\Delta)}{\Delta} - \frac{\theta^2}{\Delta R(\Delta)} \right) - \frac{3}{\Delta}g + \frac{1}{R(\Delta)} g^2.
\end{equation}

The family we are looking for is characterised by the constraint $R(\Delta)=1$. Equation \eqref{riccati} reduces to
\begin{equation}\label{equadiff-reduced} 
2t \frac{dG}{dt} + 3\left(G(t)+1 \right) - \frac{t}{6} G^2(t) = 0,\ g(t)=:\frac{\theta^2}{3} G(2\theta^2 t).
\end{equation}
It can be shown
\cite{Kupriyanov:2015} that this equation admits the following solution
\begin{equation}\label{kv-g}
G(t) = -6\sum_{n=1}^\infty \frac{2^n B_{2n}}{(2n)!}\ t^{n-1},
\end{equation}
where $B_n$'s are Bernoulli numbers. On the other hand, eq. \eqref{1st_condition} reduces to
\begin{equation}\label{kv-chi} 
\ell(\Delta)=g(\Delta).
\end{equation}
Hence,
\begin{subequations}\label{kv-star}
\begin{align}
&\hat{x}_\mu=x^\alpha\left[(1-g(\Delta)\Delta)\delta_{\alpha\mu}+g(\Delta)\partial_\alpha\partial_\mu+
i\theta\varepsilon_{\alpha\mu}^{\hspace{11pt}\rho}\partial_\rho\right]+g(\Delta)\partial_\mu, \\
&g(\Delta)=-\sum_{n=1}^\infty \frac{(2\theta)^{2n} B_{2n}}{(2n)!}\ \Delta^{n-1},\label{gkv-star}
\end{align}
\end{subequations}
which selects a subfamilies of *-representations among those given by \eqref{general_rep}.
\subsection{Deformed plane waves and Kontsevich star product.}\label{sec-kont}
The deformed plane waves associated with \eqref{kv-star} are easily obtained from eq. \eqref{volterra}. Upon combining $f=1-tg$, which is equivalent to the constraint $R(t)=1$,  with eq. \eqref{2nd_condition}, we find that $2tf'=f-f^2+\theta^2t$. It follows that
\begin{equation}
X(t)+tY(t)=\frac{f-2tf'}{f^2-\theta^2t}=1,
\end{equation}
where $X$ and $Y$ are defined in eq. \eqref{XYZ-coord}. Using this last equation in \eqref{volterra}, we obtain
\begin{subequations}
\begin{align}
&\xi_\mu(p)=p_\mu,\\
&\omega(p) = \exp \left( \int_{-p^2}^0 g(t) dt\right).
\end{align}
\end{subequations}
Now, observe that $g$, eq. \eqref{gkv-star}, formally converges to
\begin{equation}\label{closed_g}
g(\Delta) = - \Delta^{-1} \left( \theta\sqrt{\Delta} \coth(\theta\sqrt{\Delta}) - 1 \right) .
\end{equation}
Passing from hyperbolic to trigonometric functions and performing the change of variable $x=\theta \sqrt{t}$, we can rewrite $\omega$ as
\begin{equation}
\omega(p) = \exp\left(2 \int^{\theta |p|}_0 \left( \cot(x) - \frac{1}{x} \right) dx \right).
\end{equation}
Integrating this latter expression, we finally obtain the following expression for the deformed plane waves
\begin{equation} \label{pw-example}
Q(e^{ip\cdot x})=E_p(\hat{x}) = \left( \frac{\sin(\theta |p|)}{\theta |p|} \right)^2 e^{ip\cdot\hat{x}}.
\end{equation}

According to the discussion of $\S$\ref{sec-qmapsu2} together with eq. \eqref{compopw}, the corresponding star product is readily obtained from
\begin{subequations}
\begin{align}
&e^{ip\cdot x} \star_Q e^{iqx} = \mathcal{W}^2(p,q) e^{iB(p,q)\cdot x},\\
&\mathcal{W}(p,q) := \frac{|B(p,q)|}{\theta |p||q|}\frac{\sin(\theta |p|)\sin(\theta |q|)}{\sin(\theta |B(p,q)|)},\label{theweight}
\end{align}
\end{subequations}
with $B(p,q)$ stemming from the Baker-Campbell-Hausdorff formula for $\mathfrak{su}(2)$. We have
\begin{equation}
(f\star_Q g)(x) = \int \frac{d^3p}{(2\pi)^3}\frac{d^3q}{(2\pi)^3}\tilde{f}(p)\tilde{g}(q) \mathcal{W}^2(p,q) e^{iB(p,q)\cdot x},\ f,g \in \mathfrak{F}(\mathbb{R}^3).
\end{equation}

To relate the above star product to the Kontsevich product, it is convenient to define a new quantization map. Let $\mathcal{K}:\mathfrak{F}(\mathbb{R}^3) \to \mathcal{L}(\mathcal{H})$ be defined by
\begin{subequations}
\begin{align}
&\mathcal{K} := Q \circ H,\\
&H := \frac{\theta \sqrt{\Delta}}{\sinh(\theta \sqrt{\Delta})}.\label{Kontsevich}
\end{align}
\end{subequations}
The operator $H$ is such that
\begin{equation}
H(f\star_\mathcal{K}g)=H(f)\star_QH(g),\ f,g \in \mathfrak{F}(\mathbb{R}^3),
\end{equation}
which defines obviously an equivalence relation between the star products $\star_Q$ and $\star_\mathcal{K}$.\\
A standard calculation yields
\begin{equation}\label{checkpoint2}
\mathcal{K}(e^{ip\cdot x}) = \frac{\sin(\theta |p|)}{\theta |p|} e^{ip\cdot\hat{x}}. 
\end{equation}
Hence, the star product $\star_\mathcal{K}$ associated with $\mathcal{K}$, which is ($H$-)equivalent to $\star_Q$, is obtained from 
\begin{equation}\label{duflopw}
e^{ip\cdot x} \star_\mathcal{K} e^{iq\cdot x} = \mathcal{W}(p,q) e^{iB(p,q)\cdot x},
\end{equation}
and we can write
\begin{equation}\label{kontsev-product}
(f\star_\mathcal{K}g)(x)=\int \frac{d^3p}{(2\pi)^3}\frac{d^3q}{(2\pi)^3}\tilde{f}(p)\tilde{g}(q) \mathcal{W}(p,q) e^{iB(p,q)\cdot x},\ f,g \in \mathfrak{F}(\mathbb{R}^3), 
\end{equation}
where $\mathcal{W}(p,q)$ is still given by \eqref{theweight}.\smallskip

This star product $\star_\mathcal{K}$ coincides with the Kontsevich product
\cite{Kontsevich:2003}. It has been derived for $\mathbb{R}^3_\theta$ within a different approach in 
\cite{Freidel:2008,Guedes:2013,Kupriyanov:2015}, namely via the relation
\begin{equation}\label{deriv-konts}
\mathcal{K}=W\circ j^{\frac{1}{2}}(\Delta), 
\end{equation}
where $W$ is the Weyl quantization map and 
\begin{equation}\label{duflo}
j^{\frac{1}{2}}(\Delta)=\frac{\sinh(\theta \sqrt{\Delta})}{\theta \sqrt{\Delta}}, 
\end{equation}
is the Harish-Chandra map
\cite{Harish:1951,Duflo:1970,Duflo:1977}. Recall that $\star_\mathcal{K}$ is closed for the trace functional defined by the Lebesgue integral on $\mathbb{R}^3$, namely
\begin{equation}\label{starclos}
\int d^3x\ (f \star_\mathcal{K} g)(x) = \int d^3x\ f(x) g(x).
\end{equation}

Finally, comparing \eqref{Kontsevich} and \eqref{duflo}, we infer
\begin{equation}
j^{\frac{1}{2}}(\Delta)=H^{-1}.
\end{equation}
Hence, $H$ is the inverse of the Harish-Chandra map. Notice that, by using \eqref{pw-example} combined with \eqref{checkpoint2}, we have 
\begin{equation}
\mathcal{K}(e^{ip\cdot x})\triangleright 1=\frac{\theta|p|}{\sin(\theta|p|)}e^{ip\cdot x},
\end{equation}
while \eqref{checkpoint2} and \eqref{deriv-konts} yield
\begin{equation}
W(e^{ip\cdot x})=\frac{\theta|p|}{\sin(\theta|p|)}\mathcal{K}(e^{ip\cdot x})=e^{ip\cdot\hat{x}}.
\end{equation}
\part{Noncommutative quantum field theory.}\label{part-ncft}
Spacetime plays a central role in the current description of the physical phenomena as it provides the mathematical framework upon which the construction of physical models is based on.\footnote{Here, by ``spacetime" we designate either the spacetime as a whole, or space and time separately.} It is therefore to expect that modifications in the structure of spacetime, as occurring in noncommutative geometry, induce modifications in the descriptions of the physical phenomena themselves. This is all the more true as modern physics is mainly based on the notion of field, which is nothing but a mathematical object which associates with each point in space and time a physical, dynamical or not, quantity.\smallskip

It is precisely the aim of noncommutative field theory (NCFT) to study the interplay between quantum spacetime and field dynamics or, in other words, to study the possibly new (both classical and quantum) behaviours of fields on noncommutative background. Unfortunately, as far as we know, there is no canonical way of carrying out such a study, as no experimental, nor observational, evidences exist for guiding the construction of a reasonable action functional describing the dynamics of interacting fields on noncommutative background. Nevertheless, based on our current understanding in both ordinary quantum field theory and classical gravity, together with the assumption that known physical theories should be recovered in some limit of the NCFT, (what we believed to be) reasonable requirements the action functional should satisfy can be postulated to guide its derivation. This last statement will be clarified throughout this part. Note that, in the present dissertation we are only concerned with the quantum properties of NCFT, unfortunately leaving aside the study of their classical ($\hbar=0$) properties. Furthermore, although attempts to extend the canonical quantisation scheme to the context of field theory have been studied in the literature, see, e.g.,
\cite{Fiore:2010,Bu:2006,Arzano:2007}, and references therein, we here adopt the point of view of path integral quantisation. \smallskip
 
One way to investigate the quantum properties of a NCFT is to represent this latter as a matrix model. This has been done in the Moyal case, see, e.g.,
\cite{Grosse:2003,Grosse:2005,Grosse:2008,JCW:2008a,JCW:2008b,JCW:2013a}, as well as for $\mathbb{R}_\theta^3$, see, e.g.,
\cite{JCW:2013b,JCW:2015,JCW:2016}. For a review see, e.g.,
\cite{Lizzi:2014}, and references therein. One of the advantage of this approach is that, formulated in the matrix basis, the expression of the interaction potential may become very simple. On the other hand, the expression of the kinetic operator may become cumbersome and difficult to invert. Therefore, albeit powerful, it may happen that the matrix model formulation of a NCFT becomes unexploitable due to severe technical difficulties. Another alternative (widely used) framework to study properties of NCFT is to take advantage of star products and deformation theory approach, either from the standard viewpoint of formal deformations extending the quantisation approach of classical phase space, or taking advantage of underlying Hopf algebra structures and related twists. We adopt the former viewpoint in this dissertation. Examples of star products -- we are going to use to investigate the quantum properties of scalar NCFT built on $\mathbb{R}^3_\theta$ and $\kappa$-Minkowski --  have been derived in Part \ref{ch-ncst}. The use of the star product formulation of NCFT is often convenient for fast construction of reasonable action functionals but may lead to technical difficulties whenever the star product is represented by a complicated formula. This is the case, e.g., when the algebra of coordinates associated with the quantum spacetime is isomorphic to a semisimple Lie algebra; see Chap. \ref{sap-ncftsu2} for an example of such difficulties. On the contrary, whenever the algebra of coordinates is isomorphic to a nilpotent or solvable Lie algebra, the corresponding BCH formula (from which the momentum conservation law, as well, composition of plane waves, can be read off) admits a simplified expression, and the corresponding star product may admit a relatively simple expression as well.\smallskip 
 
Although quantum properties of NCFT on Moyal space, as well as $\mathbb{R}^3_\theta$, have been widely studied in the literature, it is not the case for NCFT built on $\kappa$-Minkowski. This is probably due to the very different structure of the algebra of fields modelling $\kappa$-Minkowski than that of Moyal space or $\mathbb{R}^3_\theta$. Indeed, for these two latter spaces, the corresponding groups are unimodular. This is not the case for $\kappa$-Minkowski. As we are going to see, the nonunimodularity of the Lie group underlying $\kappa$-Minkowski is reflected at the level of the action functional by the loss of cyclicity of the integral involved in it. To be more precise, requiring the action functional to be $\kappa$-Poincar\'{e} invariant conditioned the Lebesgue integral to define a twisted trace, the twist being related to the modular function, eq. \eqref{function-modulaire-4d}. As far as we know, there is only one other paper in the literature
\cite{Grosse:2006} dealing with interacting scalar field theories on $\kappa$-Minkowski. In this paper, the NCFT was built from another (albeit presumably equivalent) star product and a different kinetic operator was used. The conclusions we obtain seem to qualitatively agree with those obtained in
\cite{Grosse:2006}. However, the precise comparison between both works is drastically complicated by the technical approach used in
\cite{Grosse:2006} leading to very involved formulas.\smallskip
 
This part is organised in two independent chapters. The first chapter (Chap. \ref{ch-ncft}) is devoted to the study of the quantum properties of various models of $\kappa$-Poincar\'{e} invariant noncommutative field theories. We give the full derivation of the one-loop order corrections to both the 2-point and 4-point functions for different families of (quartic) interactions and kinetic operators. As we are going to see, the relatively simple expression of the star product \eqref{star-4d} makes the computation of the various contributions very tractable. In Chap. \ref{sap-ncftsu2}, we use the Kontsevich star product derived in $\S$\ref{sec-kont} to study the one-loop 2-point function for two models of noncommutative field theory with quartic interactions. In this case, we find that the deformation parameter $\theta$ provides a natural UV cutoff which regularises both the UV and the IR. On the contrary, $\kappa$ does not play a role similar to $\theta$ for the NCFT built from \eqref{starinvol4d}. We find that, for the three propagators considered, the NCFT on $\kappa$-Minkowski are not finite. In one case however, the one-loop 2-point function has milder behaviour than its commutative counterpart, and the one-loop 4-point function is found to be finite. These results are summarised in Table \ref{tableau2}. Finally, we give an interpretation of the noncyclicity of the Lebesgue integral as reflecting the occurrence of a KMS condition at the level of the algebra of fields. This is discussed $\S$\ref{sec-KMS}.\smallskip

\noindent In the following, we work with Euclidean signature.
\chapter{\texorpdfstring{$\kappa$}{k}-Poincar\'{e} invariant scalar field theories.}\label{ch-ncft}
\section{Construction of real action functionals from KMS weight.}\label{sec-action}
The presentation of the $\kappa$-Minkowski space $\mathcal{M}_\kappa$ given in Chap. \ref{sec-Minkowski} provides us with all the needed ingredients for constructing physically reasonable expressions for an action functional $\mathcal{S}_{\kappa,\star}$ aiming to describe the dynamics of interacting complex scalar fields on noncommutative $\kappa$-Minkowski background. In this framework, the (classical) fields are merely the elements of the group C*-algebra $C^*(\mathcal{G}_4)$ modelling $\kappa$-Minkowski, while a canonical measure of integration to be involved in the action functional is provided by the (right invariant) Haar measure of the (nonunimodular) locally compact Lie group $\mathcal{G}_4=\mathbb{R}\ltimes\mathbb{R}^{3}$. Recall that this measure of integration coincides in the group parametrisation \eqref{group4d-parametrization} with the Lebesgue measure on $\mathbb{R}^4$, \textit{i.e.} $\int d^4x$. Nevertheless, the construction of the action functional remains  equivocal and additional assumptions are needed to guide its full derivation.
\subsection{Preliminary considerations.}
\noindent In this purpose, we demand the action functional to obey the following two conditions:
\begin{description}
\item[{\footnotesize(SP)}]{First, to be invariant under the action of the $\kappa$-Poincar\'{e} algebra;}
\item[{\footnotesize(CP)}]{Next, to reduce to a known field theory in the commutative (low energy) limit.}
\end{description}

The former condition (SP) constitutes the core principle upon which our construction of noncommutative field theories is based. It is physically motivated by the important role played by the Poincar\'{e} invariance in ordinary quantum field theory together with the fact that the $\kappa$-Poincar\'{e} algebra can be viewed as describing the (quantum) symmetries of the $\kappa$-Minkowski space; see Appendix \ref{sap-poincare}. Hence, a reasonable requirement for an action functional aiming to describe the dynamics of elementary particles on $\kappa$-Minkowski background is to be compatible with the symmetries of this deformed spacetime.\\
Straightforward computations show that the Lebesgue measure is invariant under the action of the $\kappa$-Poincar\'{e} Hopf algebra $\mathfrak{P}_\kappa$ in the sense that\footnote{Since every element of $\mathfrak{P}_\kappa$ can be written as a linear combination of the generators $\mathcal{E}$, $P_i$, $M_i$, and $N_i$, it is sufficient to check the invariance of the measure under the action of these generators to prove its $\kappa$-Poincar\'{e} invariance.} 
\begin{equation}\label{measure-invariance}
\int d^4x\left(h\triangleright f\right)(x)=\epsilon(h)\int d^4x\ f(x),\ \forall h\in\mathfrak{P}_\kappa,\ \forall f\in\mathcal{M}_\kappa,
\end{equation}
where the action $\triangleright$ is given by eq. \eqref{module-action} and $\epsilon$ is the counit, eq. \eqref{hopf3}, of $\mathfrak{P}_\kappa$.\smallskip

Since the (star) product of two any functions $f,g\in\mathcal{M}_\kappa$ still belongs to $\mathcal{M}_\kappa$, as well as $f^\ddagger\in\mathcal{M}_\kappa$ and $\mathcal{O}f\in\mathcal{M}_\kappa$ for any operator $\mathcal{O}:\mathcal{M}_\kappa\to\mathcal{M}_\kappa$ with dense domain in $\mathcal{M}_\kappa$, it follows from the $\kappa$-Poincar\'{e} invariance of the Lebesgue measure, eq. \eqref{measure-invariance}, that any action functional of the form
\begin{equation}
S_{\kappa,\star}[\phi,\phi^\ddagger]:=\int d^4x\ \mathcal{L}_\kappa[\phi,\phi^\ddagger](x),
\end{equation}
is $\kappa$-Poincar\'{e} invariant provided that the Lagrangian density $\mathcal{L}_\kappa$ is made of polynomials in the fields $\phi$ and $\phi^\ddagger$, together with terms of the form $\mathcal{O}\phi$ and $\mathcal{O}\phi^\ddagger$. Namely,
\begin{equation}\label{action-invariance}
h\blacktriangleright S_{\kappa,\star}[\phi,\phi^\ddagger]:=\int d^4x\left(h\triangleright\mathcal{L}_\kappa[\phi,\phi^\ddagger]\right)(x)=\epsilon(h)S_{\kappa,\star}[\phi,\phi^\ddagger],\ \forall h\in\mathfrak{P}_\kappa,
\end{equation}
which is a Hopf algebraic formulation of the more familiar $\delta S=0$; see, e.g.,
\cite{Agostini:2004}.\smallskip

The second condition (CP) is a guideline to offset the lack of observational and experimental data which would ideally constrain the admissible expressions for $\mathcal{S}_{\kappa,\star}$. This requirement is supported by the fact that the $\kappa$-deformed Minkowski space (resp. Poincar\'{e} algebra) is obtained from (smooth) deformation of the classical Minkowski space-time (resp. Poincar\'{e} algebra), the real, positive, dimensionful parameter $\kappa^{-1}$ increasing from zero to go from commutative to noncommutative. Consequently, any $\kappa$-Poincar\'{e} invariant noncommutative field theory can be interpreted as providing a high energy extension of some Poincar\'{e} invariant field theory we should recover when $\kappa^{-1}\to0$.\\
In the following, we restrict our analysis to $\kappa$-Poincar\'{e} invariant action functional describing the dynamics of $\mathbb{C}$-valued scalar field, with quartic interactions, admitting the ordinary $\vert\phi\vert^4$ model as commutative limit, namely
\begin{equation}\label{action-limit}
\lim\limits_{\kappa\to\infty}S_{\kappa,\star}[\phi,\phi^\ddagger]=\int d^4x\left(\frac{1}{2}\bar{\phi}(x)(-\partial_\mu\partial^\mu+\bar{m}_0^2)\phi(x)+\frac{\bar{g}}{4!}\vert\phi(x)\vert^4\right),
\end{equation}
where $\bar{m}_0$ is the (bare) rest mass of the complex scalar field $\phi$ and $\bar{g}$ the corresponding (bare) coupling constant.\bigskip

The determination of admissible expressions for $S_{\kappa,\star}$, satisfying all the above-mentioned requirements, is further facilitated by the introduction of a Hilbert product on $\mathcal{M}_\kappa$. Let $\langle\cdot,\cdot\rangle_\star:\mathcal{M}_\kappa\times\mathcal{M}_\kappa\to\mathbb{C}$ be a positive definite Hermitian form defined, $\forall f,g\in\mathcal{M}_\kappa$, by
\begin{equation}\label{hilbert-def2}
\langle f,g\rangle_\star:=\int d^4x\left(f\star g^\ddagger\right)(x).
\end{equation}
Upon developing the expressions of the star product and involution, eq. \eqref{starinvol4d}, in the left-hand-side of the above expression, we easily obtain 
\begin{equation}\label{hilbert-prop2}
\int d^4x \left(f\star g^\ddagger\right)(x)=\int d^4x f(x)\bar{g}(x),\ \int d^4x f^\ddagger(x)=\int d^4x \bar{f}(x).
\end{equation}
From this, it follows that
\begin{equation}
\| f\|^2_\star:=\langle f,f\rangle_\star=\int d^4x \vert f(x)\vert^2=\| f\|^2_2\geq0,\ \forall f\in\mathcal{M}_\kappa,
\end{equation}
with equality if, and only if, $f=0$, while $\langle g,f\rangle_\star=\overline{\langle f,g\rangle}_\star$ where we have used \eqref{antivolution}.\\
The Hilbert product \eqref{hilbert-def2} provides an efficient tool to control the reality of the action functional, including the properties that the kinetic operator have to satisfy to ensure this reality condition. We easily deduce that a sufficient condition to ensure the reality of the action functional consists in considering terms of the form $\langle f,f\rangle_\star$ and $\langle f,\mathcal{O}f\rangle_\star$ provided $f\in\mathcal{M}_\kappa$ is any (star) polynomial in the fields $\phi$, $\phi^\ddagger$, and $\mathcal{O}:\mathcal{M}_\kappa\to\mathcal{M}_\kappa$ is selfadjoint. More details on the actual expressions of $f$ and $\mathcal{O}$ are discussed in sections $\S$\ref{sec-kinetic} and $\S$\ref{sec-kinetic}. For convenience, we further assume the action functional decomposes as usual into a kinetic term and an interaction one, \textit{i.e.} 
\begin{equation}\label{actiondecomposition}
S_{\kappa,\star}[\phi,\phi^\ddagger]=S^\text{kin}_{\kappa,\star}[\phi,\phi^\ddagger]+S^\text{int}_{\kappa,\star}[\phi,\phi^\ddagger].
\end{equation}
Anticipating the ensuing derivation of $S_{\kappa,\star}$, note that, for the theories under consideration, the mass dimension of the fields and parameters are $[\phi]=[\phi^\ddagger]=1$, $[g]=0$ and $[m]=1$, where $g$ (resp. $m$) denotes generically a coupling constant (resp. a mass).\smallskip

\noindent Before proceeding to this analysis, some comments are in order.
\begin{enumerate}[label={$\textit{(\roman*)}$}]
\item{First of all, making use of the $\kappa$-Poincar\'{e} invariance of the Lebesgue integral together with \eqref{dag-hopfoperat} and \eqref{derivtwist}, we can show that both $\mathcal{E}$ and the $P_i$'s are selfadjoint with respect to \eqref{hilbert-def2}, \textit{i.e.} $\langle f,t^\dag g\rangle_\star:=\langle tf, g\rangle_\star=\langle f,t g\rangle_\star$, for any $t\in \mathfrak{T}_\kappa$.\\
For example, we compute
\begin{align}
\langle P_i f, g\rangle_\star&=-\int d^4x\ (\mathcal{E}^{-1}P_i\triangleright f^\ddagger)\star g=-\int d^4x\  (P_i\triangleright f^\ddagger)\star (\mathcal{E}\triangleright g)\\
&=\int d^4x\  (\mathcal{E}\triangleright f^\ddagger)\star (P_i\mathcal{E}\triangleright g)=\int d^4x\  f^\ddagger\star (P_i\triangleright g)=\langle f,P_ig\rangle;\nonumber
\end{align}
}
\item{Next, we see that our construction involves both $\phi$ and $\phi^\ddagger$ as primary objects. Some might object that the action functional should rather involved $\phi$ and $\bar{\phi}$ as primary objects, as it is the case in ordinary quantum field theory (QFT) or even in most of the NCFT in the literature. To this objection, we answer that, unlike in ordinary QFT or NCFT built on Moyal or $\mathbb{R}^3_\theta$, the group underlying the construction of the $\kappa$-Minkowski space is nonunimodular. Recall that, in this context, the involution compatibles with the structure of group C*-algebra underlying the quantum spaces, is defined, eq. \eqref{involution2-ap} and \eqref{invol-general}, by 
\begin{equation}
f^\ddagger:=\mathcal{F}^{-1}\Big(\big(\Delta_{\mathcal{G}_4}\overline{\mathcal{F}f}\hspace{2pt}\big)^\flat\Big).
\end{equation}
Hence, as already mentioned in Part \ref{ch-ncst}, the fact that for NCFT built on Moyal of $\mathbb{R}^3_\theta$ a natural involution is provided by the complex conjugation $f\mapsto\bar{f}$ simply reflects the fact that the underlying groups are unimodular, namely $\Delta_\mathcal{G}=1$. This is obviously not the case for $\kappa$-Minkowski. For these reasons, together with the compatibility of $\ddagger$ with the star product, it is necessary to incorporate $\phi^\ddagger$ in the expression of the action functional to ensure its reality. Note that, the complex conjugation is still needed since the fields are $\mathbb{C}$-valued functions;}
\item{Finally, straightforward computations show that the Lebesgue integral is not cyclic. Indeed,
\begin{equation}\label{twistedtrace}
\int d^4x\ f\star g=\int d^4x\ (\sigma\triangleright g)\star f,
\end{equation}
where we have defined for convenience
\begin{equation}\label{twist-def}
\sigma\triangleright f := \mathcal{E}^3 \triangleright f = e^{-3P_0/\kappa}\triangleright f,
\end{equation}
with $\mathcal{E}$ given by eq. \eqref{twist0}. Hence, the Lebesgue integral cannot define a trace, but instead, it defines a twisted trace. The next subsection is devoted to this important issue.\label{remark}}
\end{enumerate}
\subsection{Trading cyclicity for a KMS condition.}\label{sec-KMS}
In the previous section, we have discussed some of the general properties a reasonable action functional describing the dynamics of interacting scalar fields on $\kappa$-Minkowski background must satisfy. We found that requiring both the action functional to be $\mathbb{R}$-valued and invariant under the action of the $\kappa$-Poincar\'{e} algebra strongly constrain its admissible expressions. This has conditioned the definition of the Hilbert product \eqref{hilbert-def2}. More drastically, we found that these requirements lead to the loss of cyclicity of the Lebesgue integral with respect to the star product, eq. \eqref{twistedtrace}. Instead of being a trace, $\text{Tr}(a\star b)=\text{Tr}(b\star a)$, the Lebesgue integral defines a twisted trace, $\text{Tr}(a\star b)=\text{Tr}(\sigma b\star a)$, on the algebra of fields, the twist being given by $\sigma$.\smallskip

Although this loss of cyclicity is known for a long time in the physics literature, it has often been considered as a troublesome feature of $\kappa$-Poincar\'{e} invariant field theories; this having probably discouraged the pursue of many studies of their properties at the quantum level. On the other hand, probably as an attempt to avoid this difficulty, a lot of investigations have been undertaken working with momenta instead of spacetime variables. But also in this case, it has been noted that depending on the ordering chosen to define the deformed plane waves (in the sense of Chap. \ref{sec-products}), various candidates for a measure of integration (over the momenta) are possible. This has sometimes been regarded
\cite{Amelino:2000b} as constituting an ambiguity for deriving an expression for the action functional. Recall that, in view of \eqref{group-elem2}, a choice of ordering merely reflects a choice of parametrisation of the group elements, \textit{i.e.} a redefinition of the momenta. 

It must be emphasised that this loss of cyclicity merely reflects the nonunimodularity of the Lie group underlying the construction of the C*-algebra of fields modelling $\kappa$-Minkowski. We can be convinced by comparing the expression of the twist, \textit{i.e.} $\sigma=e^{-3P_0/\kappa}$, with the expression of the modular function, \textit{i.e.} $\Delta_{\mathcal{G}_4}(k)=e^{3k^0/\kappa}$. In fact, the two objects are related via $\mathcal{F}(\sigma\triangleright f)(k)=\Delta_{\mathcal{G}_4}^{-1}(k)\mathcal{F}(f)(k)$. Thus, the two above issues of the integration's measure (in both spacetime and momentum space) are actually the two sides of the same coin. \smallskip

In the following, we show that eq. \eqref{twistedtrace} can be interpreted as reflecting a Kubo-Martin-Schwinger  (KMS) condition for the positive linear functional defined by
\begin{equation}\label{zeta}
\zeta(f):=\int d^4x f(x),\ f\in\mathcal{M}_\kappa,
\end{equation}
thus giving a positive interpretation/solution to what we discussed above. Observe that $\zeta$ and the Hilbert product \eqref{hilbert-def2} are related by $\langle f,g\rangle_\star=\zeta(f\star g^\ddagger)$.
\subsubsection{KMS condition in quantum statistical mechanics.}
The KMS condition has been introduced a long time ago in the context of quantum statistical mechanics
\cite{Kubo:1957a,Martin:1959,Haag:1967} as a tool to characterise equilibrium temperature states of quantum systems; see also, e.g.,
\cite{Bratteli:1981,Haag:2012}. It can be (schematically) illustrated as follow.\smallskip

Let us consider an arbitrary quantum mechanical system with time independent Hamiltonian $H$. Let $A\in\mathcal{B}(\mathcal{H})$ be a (bounded) observable acting on the Hilbert space $\mathcal{H}$. In general, a (mixed) state is described by a density matrix, say $\rho$, such that the expectation value of $A$ in this state is given by
\begin{equation}\label{statekms0}
\langle A\rangle=\text{Tr}(\rho A),\ A\in\mathcal{B}(\mathcal{H}).
\end{equation}
If the system is in thermal equilibrium at finite temperature $T=\beta^{-1}$, a thermal state is defined by $\rho=e^{-\beta H}$ and eq. \eqref{statekms0} becomes (Gibbs formula)
\begin{equation}\label{statekms1}
\langle A\rangle_\beta=Z_\beta^{-1}\text{Tr}\big(e^{-\beta H} A\big),\ Z_\beta^{-1}:=\text{Tr}\big(e^{-\beta H}\big).
\end{equation}

Now, let us define the following objects
\cite{Fulling:1987}
\begin{subequations}\label{corrKMS}
\begin{align}
&G_{+}^\beta(t;A,B):=\langle \Sigma_t(A)B\rangle_\beta,\\
&G_{-}^\beta(t;A,B):=\langle B\Sigma_t(A)\rangle_\beta,\ A,B\in\mathcal{B}(\mathcal{H}),
\end{align}
\end{subequations}
where $\Sigma_t(A)$ gives the time translate of $A$ in the Heisenberg picture, \textit{i.e.}
\begin{equation}\label{timeA}
\Sigma_t(A):=e^{itH}Ae^{-itH},\ t\in\mathbb{R}.
\end{equation}
Making use of the cyclicity of $\text{Tr}$ in \eqref{statekms1}, together with $[e^{itH},e^{-\beta H}]=0$, we compute
\begin{align}
G_{+}^\beta(t;A,B)&= Z_\beta^{-1}\text{Tr}\big(e^{-\beta H} e^{itH}Ae^{-itH}B\big)\\
&=Z_\beta^{-1}\text{Tr}\big(e^{-\beta H} Ae^{-itH}Be^{itH}\big)=G_{-}^\beta(-t;B,A).\nonumber
\end{align}
Equations \eqref{corrKMS} can formally be extended to the complex plane, $z\in\mathbb{C}$, via
\begin{subequations}\label{corrKMS2}
\begin{align}
&G_{+}^\beta(z;A,B)=Z_\beta^{-1}\text{Tr}\big(e^{i(z+i\beta)H}Ae^{-izH}B\big),\label{corrKMS2a}\\
&G_{-}^\beta(z;A,B)=Z_\beta^{-1}\text{Tr}\big(B e^{izH} Ae^{-i(z-i\beta)H}\big).\label{corrKMS2b}
\end{align}
\end{subequations}
To ensure the exponents of the exponentials involved in \eqref{corrKMS2a} (resp. \eqref{corrKMS2b}) to decay, the imaginary part of $z=t+is$ have to satisfy the following bounds $-\beta<s<0$ (resp. $0<s<\beta$). We conclude that $G^\beta_\pm$ define holomorphic functions in these respective strips, and we have, for any $z\in\mathbb{C}$ such that $0\leq \text{Im}(z)\leq\beta$,
\begin{equation}
G_{-}^\beta(z;A,B)=G_{+}^\beta(z-i\beta;A,B).
\end{equation}
Written in term of expectation value $\langle\cdot\rangle_\beta$, we finally obtain the celebrated KMS condition at temperature $\beta^{-1}$
\begin{equation}\label{trueKMS}
\langle B\Sigma_z(A)\rangle_\beta=\langle\Sigma_{z-i\beta}(A)B\rangle_\beta,
\end{equation}
which can be related to a periodicity property for the thermal (2-point) correlation function.\smallskip

We can link this derivation to our case of study by formally rewriting eq. \eqref{statekms1} 
\begin{align}
\omega(A):=\langle A\rangle_\beta.
\end{align}
Hence, eq. \eqref{trueKMS} becomes
\begin{equation}
\omega\big(B\Sigma_z(A)\big)=\omega\big(\Sigma_{z-i\beta}(A)B\big),
\end{equation}
which for $z=0$ reads 
\begin{equation}
\omega(BA)=\omega(\Sigma_{-i\beta}(A)B).
\end{equation}
Recall that, using $\zeta$, eq. \eqref{zeta}, the twisted trace property of the Lebesgue integral, eq. \eqref{twistedtrace} becomes
\begin{equation}\label{twistedtrace2}
\zeta\big(f\star g\big)=\zeta\big((\sigma\triangleright g)\star f\big).
\end{equation}
Hence, identifying $\zeta$ with $\omega$, and $\sigma$ with $\Sigma_{-i\beta}$, we find that eq. \eqref{twistedtrace} looks like a KMS condition for $\zeta$. We now characterised this assertion in more details.
\subsubsection{KMS weight on \texorpdfstring{$\mathcal{M}_\kappa$}{k-Minkowski}.}
To show that eq. \eqref{twistedtrace2} actually reflects the occurrence of a KMS condition, we can show that $\zeta$ defines a KMS weight on $\mathcal{M}_\kappa$ for the one-parameter group of *-automorphisms $\lbrace\sigma_t\rbrace_{t\in\mathbb{R}}$, $\sigma_t\in\text{Aut}(\mathcal{M}_\kappa)$, defined by
\begin{subequations}
\begin{equation}\label{defautomorphism}
\sigma_t(f):=e^{\frac{3t}{\kappa}\partial_0}\triangleright f,\ t\in\mathbb{R},\ f\in\mathcal{M}_\kappa.
\end{equation} 
Let us first characterised $\lbrace\sigma_t\rbrace_{t\in\mathbb{R}}$. Standard computations show that 
\begin{equation}\label{modulargroup}
\sigma_{t_1}\circ\sigma_{t_2}=\sigma_{t_1+t_2},\ \sigma_t^{-1}=\sigma_{-t},\ \forall t,t_1,t_2\in\mathbb{R},
\end{equation}
where $\circ$ is the composition of functions, \textit{i.e.} $\sigma_{t_1}\circ\sigma_{t_2}(f)=\sigma_{t_1}\big(\sigma_{t_2}(f)\big)$, together with
\begin{equation}\label{modular-sigma}
\sigma_t (f\star g)=\sigma_t(f) \star \sigma_t(g),\ \sigma_t(f^\ddagger)=\sigma_t(f)^\ddagger,\ \forall t\in\mathbb{R},\ f,g\in\mathcal{M}_\kappa.
\end{equation}
\end{subequations}
Thus, $\lbrace\sigma_t\rbrace_{t\in\mathbb{R}}$ defines a group of *-automorphisms of $\mathcal{M}_\kappa$.\smallskip

In order to link together $\sigma_t$ and the twist $\sigma$ appearing in eq. \eqref{twistedtrace}, we have to extend $t\mapsto\sigma_t$ to the complex plane. We define
\begin{subequations}
\begin{equation}\label{sigmat-modul-z}
\sigma_z(f):=e^{\frac{3z}{\kappa}\partial_0}\triangleright f,\ \forall z\in\mathbb{C},\ f\in\mathcal{M}_\kappa,
\end{equation}
such that \eqref{modulargroup} and \eqref{modular-sigma} extend respectively to
\begin{align}
&\sigma_{z_1}\sigma_{z_2}=\sigma_{z_1+z_2},\ \sigma^{-1}_z=\sigma_{-z},\ \forall z,z_1,z_2\in\mathbb{C}, \label{prop-modulargroup-z}\\
&\sigma_z(f\star g)=\sigma_z(f)\star\sigma_z(g),\ \forall z\in\mathbb{C},\label{morphalg-modul-z}
\end{align}
while $\sigma_z$ is no longer an automorphism of $^*$-algebra. Instead, we have
\begin{equation}\label{modulargroup-z}
\sigma_z(f^\ddagger)=\sigma_{\bar{z}}(f)^\ddagger,\ \forall z\in\mathbb{C}.
\end{equation}
\end{subequations}
In particular, the twist $\sigma$, eq. \eqref{twist-def}, is recovered for $z=i$, \textit{i.e.}
\begin{equation}\label{b10}
\sigma=\sigma_{z=i},
\end{equation}
and one has $\sigma(f^\ddagger)=\sigma^{-1}(f)^\ddagger$. This type of automorphism is known as a regular automorphism and occurs in the framework of twisted spectral triples. It has been introduced in
\cite{Connes:2008} in conjunction with the assumption of the existence of a distinguished group of *-automorphisms of the algebra indexed by one real parameter, says $t$, \textit{i.e.} the modular group, such that the analytic extension $\sigma_{t=i}$ coincides precisely with the regular automorphism. Here, the modular group linked with the twisted trace is defined by $\lbrace\sigma_t\rbrace_{t\in\mathbb{R}}$ while the twist $\sigma=\sigma_{t=i}$ defines the related regular automorphism.\smallskip

Recall that a KMS weight on a C*-algebra $\mathcal{A}$ for a modular group of $^*$-automorphisms $\{\sigma_t\}_{t\in\mathbb{R}}$ is defined
\cite{Kustermans:1997} as a (densely defined lower semi-continuous) linear map $\zeta:\mathcal{A}_+\to\mathbb{R}^+$, where $\mathcal{A}_+$ is the set of positive elements of $\mathcal{A}$, such that $\lbrace\sigma_t\rbrace_{t\in\mathbb{R}}$ admits an analytic extension, still a (norm continuous) one-parameter group, $\lbrace\sigma_z\rbrace_{z\in\mathbb{C}}$, acting on $\mathcal{A}$ satisfying the following two conditions 
\begin{subequations}\label{prop-kmsweight}
\begin{align}
&\zeta\circ\sigma_z=\zeta,\label{prop-kmsweighta}\\
&\zeta(a^\ddagger \star a)=\zeta\big(\sigma_{\frac{i}{2}}(a)\star\sigma_{\frac{i}{2}}(a)^\ddagger\big),\ a\in\text{Dom}\big(\sigma_{\frac{i}{2}}\big).\label{prop-kmsweightb}
\end{align}
\end{subequations}
The characterization of the relevant C*-algebra has been discussed in Chap. \ref{sec-Minkowski}, see also Appendix \ref{sap-harmonic}. For our purpose, it will be sufficient to keep in mind that it involves $\mathcal{M}_\kappa$ as a dense *-subalgebra. For more mathematical details on KMS weights see, e.g.,
\cite{Kustermans:1997}. Note that the notion of KMS weight related to the present twisted trace has been already used in
\cite{Matassa:2013,Matassa:2014} to construct a modular spectral triple for $\kappa$-Minkowski space.\smallskip

On the one hand, the first property, eq. \eqref{prop-kmsweighta}, is found to be satisfied by $\zeta$ as a mere consequence of its $\kappa$-Poincar\'{e} invariance (in the sense of \eqref{measure-invariance}). We compute
\begin{align}
\zeta\big(\sigma_z(f)\big)&=\int d^4x\ \sigma_z(f)(x)=\int d^4x \ (e^{\frac{3z}{\kappa}\partial_0}\triangleright f)(x)\\
&=\int d^4x \ (\mathcal{E}^{-3iz}\triangleright f)(x)=\epsilon(\mathcal{E})^{-3iz}\int d^4x f(x)=\zeta(f)\nonumber.
\end{align}
On the other hand, using eq. \eqref{morphalg-modul-z}, \eqref{modulargroup-z}, and \eqref{twistedtrace}, we find 
\begin{align}\label{proof-kmsweight}
\zeta(\sigma_{\frac{i}{2}}(f)\star\sigma_{\frac{i}{2}}(f)^\ddagger)&=\int d^4x\ \sigma_{\frac{i}{2}}(f)\star\sigma_{-\frac{i}{2}}(f^\ddagger)=\int d^4x\ \sigma_{\frac{i}{2}}(f\star\sigma_{-i}(f^\ddagger))\\
&=\int d^4x\ f\star\sigma_{-i}(f^\ddagger)=\int d^4x\ \sigma(\sigma_{-i}(f^\ddagger))\star f
=\zeta(f^\ddagger\star f).\nonumber
\end{align}
This shows that the two properties \eqref{prop-kmsweight} are satisfied by \eqref{zeta}. Hence, $\zeta$ defines a KMS weight on $\mathcal{M}_\kappa$.

Now, the Theorem 6.36 of
\cite{Kustermans:1997} guarantees, for each pair $(a,b)\in\mathcal{A}$, the existence of a bounded continuous function $f:\Sigma\to\mathbb{C}$, where $\Sigma$ is the strip defined by $\{z\in\mathbb{C},\ 0\le\textrm{Im}(z)\le 1\}$, such that one has
\begin{equation}\label{KMS-abst}
f(t)=\zeta(\sigma_t(a)\star b),\ \ f(t+i)=\zeta(b\star\sigma_t(a)),
\end{equation}
which is nothing but an abstract version of the KMS condition introduced in eq. \eqref{trueKMS}. Therefore, the requirement of $\kappa$-Poincar\'{e} invariance trades the cyclicity of the Lebesgue integral for a KMS condition.\smallskip

Finally, note that $\sigma_t$, eq. \eqref{defautomorphism}, defines ``time translations" since we have $\sigma_t(\phi)(x_0,\vec{x})=\phi(x_0+\frac{3t}{\kappa},\vec{x})$. We can show that this evolution is transferred at the level of the operators stemming from the Weyl quantisation map $Q$, eq. \eqref{Weyl-Minkowski}. To see it, we introduce the Gelfand-Naimark-Segal (GNS) representation of $\mathcal{M}_\kappa$, $\pi_{GNS}:\mathcal{F}(\mathcal{S}_c)\to\mathcal{B}(\mathcal{H})$, defined as usual by $\pi_{GNS}(\phi)\cdot v=\phi\star v$ for any $v\in\mathcal{H}$. Then, we compute 
\begin{align}\label{compo2}
\pi_{GNS}(\sigma_t\phi)\cdot\omega&=(\sigma_t\phi)\star\omega=\sigma_t(\phi\star(\sigma_t^{-1}\omega))
=(\sigma_t\circ\pi_{GNS}(\phi)\circ\sigma_t^{-1})\cdot\omega\\
&=\big((\Delta_T)^{it}\circ\pi_{GNS}(\phi)\circ(\Delta_T)^{-it}\big)\cdot\omega,\nonumber
\end{align}
for any $\omega\in\mathcal{H}$ and any $\phi\in\mathcal{M}_\kappa$, and $\Delta_T$ is the Tomita operator given by
\begin{equation}
\Delta_T:=e^{\frac{3P_0}{\kappa}},
\end{equation}
which coincides with \eqref{function-modulaire-4d} and such that $\sigma_t=(\Delta_T)^{it}$. Equation \eqref{compo2} indicates that the modular group defined by $\lbrace\sigma_t\rbrace_{t\in\mathbb{R}}$ generates an evolution for the operators stemming from the Weyl quantisation map $Q$.
\subsection{Derivation of admissible kinetic terms.}\label{sec-kinetic}
We begin with the construction of the kinetic term $\mathcal{S}_{\kappa,\star}^\text{kin}[\phi,\phi^\ddagger]$ encoding the dynamics of two \textit{independent} massive \textit{free} $\mathbb{C}$-valued scalar fields $\phi$ and $\phi^\ddagger$ we assume to be described by the same kinematic, although they may have different masses which should differ by terms of order (at least) $\kappa^{-1}$ in view of (CP) and eq. \eqref{action-limit}.
\subsubsection{General structure.}
Making use of the Hilbert product, eq. \eqref{hilbert-def2}, we find that admissible expressions for $\mathcal{S}_{\kappa,\star}^\text{kin}$ are given by terms of the form $\langle\phi_a,T_a\phi_a\rangle_\star$ where $a$ labels the nature of the field (i.e. either $\phi$ or $\phi^\ddagger$) and $T_a:\mathcal{M}_\kappa\to\mathcal{M}_\kappa$, $T_a:=K+m_a$, denotes any selfadjoint kinetic operator $K:\mathcal{M}_\kappa\to\mathcal{M}_\kappa$, with dense domain in $\mathcal{M}_\kappa$, supplemented by a mass-like term $m_a$. The reality of the action functional follows as a mere consequence of the selfadjointness of $K$. Indeed, $\langle\phi_a,T_a\phi_a\rangle_\star=:\langle T_a^\dag\phi_a,\phi_a\rangle_\star=\langle T_a\phi_a,\phi_a\rangle_\star=\overline{\langle\phi_a,T_a\phi_a\rangle}_\star$.\\
The general expression for the kinetic term we consider in the present dissertation is then given by
\begin{equation}\label{action-kin}
\mathcal{S}^\text{kin}_{\kappa,\star}[\phi,\phi^\ddagger]=\frac{1}{4}\langle\phi,T_1\phi\rangle_\star+\frac{1}{4}\langle\phi^\ddagger,T_2\phi^\ddagger\rangle_\star.
\end{equation}
Notice that, we cannot turn off one of the two terms in eq. \eqref{action-kin} since the omitted term would eventually reappeared at the quantum level. This might signal the occurrence of an internal symmetry for the action functional, namely under the exchange $\phi\leftrightarrow\phi^\ddagger$. More evidence pointing in this direction will be given when studying the one-loop quantum properties of the various NCFT considered in next section.\smallskip

We further assume the kinetic operator $K$ to be a differential operator we identify with a function of the generators $\mathcal{E}$, $P_i$, of the translations' Hopf subalgebra $\mathfrak{T}_\kappa$.\footnote{Recall that the action of $P_\mu$ on the elements of $\mathcal{M}_\kappa$ is defined through the representation $P_\mu=-i\partial_\mu$. For explicit expressions see eq. \eqref{module-action} in Appendix \ref{sap-poincare}.} Formally, we have $K(\partial)\to K(P)$, which leads to the following expression
\begin{equation}\label{kin-kernel}
\left(Kf\right)(x)=\int d^4y \left(\int \frac{d^4k}{(2\pi)^4}\ \tilde{K}(k)e^{ik\cdot(x-y)}\right) f(y),
\end{equation}
valid for any $f\in\mathcal{M}_\kappa\cap\text{Dom}(K)$. Hence, the selfadjointness of $K$ requires its symbol $\tilde{K}$ to be real,\footnote{Indeed, as a differential operator $K$ defines an integral transform $(Kf)(x):=\int d^4y\ K(x,y)f(y)$ whose kernel has to satisfy $K(x,y)=\overline{K(y,x)}$ to ensure the selfadjointness of $K$. This can be shown by developing both expressions $\langle\phi,K\phi\rangle_\star$ and $\langle K\phi,\phi\rangle_\star$. The reality of $\tilde{K}$ follows directly from \eqref{kin-kernel}.} while the compatibility condition, eq. \eqref{pairing-involution}, among the various involutions yields
\begin{equation}\label{kin-involution}
\left(K(P)\triangleright\phi_a\right)^\ddagger=S\big(K(P)\big)^\dag\triangleright\phi_a^\ddagger=K\big(S(P)\big)\triangleright\phi_a^\ddagger,
\end{equation}
where we have used the fact that the antipode $S$ defines an algebra (linear) anti-homomorphism together with $[P_\mu,P_\nu]=0$ and \eqref{dag-hopfoperat}.

Now, straightforward computations show that
\begin{equation}\label{hilbert-prop}
\langle f^\ddagger,f^\ddagger\rangle_\star=\langle \sigma f,f\rangle_\star,
\end{equation}
which combined with \eqref{kin-involution} in the second term of \eqref{action-kin} yield
\begin{equation}
\langle\phi^\ddagger,T_2\phi^\ddagger\rangle_\star=\langle\sigma\left(T_2\phi^\ddagger\right)^\dag,\phi\rangle_\star=\langle\sigma S(T_2)\phi,\phi\rangle_\star=\langle\phi,\sigma S(T_2)\phi\rangle_\star.
\end{equation}
Hence, the kinematic for the $\phi^\ddagger$'s can actually be described in term of the $\phi$'s by appropriately adjusting the kinetic operator $T_2\to\sigma S(T_2)$. It follows that
\begin{equation}\label{action-kin2}
\mathcal{S}^\text{kin}_{\kappa,\star}[\phi,\phi^\ddagger]=\frac{1}{4}\langle\left(T_1+S(T_2)\sigma\right)\phi,\phi\rangle_\star.
\end{equation}
As we are going to see throughout this chapter, thanks to the integral representation of both the star product \eqref{star-4d} and the involution \eqref{invol-4d}, any $\kappa$-Poincar\'{e} invariant NCFT involving $\phi$, $\phi^\ddagger$, and the star product can be conveniently represented as an ordinary, albeit nonlocal, complex scalar field theory depending on $\phi$, $\bar{\phi}$, and the pointwise (commutative) product among functions. This will be formally achieved upon identifying $\mathcal{S}_{\kappa,\star}[\phi,\phi^\ddagger]$ with $\mathcal{S}_\kappa[\phi,\bar{\phi}]$. Although trivial, this identification will lead to great simplifications in the computation of the propagator as well as in the analysis of the quantum properties of the various NCFT under consideration as it will enable us to make use of standard techniques from path integral quantisation and perturbation theory, reducing the analysis to ordinary quantum field theory computations.\smallskip

Combining the expression of the Hilbert product \eqref{hilbert-def2} with \eqref{hilbert-prop2} in \eqref{action-kin2} yields
\begin{subequations}\label{kinetic-map}
\begin{equation}
\mathcal{S}^\text{kin}_{\kappa,\star}[\phi,\phi^\ddagger]\to\mathcal{S}^\text{kin}_{\kappa}[\phi,\bar{\phi}]=\frac{1}{2}\int d^4x_1d^4x_2\ \bar{\phi}(x_1)\mathcal{K}(x_1,x_2)\phi(x_2),
\end{equation}
where $\mathcal{K}$ denotes the (nontrivial) kinetic operator for the complex scalar field theory characterised by $\mathcal{S}_{\kappa}[\phi,\bar{\phi}]$, illustrating the above discussion.\footnote{Note that $\mathcal{K}$ is symmetric with respect to the (canonical) Hilbert product on $L^2(\mathbb{R}^4)$.} It is defined by
\begin{align}
&\mathcal{K}(x_1,x_2):=\int\frac{d^4k}{(2\pi)^4}\ \tilde{\mathcal{K}}(k)e^{ik\cdot(x_1-x_2)},\\
&\tilde{\mathcal{K}}(k):=\frac{1}{2}(\tilde{K}(k)+m^2_1)+\frac{1}{2}(\tilde{K}\big(S(k)\big)+m^2_2)e^{-3k^0/\kappa}.\label{kinetic-mapb}
\end{align}
\end{subequations}
From this expression can easily be computed the expression of the Feynman propagator $\Delta_F$ associated with such NCFT. This is achieved as usual by mere inversion of the kinetic operator $\mathcal{K}$, namely by solving, for any suitable test function $f$,
\begin{equation}
\int d^4x_2 d^4y\ \Delta_F(x_1,y)\mathcal{K}(y,x_2) f(x_2)=\int d^4 x_2\ \delta^{(\hspace{-1pt}4\hspace{-1pt})}(x_1-x_2)f(x_2).
\end{equation}
Standard computations yield
\begin{equation}\label{gen-propagator}
\Delta_F(x_1,x_2)=\int\frac{d^4k}{(2\pi)^4}\ \Delta_F(k)e^{ik\cdot(x_1-x_2)},\ \Delta_F(k):=1/\tilde{\mathcal{K}}(k).
\end{equation}

Assuming that $S(K)=K$, a condition that is fulfilled by the models considered in the present dissertation, the symbol of the above kinetic operator, eq. \eqref{kinetic-mapb}, takes the convenient form
\begin{subequations}\label{kin-hyp}
\begin{align}
&\tilde{\mathcal{K}}(k)=\frac{1}{2}\left(1+e^{-3k^0/\kappa}\right)\left(\tilde{K}(k)+M^2\right),\label{kin-hypa}\\
&M^2(k^0;m_1,m_2):=\frac{m_1^2+m_2^2e^{-3k^0/\kappa}}{1+e^{-3k^0/\kappa}},\label{full-mass}
\end{align}
\end{subequations}
where the masslike (energy dependent) term, $M^2(k^0)$, is bounded with (energy independent) bounds given by
\begin{equation}\label{bounds-mass}
\min(m_1^2,m_2^2)\leq M^2(k^0)\leq\max(m_1^2,m_2^2).
\end{equation}
This last result indicates that it is sufficient to consider the case $M=m(\kappa)\in\mathbb{R}$ constant to study the general quantum behaviour of such models of NCFT. Note that this condition is automatically satisfied if $m_1=m_2=m$, in which case eq. \eqref{full-mass} reduces to $M^2=m^2$. Under this assumption, the symbol of the propagator, eq. \eqref{gen-propagator}, becomes
\begin{equation}
\Delta_F(k):=\frac{2}{\left(1+e^{-3k^0/\kappa}\right)\left(\tilde{K}(k)+m^2\right)}.
\end{equation}

Before proceeding to the presentation of the kinetic operators used to investigate the quantum properties of various models of NCFT, two remarks are in order.
\begin{enumerate}[label={$\textit{(\roman*)}$}]
\item{First, the kinetic operators considered in the present study will be assumed to be square of Dirac operators, \textit{i.e.} $K=\mathcal{D}_\mu \mathcal{D}^\mu$. This choice is motivated by the early proposal
\cite{Chamseddine:1997,Chamseddine:1996,Chamseddine:2007} that a Hilbert-Einstein-Yang-Mills action functional for describing fundamental physics might be provided by the spectral action associated with some spectral triple.\footnote{Recall that the spectral action is essentially a regularized heat kernel expansion of the Dirac operator defining the spectral triple.}$^{,}$\footnote{Note however that, as far as we now, no spectral triple (enjoying all of the required properties to defined a spectral triple) \textit{\`{a} la} Connes for $\kappa$-Minkowski has been constructed until now.} The kinetic operator $K(P)$ being a function of $\mathcal{E}$ and $P_i$, so are the $\mathcal{D}_\mu$'s. From this follows the selfadjointness of the $\mathcal{D}_\mu$'s. Indeed, assuming
\begin{equation}
\mathcal{D}_\mu(P)=\sum_{\alpha_0,\cdots,\alpha_3}\lambda_{\alpha_0,\cdots,\alpha_3;\mu}\ \mathcal{E}^{\alpha_0}P_1^{\alpha_1}\cdots P_3^{\alpha_3},\ \ \lambda_{\alpha_0,\cdots,\alpha_3;\mu}\in\mathbb{R},
\end{equation}
we easily find that, for any $f,g\in\mathcal{M}$,
\begin{equation}
\langle f,\mathcal{D}_\mu g\rangle_\star=\sum_{\alpha_0,\cdots,\alpha_3}\lambda_{\alpha_0,\cdots,\alpha_3;\mu}\ \langle f,\mathcal{E}^{\alpha_0}P_1^{\alpha_1}\cdots P_3^{\alpha_3} g\rangle_\star=\langle \mathcal{D}_\mu f,g\rangle_\star,
\end{equation}
where we have used the selfadjointness of $\mathcal{E}$ and $P_i$ together with $[P_\mu,P_\nu]=0$ to show the last equality. Finally, observe that from the selfadjointness of $\mathcal{D}_\mu$ follows the pleasant property
\begin{equation}
\langle f,K_\kappa g\rangle_\star=\langle f,\mathcal{D}^\mu\mathcal{D}_\mu g\rangle_\star=\langle\mathcal{D}^\mu f,\mathcal{D}_\mu g\rangle_\star,\ \ \forall f,g\in\mathcal{M}_\kappa.
\end{equation}
However, in view of eq. \eqref{derivtwist}, there is no reason for the $\mathcal{D}_\mu$'s to be derivations of the algebra $\mathcal{M}_\kappa$ since, in general, $\mathcal{D}_\mu(f\star g)\neq\mathcal{D}_\mu(f)\star g+f\star\mathcal{D}_\mu(g)$, $f,g\in\mathcal{M}_\kappa$.}
\item{Next, we mention that a full family of kinetic operators, hence propagators, associated with $K$, which are still compatible with the desired properties for $S^\text{kin}_{\kappa,\star}$, can be obtained upon substituting
\begin{equation}\label{fields-family}
\phi\to\phi_{\alpha}:=\mathcal{E}^{\alpha}\phi,\ \ \phi^\ddagger\to\phi^\ddagger_{\alpha}:=\mathcal{E}^{-\alpha}\phi^\ddagger,\ \ \alpha\in\mathbb{R},
\end{equation}
in \eqref{action-kin}. This family of fields, labelled by $\alpha$, are formally obtained from the action of some power of the twist factor \eqref{twist-def} on $\phi$ and $\phi^\ddagger$, all of them admitting the same commutative limit.\footnote{Indeed, setting $\alpha=3\alpha'$ in \eqref{fields-family}, these transformations read $\phi_{\alpha}=\sigma^{\alpha'}\phi$ and $\phi^\ddagger_{\alpha}=\sigma^{-\alpha'}\phi^\ddagger$. }$^,$\footnote{Note that the respective powers of the twist factor in front of $\phi$ and $\phi^\ddagger$ are not independent. This is essential to ensure consistency of relations of the form $\langle\phi,\phi\rangle=\langle\sigma\phi^\ddagger,\phi^\ddagger\rangle$.} Now, let $\mathcal{O}:\mathcal{M}_\kappa\to\mathcal{M}_\kappa$ be a selfadjoint operator with dense domain in $\mathcal{M}_\kappa$ and let $f_\alpha:=\mathcal{E}^\alpha f$ for any $f\in\mathcal{M}_\kappa$. Then,
\begin{equation}
\langle f_\alpha,\mathcal{O}f_\alpha\rangle_\star=\langle\mathcal{E}^\alpha f,\mathcal{O}\mathcal{E}^\alpha f\rangle_\star=\langle f,\mathcal{O}_\alpha f\rangle_\star,
\end{equation}
where we have used the selfadjoitness of $\mathcal{E}$ to obtain the last equality and defined
\begin{equation}\label{op-family}
\mathcal{O}_\alpha:=\mathcal{E}^\alpha\mathcal{O}\mathcal{E}^\alpha,\ \forall\alpha\in\mathbb{R}.
\end{equation}
Upon applying this latter relation to \eqref{action-kin2}, mere adaptation of the derivation leading to \eqref{kinetic-mapb} yield
\begin{equation}\label{kin-family}
\tilde{\mathcal{K}}_\alpha(k):=e^{-2\alpha k^0/\kappa}\tilde{\mathcal{K}}(k),\ \alpha\in\mathbb{R}.
\end{equation}
Assuming $S(K)=K$, the previous factorisation \eqref{kin-hyp}, and related discussions still applied and the family of corresponding propagator is finally found to be given by
\begin{equation}\label{family-propagator}
\Delta_{F,\alpha}(k):=e^{2\alpha k^0/\kappa}\Delta_{F}(k),\ \alpha\in\mathbb{R}.
\end{equation}
}
\end{enumerate}
\subsubsection{Casimir kinetic operator.}
The simplest (natural) example of kinetic operator we can think about to generalise the ordinary scalar field theory is that of the first Casimir operator $\mathcal{C}_\kappa$ of the $\kappa$-Poincar\'{e} algebra. This latter is given, in the Majid-Ruegg basis, by
\begin{equation}\label{Casimir0}
\mathcal{C}_\kappa(P):=4\kappa^2\sinh^2\left(\frac{P_0}{2\kappa}\right) + e^{P_0/\kappa}P_iP^i.
\end{equation}
For latter convenience, it is useful to rewrite the Casimir operator as
\begin{equation}\label{Casimir}
\mathcal{C}_\kappa(P)=\mathcal{E}^{-1}\left(\mathcal{P}_0^2+P_iP^i\right),\ \mathcal{P}_0:=\kappa\left(1-\mathcal{E}\right).
\end{equation}
From these expressions, we readily infer that $S(\mathcal{C}_\kappa)=\mathcal{C}_\kappa$, while $\mathcal{C}_\kappa\to P_\mu P^\mu$ in the limit $\kappa\to\infty$. Actually, any polynomial in $\mathcal{C}_\kappa$ satisfies these properties and could, in principle, be used as kinetic operator. Observe that, in view of the expression of the first Casimir of the (ordinary) Poincar\'{e} algebra, \textit{i.e.} $\mathcal{C}(P):=P_\mu P^\mu$, it seems natural to interpret the quantity $\mathcal{P}_0$ appearing in eq. \eqref{Casimir}, which by the way reduces to $P_0$ in the low energy limit, as the natural quantity replacing the standard (undeformed) energy, ${P}_0$, in the context of $\kappa$-deformed theories involving deformed dispersion relation. This is supported by the role already played by $\mathcal{E}$ in the description of the $\kappa$-deformed translations' Hopf subalgebra encoding some of the symmetries of $\kappa$-Minkowski. Finally, $\mathcal{C}_\kappa$ can be put into the form $\mathcal{C}_\kappa(P)=\mathcal{D}_0^2+\mathcal{D}_i\mathcal{D}^i$, where the selfadjoint operators $\mathcal{D}_\mu$ are defined by
\begin{equation}
\mathcal{D}_0:=\kappa \mathcal{E}^{-1/2}(1-\mathcal{E}),\ \mathcal{D}_i:= \mathcal{E}^{-1/2}P_i.
\end{equation}

Identifying $\tilde{K}(k)$ with $\mathcal{C}_\kappa(k)$ in \eqref{family-propagator} leads to the following expression for the Feynman propagator associated with the Casimir kinetic operator
\begin{subequations}\label{c-propagator}
\begin{align}
&\Delta^c_{F;\alpha}(k)=\frac{e^{(2\alpha-1)k^0/\kappa}}{\left(1+e^{-3k^0/\kappa}\right)}\frac{2}{\|\vec{k}\|^2+\kappa^2\mu_c^2(k^0)},\\
&\mu_c^2(k^0;m):=\left(1-e^{-k^0/\kappa}\right)^2+\left(m/\kappa\right)^2e^{-k^0/\kappa}.
\end{align}
\end{subequations}
Let us investigate in more details the properties of the above (positive) propagator.\\
On the one hand, its decay properties can be studied by taking the large momentum limit in eq. \eqref{c-propagator}. Namely, keeping the energy $k^0$ fixed, we find
\begin{subequations}\label{cuv-propagator}
\begin{equation}
\lim_{\|\vec{k}\|\to\infty}\Delta^c_{F;\alpha}(k)=\lim_{\|\vec{k}\|\to\infty}\frac{1}{\|\vec{k}\|^2}=0,
\end{equation}
while, keeping $\|\vec{k}\|$ fixed, we find
\begin{align}
&\lim_{k^0\to+\infty}\Delta^c_{F;\alpha}(k)=\|\vec{k}\|^{-2}\lim_{k^0\to+\infty}e^{(2\alpha-1)k^0/\kappa},\label{casimir1c}\\
&\lim_{k^0\to-\infty}\Delta^c_{F;\alpha}(k)=\lim_{k^0\to-\infty}e^{(4+2\alpha)k^0/\kappa}.\label{casimir2c}
\end{align}
\end{subequations}
We can infer from the above results that the propagator vanishes at large (infinite) momenta if, and only if,
\begin{equation}
-2<\alpha\leq\frac{1}{2},
\end{equation}
hence restricting the number of admissible transformations \eqref{fields-family}, then kinetic operators \eqref{kin-family}. On the other hand, the massless case ($m=0$) is singular in the infrared as its commutative counterpart. This is apparent from
\begin{equation}
\Delta^c_F(k)=\frac{e^{-k^0/\kappa}}{1+e^{-3k^0/\kappa}}\ \frac{2}{\mathcal{P}_0^2(k^0)+\|\vec{k}\|^2},
\end{equation}
which diverges when  $\mathcal{P}_0$ and $\|\vec{k}\|$ are taken simultaneously to 0. Since $\mathcal{P}_0:\mathbb{R}\to]-\infty,\kappa[$ is in one-to-one correspondence with $k^0$, we concludes that the assertion ``infrared singular" can be understood in its ordinary sense.
\subsubsection{Equivariant kinetic operator}
A second natural choice for the kinetic operator is provided by the square of the $\mathcal{U}_\kappa(\text{iso}(4))$-equivariant Dirac operator appearing in the construction of an equivariant spectral triple aiming to encode the geometry of $\kappa$-Minkowski
\cite{dAndrea:2006}.\footnote{Note however that, as mentioned by the author of
\cite{dAndrea:2006}, this Dirac operator does not satisfy the axioms for defining a spectral triple.} It is defined by
\begin{equation}
\mathcal{D}^\text{eq}_0=\frac{\mathcal{E}^{-1}}{2\kappa}\big(\kappa^2(1-\mathcal{E}^2)-P_iP^i\big),\ \mathcal{D}^\text{eq}_i=\mathcal{E}^{-1}P_i.
\end{equation}

A useful factorisation of the equivariant kinetic operator $K^\text{eq}$, when supplemented by a masslike term $m$, is given, assuming $m^2\leq\kappa^2$, by
\begin{subequations}
\begin{align}
&\tilde{K}^\text{eq}(k)+m^2 = \frac{e^{2k^0/\kappa}}{4\kappa^2}\left(\|\vec{k}\|^2+\kappa^2\mu^2_{+}(k^0)\right)\left(\|\vec{k}\|^2+\kappa^2\mu^2_{-}(k^0)\right),\\
&\mu^2_{\pm}(k^0;m):=1+ e^{-2k^0/\kappa}\pm 2e^{-k^0/\kappa} \sqrt{1-\left(m/\kappa\right)^2}.
\end{align}
\end{subequations}
This leads to the following expression for the Feynman propagator
\begin{equation}\label{eq-propagator}
\Delta_{F;\alpha}^\text{eq}(k)=\frac{e^{2(\alpha-1)k^0/\kappa}}{1+e^{-3k^0/\kappa}}\ \frac{8\kappa^2}{(\|\vec{k}\|^2+\kappa^2\mu^2_{+})(\|\vec{k}\|^2+\kappa^2\mu^2_{-})},
\end{equation}
whose decay properties are given by
\begin{subequations}\label{eq-decay}
\begin{align}
&\lim_{\|\vec{k}\|\to\infty}\Delta^\text{eq}_{F;\alpha}(k)=\lim_{\|\vec{k}\|\to\infty}\frac{1}{\|\vec{k}\|^4}=0,\\
&\lim_{k^0\to+\infty}\Delta^\text{eq}_{F;\alpha}(k)=\|\vec{k}\|^{-4}\lim_{k^0\to+\infty}e^{2(\alpha-1)k^0/\kappa},\\
&\lim_{k^0\to-\infty}\Delta^\text{eq}_{F;\alpha}(k)=\lim_{k^0\to-\infty}e^{(5+2\alpha)k^0/\kappa},
\end{align}
\end{subequations}
indicating that, this time, the propagator vanishes at large momenta if, and only if,
\begin{equation}
-\frac{5}{2}<\alpha\leq1.
\end{equation}
Again, the propagator is IR singular as it is apparent from eq. \eqref{eq-Casimir} below.\smallskip

It is interesting to notice that the equivariant kinetic operator is related to the Casimir operator through the relation
\begin{equation}\label{eq-Casimir}
K^\text{eq}=\mathcal{C}_\kappa\left(1+\frac{1}{4\kappa^2}\mathcal{C}_\kappa\right).
\end{equation}
It follows from this observation that the ``equivariant propagator," eq. \eqref{eq-propagator}, can actually be regarded as a Pauli-Villars regularised version of the ``Casimir propagator," eq. \eqref{c-propagator}. This is obvious when considering the massless theory, $m=0$, namely
\begin{subequations}\label{c-Villars}
\begin{align}\label{c-Pauli}
\Delta_F^\text{eq}(k)=\frac{2}{1+e^{-3k^0/\kappa}}\left(\frac{1}{\mathcal{C}_\kappa(k)}-\frac{1}{\mathcal{C}_\kappa(k)+4\kappa^2}\right),
\end{align}
$2\kappa$ playing the role of a Pauli-Villars cutoff. A similar interpretation is still possible in the massive case. To do so, it is first convenient to use a partial fraction decomposition to write the propagator \eqref{eq-propagator} as 
\begin{align}
&\Delta_F^\text{eq}(k)=\frac{2\kappa}{\sqrt{\kappa^2-m^2}\left(1+e^{-3k^0/\kappa}\right)}\left(\frac{1}{C_\kappa+m_{-}^2}-\frac{1}{C_\kappa+m_{+}^2}\right),\\
&m_\pm(m,\kappa):=2\kappa^2\left(1\pm\sqrt{1-(m/\kappa)^2}\right),
\end{align}
then to identify $m_{-}$ with the bare mass for the NCFT. In that case, the initial parameter $m$ becomes $m=m_{-}\sqrt{1-(m_{-}/2\kappa)^2}$ and $m_{+}=4\kappa^2-m_{-}$, such that
\begin{equation}
\Delta_F^\text{eq}(k)=\frac{4\kappa^2}{(2\kappa^2-m_{-}^2)\left(\lambda_1+\lambda_2e^{-3k^0/\kappa}\right)}\left(\frac{1}{C_\kappa+m_{-}^2}-\frac{1}{C_\kappa+4\kappa^2-m_{-}^2}\right),
\end{equation}
\end{subequations}
where the cutoff is now $\sqrt{4\kappa^2-m_{-}^2}$ and \eqref{c-Pauli} is recovered in the limit $m_{-}\to0$.
\subsubsection{Modular kinetic operator.}
A third example of kinetic operator is provided by the square of the Dirac operator introduced in
\cite{Matassa:2014,Matassa:2013} in the attempt to construct a modular spectral triple for $\kappa$-Minkowski. This Dirac operator is characterised by
\begin{equation}
\mathcal{D}^m_0:=\kappa(1-\mathcal{E}),\ \mathcal{D}^m_i:= P_i,
\end{equation}
such that $K^m(P)=\mathcal{E}\mathcal{C}_\kappa(P)$. Hence, unlike both the Casimir and equivariant kinetic operators, the modular kinetic operator does not satisfy the relation $S(K)=K$. Instead,
\begin{equation}\label{lost}
S(K^m)=S(\mathcal{C}_\kappa)S(\mathcal{E})=\mathcal{E}^{-1}\mathcal{C}_\kappa=\mathcal{E}^{-2}K^m(P),
\end{equation}
and the factorisation \eqref{kin-hyp} does not hold anymore. Going back to eq. \eqref{kinetic-map}, we find
\begin{subequations}
\begin{equation}
\tilde{\mathcal{K}}^m(k)=\frac{1}{2}\left(1+e^{-k^0/\kappa}\right)\left(\tilde{K}^m(k)+M^2(k^0)\right),
\end{equation}
where $M^2$ is now given, assuming $m_1=m_2=m$, by\footnote{We made used of the decomposition $1+y^3=(1+y)(1-y+y^2)$.}
\begin{equation}\label{massmod}
M^2(y;m):=(1-y+y^2) m^2.
\end{equation}
\end{subequations}
Simple inspection of eq. \eqref{massmod} shows that no bounds such as those previously found in eq. \eqref{bounds-mass} can be used in the present case to simplify the computations. In particular, we can no longer treat $M$ as a constant of the energy.\footnote{This conclusion would have been the same in the more general case $m_1\neq m_2$.}

The corresponding family of propagators is given by
\begin{subequations}
\begin{align}
&\Delta^m_{F;\alpha}(k)=\frac{e^{2\alpha k^0/\kappa}}{1+e^{-k^0/\kappa}}\ \frac{2}{\|\vec{k}\|^2+(\kappa^2+m^2)\mu_\text{m}^2(k^0)},\\
&\mu_\text{m}^2(k^0;m):=1-\left(1+\frac{\kappa^2}{\kappa^2+m^2}\right)e^{-k^0/\kappa}+e^{-2k^0/\kappa},
\end{align}
\end{subequations}
whose decay properties are given by
\begin{subequations}
\begin{align}
&\lim_{\|\vec{k}\|\to\infty}\Delta^m_{F;\alpha}(k)=\lim_{\|\vec{k}\|\to\infty}\frac{1}{\|\vec{k}\|^2}=0,\\
&\lim_{k^0\to+\infty}\Delta^m_{F;\alpha}(k)=\|\vec{k}\|^{-2}\lim_{k^0\to+\infty}e^{2\alpha k^0/\kappa},\label{mod1c}\\
&\lim_{k^0\to-\infty}\Delta^m_{F;\alpha}(k)=\lim_{k^0\to-\infty}e^{(3+2\alpha)k^0/\kappa},\label{mod2c}
\end{align}
\end{subequations}
indicating that this propagator vanishes at large momentum if, and only if,
\begin{equation}
-\frac{3}{2}<\alpha\leq0.
\end{equation}

Let us compare the decay properties of the modular propagator with those of the Casimir propagator. We first observe that they have the same dependence in the spacelike variable, $\vec{k}$. Comparing their respective behaviour at large energy (large $k^0$), namely \eqref{casimir1c} with \eqref{mod1c}, and \eqref{casimir2c} with \eqref{mod2c}, we find
\begin{subequations}
\begin{align}
&\lim_{k^0\to+\infty}\frac{\Delta^m_{F;\alpha}}{\Delta^c_{F;\alpha}}(k)=\lim_{k^0\to+\infty}e^{k^0/\kappa},\\
&\lim_{k^0\to-\infty}\frac{\Delta^m_{F;\alpha}}{\Delta^c_{F;\alpha}}(k)=\lim_{k^0\to-\infty}e^{-k^0/\kappa}.
\end{align}
\end{subequations}
From the above results, it is to expect the UV behaviour of the NCFT built from the modular kinetic operator to be worth (\textit{i.e.} more divergent) than the UV behaviour of NCFT built from the Casimir kinetic operator. We have checked that it is indeed the case. This propagator also has a pole at $(0,\vec{0})$ in the massless case. For these reasons, and to not overload the presentation, we will not study this model further. 
\subsection{Derivation of admissible interaction potentials.}\label{sec-interaction}
We now turn to the analysis of the interaction term $S_{\kappa,\star}^\text{int}$. As already mentioned at the beginning of this section, a sufficient condition to ensure the reality of the action functional $S_{\kappa,\star}$ consists in considering interaction terms of the form $\langle f,f\rangle_\star$ where $f\in\mathcal{M}_\kappa$ is any (star) polynomial in the fields $\phi$ and $\phi^\ddagger$. It follows that, in contrast with the commutative $\vert\phi\vert^4$ model for which there exists only one (local) interaction, one can easily exhibit for its ($\kappa$-Poincar\'e invariant) noncommutative counterpart four different (nonlocal) interactions which result from the noncommutativity of the star product together with the noncyclicity of the integral involved in $S_{\kappa,\star}$.\smallskip

According to the terminology of NCFT we distinguish two \textit{orientable interactions}
\begin{subequations}\label{interactions}
\begin{align}\label{o-int}
S_{\kappa,\star}^{\text{int},\textit{o}}[\phi,\phi^\ddagger]:=&\ \frac{g_1}{4!}\langle\phi^\ddagger\star\phi,\phi^\ddagger\star\phi\rangle_\star+\frac{g_2}{4!}\langle\phi\star\phi^\ddagger,\phi\star\phi^\ddagger\rangle_\star,\ \ g_1,g_2\in\mathbb{R},
\end{align}
where the $\phi$'s alternate with the $\phi^\ddagger$'s, and two \textit{nonorientable interactions}\footnote{For more technical details on the diagrammatic associated with orientable/nonorientable interactions see, e.g.,
\cite{Vignes:2006,Vignes:2007} for NCFT on Moyal space and
\cite{moi:2016} for the $\mathbb{R}^3_\theta$ case and references therein.}
\begin{align}\label{no-int}
S_{\kappa,\star}^{\text{int},\textit{no}}[\phi,\phi^\ddagger]:=&\ \frac{g_3}{4!}\langle\phi\star\phi,\phi\star\phi\rangle_\star+\frac{g_4}{4!}\langle\phi^\ddagger\star\phi^\ddagger,\phi^\ddagger\star\phi^\ddagger\rangle_\star,\ \ g_3,g_4\in\mathbb{R}.
\end{align}
\end{subequations}
Now, identifying $S_{\kappa,\star}^{\text{int},\textit{o}}[\phi,\phi^\ddagger]\to S_{\kappa}^{\text{int},\textit{o}}[\phi,\bar{\phi}]$, the ``orientable model" reduces to
\begin{subequations}\label{o-int2}
\begin{equation}
S_{\kappa}^{\text{int},\textit{o}}(\bar{\phi},\phi)=\frac{1}{4!}\int \prod_{\ell=1}^4\left[\frac{d^4k_\ell}{(2\pi)^4}\right] \bar{\phi}(k_1)\phi(k_2)\bar{\phi}(k_3)\phi(k_4)\mathcal{V}_\textit{o}(k_1,k_2,k_3,k_4),
\end{equation}
where the nonlocal 4-vertex function $\mathcal{V}_\textit{o}$ is defined by
\begin{align}
&\mathcal{V}_\textit{o}(k_1,k_2,k_3,k_4):=(2\pi)^4\left(g_1+g_2e^{3k_1^0/\kappa}\right)V_\textit{o}(k_1,k_2,k_3,k_4),\label{o-vertex}\\
&V_\textit{o}(k_1,k_2,k_3,k_4):=\delta(k_4^0-k_3^0+k_2^0-k_1^0)\delta^{(\hspace{-1pt}3\hspace{-1pt})}((\vec{k}_4-\vec{k}_3)e^{k_4^0/\kappa}+(\vec{k}_2-\vec{k}_1)e^{k_1^0/\kappa}).\label{o-delta}
\end{align}
\end{subequations}
In the same way, we find for the ``nonorientable model"
\begin{subequations}\label{no-int2}
\begin{align}
&S_{\kappa}^{\text{int},\textit{no}}(\bar{\phi},\phi)=\frac{1}{4!}\int \prod_{\ell=1}^4\left[\frac{d^4k_\ell}{(2\pi)^4}\right] \bar{\phi}(k_1)\phi(k_2)\bar{\phi}(k_3)\phi(k_4)\mathcal{V}_\textit{no}(k_1,k_2,k_3,k_4),\\
&\mathcal{V}_\textit{no}(k_1,k_2,k_3,k_4):=(2\pi)^4\left(g_3+g_4e^{-3(k_1^0+k_3^0)/\kappa}\right)V_\textit{no}(k_1,k_2,k_3,k_4),\label{no-vertex}\\
&V_\textit{no}(k_1,k_2,k_3,k_4):=\delta(k_4^0-k_3^0+k_2^0-k_1^0)\delta^{(\hspace{-1pt}3\hspace{-1pt})}(\vec{k}_4-\vec{k}_3+e^{-k_4^0/\kappa}\vec{k}_2-e^{-k_3^0/\kappa}\vec{k}_1).\label{no-delta}
\end{align}
\end{subequations}

We are now in position to compute radiative corrections to both the 2-point and 4-point functions for all of the various models presented above. Before proceeding to the analysis, let us comment on the admissible expressions for the interaction term. 
\begin{enumerate}[label={$\textit{(\roman*)}$}]
\item{First, taking the limit $\kappa\to\infty$ in \eqref{interactions}, we recover the expected commutative limit, \textit{i.e.} $(\bar{g}/4!)\int d^4x\vert\phi(x)\vert^4$, provided the coupling constants $g_i$ differ from their commutative counterpart $\bar{g}$ by terms of order at least $\kappa^{-1}$;}
\item{Next, eq. \eqref{o-delta} and \eqref{no-delta} exhibit the energy-momentum conservation laws for each of these theories. As expected, the conservation law for the energy (timelike momenta) sector is the standard one, while the 3-momentum conservation law is nonlinear. This merely reflects the semidirect product structure of the group underlying the noncommutative C*-algebra of fields modelling $\kappa$-Minkowski, as well as the deformed Hopf algebraic (coproduct) structure of the $\kappa$-Poincar\'{e} algebra underlying its (quantum) symmetries. Note this has been sometimes geometrically interpreted (for instance in the context of relative locality
\cite{Amelino:2011,Gubitosi:2013,Amelino:2013}) as reflecting the curvature of the energy-momentum space at very high (i.e. of order $\kappa$) energy;}
\item{Finally, as in $\S$\ref{sec-kinetic}, full families of orientable and nonorientable interactions can be obtained upon performing the substitution \eqref{fields-family} in \eqref{interactions}. The family of 4-vertex functions corresponding to this new set of interactions are then obtained by substituting
\begin{subequations}
\begin{align}
&\mathcal{V}_\textit{o}(k_1,k_2,k_3,k_4)&&\hspace{-3cm}\to e^{-2\alpha(k_1^0+k_3^0)/\kappa}\mathcal{V}_\textit{o}(k_1,k_2,k_3,k_4)\label{ov-gen},\\
&\mathcal{V}_\textit{no}(k_1,k_2,k_3,k_4)&&\hspace{-3cm}\to e^{-4\alpha(k_1^0+k_3^0)/\kappa}\mathcal{V}_\textit{no}(k_1,k_2,k_3,k_4),\label{nov-gen}
\end{align}
\end{subequations}
in \eqref{o-vertex} and \eqref{no-vertex} respectively.}
\end{enumerate}
\section{One-loop 2-point functions.}\label{sec-2point}
In the previous section, we have discussed the admissible expressions for a $\kappa$-Poincar\'{e} invariant action functional aiming to describe the dynamics of a self-interacting $\mathbb{C}$-valued scalar field $\phi$ on $\kappa$-Minkowski background. We now turn to the study of the quantum properties of various models of NCFT built from the material presented in there, each model being characterised by one of the kinetic operator given in $\S$\ref{sec-kinetic} together with a quartic interaction potential to be chosen among the interactions given in $\S$\ref{sec-interaction}.\\
In order to clarify the presentation, we treat separately the models with orientable interactions from those with nonorientable interactions, even though we will be able to gather the various results in the end. This is essentially because to each of these families of interactions correspond nonequivalent conservation laws between the 3-momenta as it is apparent from eq. \eqref{o-delta} and \eqref{no-delta}. The extremely different structure in the expressions for the corresponding vertex functions actually gives rise to very different quantum behaviours already at the level of the one-loop 2-point function. Indeed, the former family of interactions leads only to planar contributions to the one-loop 2-point function while nonorientable interactions lead to nonplanar contributions as well. We find that the planar contributions diverge in the UV for all of the models while the nonplanar contributions diverge only at zero external momenta albeit finite otherwise, likely indicating the occurrence of UV/IR mixing for some of the models.\smallskip

To deal with the perturbative expansion, we follow the usual route taken in most of the studies of NCFT, which we briefly recall now. The essential point of this derivation lies in the fact that any $\kappa$-Poincar\'e invariant action functional $S_{\kappa,\star}[\phi,\phi^\ddagger]$ involving the star product can be represented as an ordinary, albeit nonlocal, action functional $S_{\kappa}[\phi,\bar{\phi}]$ involving the commutative pointwise product among functions, hence describing the dynamics of an ordinary (self-interacting) complex scalar field. This has already been discussed at length in previous section $\S$\ref{sec-action} and we shall not argue more about it here except by recalling that this results essentially from the existence of an integral expression for the star product involved in the construction of the action functional. Accordingly, the perturbative expansion related to the NCFT is nothing but the usual perturbative expansion for an ordinary complex scalar field theory and is obtained upon expanding (up to the desired order) the generating functional of connected correlation functions $W$. This latter is defined from the partition function
\begin{equation}
\mathcal{Z}[J,\bar{J}]:=\int d\bar{\phi}d\phi \ e^{-S_\kappa[\phi,\bar{\phi}]+\int d^4x \left( \bar{J}(x)\phi(x) + J(x)\bar{\phi}(x)\right)},
\end{equation}
via the relation $W[J,\bar{J}]:=\ln\left(\mathcal{Z}[J,\bar{J}]\right)$.\smallskip

Note that the functional measure appearing in the expression of $\mathcal{Z}$ is merely the ordinary functional measure for a complex scalar field theory implementing formally the integration over the field configurations $\phi$ and $\bar{\phi}$. The correlation functions built from $\phi$ and $\bar{\phi}$ are then generated by the repeated action of standard functional derivatives with respect to $\bar{J}$ and $J$ which satisfy the usual functional rules
\begin{equation}
\frac{\delta J_a(p)}{\delta J_b(q)}=\delta_{ab}\delta^{(\hspace{-1pt}4\hspace{-1pt})}(p-q),
\end{equation}
where $a$ and $b$ label the nature of the source, \textit{i.e.} either $J$ or $\bar{J}$.\\
As a mere consequence of the above discussion, it follows that there is no need to introduce a notion of noncommutative (star) functional derivative in the present approach.\footnote{For another approach dealing with ``star functional derivatives" in the context of $\kappa$-Minkowski see, e.g.,
\cite{Amelino:2002b}.}\smallskip

To complete the derivation of the one-loop 2-point function, it remains to define the effective action $\Gamma$ as the Legendre transform of $W$ and read the expressions for the various contributions to the one-loop 2-point function from the expansion of $\Gamma$. These last steps of calculation are recall in details (as well as the derivation of the one-loop 4-point function) in Appendix \ref{sap-perturbation} to which we refer. The expression of the one-loop quadratic part of the effective action is finally found to be given by
\begin{subequations}
\begin{align}
&\Gamma^{(2)}_1[\phi,\bar{\phi}]:=\frac{1}{2}\int\frac{d^4k_1}{(2\pi)^4}\frac{d^4k_2}{(2\pi)^4}\  \bar{\phi}(k_1)\phi(k_2) \Gamma^{(2)}_1(k_1,k_2),\\
&\Gamma^{(2)}_1(k_1,k_2)=\frac{1}{(2\pi)^4}\int\frac{d^4k_3}{(2\pi)^4}\ \Delta_F(k_3)\Big[\mathcal{V}_{3312}+\mathcal{V}_{1233}+\mathcal{V}_{1332}+\mathcal{V}_{3213}\Big],\label{gamma-2pts}
\end{align}
\end{subequations}
where $\Delta_F$ and $\mathcal{V}$ denote respectively any of the propagators and vertex functions presented in $\S$\ref{sec-action}. Also, we have introduced the notation $\mathcal{V}_{abcd}:=\mathcal{V}(k_a,k_b,k_c,k_d)$.
\subsection{Model with Casimir kinetic operator.}\label{sec-2pt-Casimir}
We begin our study of the quantum properties of $\kappa$-Poincar\'{e} invariant NCFT by considering models which are characterised by the Casimir kinetic operator \eqref{c-propagator}.
\subsubsection{Orientable model.}
The various contributions to the one-loop 2-point function are easily obtained by combining \eqref{o-vertex} with \eqref{gamma-2pts}. Straightforward computations show that all of the Wick contracted vertex functions $\mathcal{V}_\textit{o}$ involved in \eqref{gamma-2pts} reduce to the ordinary (linear) delta of conservation between external momenta, \textit{i.e.} $\delta^{(\hspace{-1pt}4\hspace{-1pt})}(k_2-k_1)$, times some power of the twist factor (which may differ from one case to the other, however). After some trivial manipulations, we find that the one-loop quadratic part of the effective action reduces to
\begin{subequations}\label{2-point}
\begin{equation}
\Gamma_1^{(2)}[\phi,\bar{\phi}]=\frac{1}{2}\int \frac{d^4k}{(2\pi)^4}\ \bar{\phi}(k)\left(\omega_1+\omega_2e^{-3k^0/\kappa}\right)\phi(k),
\end{equation}
thus indicating that the tree-level structure of the mass operator, \textit{i.e.} $m_1^2+m_2^2\sigma$,\footnote{In the following, we shall refer to the term proportional to the twist factor $\sigma$ as the ``twisted mass" or ``twisted component of the mass operator," while the other term will be called ``ordinary (component of the) mass (operator)."} is preserved by radiative corrections, at least at first order in $\hbar$.\smallskip

The corrections are given by
\begin{align}
&\omega_1:=\int\frac{d^4k}{(2\pi)^4}\ \left(3g_2+g_1e^{-3k^0/\kappa}\right)e^{-2\alpha k^0/\kappa}\Delta_{F;\alpha}(k),\label{2-pointa}\\
&\omega_2:=\int\frac{d^4k}{(2\pi)^4}\ \left(3g_1+g_2e^{3k^0/\kappa}\right)e^{-2\alpha k^0/\kappa}\Delta_{F;\alpha}(k),\label{2-pointb}
\end{align}
\end{subequations}
where, at this stage, $\Delta_{F;\alpha}$ denotes any of the propagators belonging to one of the family \eqref{c-propagator} or \eqref{eq-propagator}. In fact, $e^{-2\alpha k^0/\kappa}\Delta_{F;\alpha}(k)=\Delta_{F}(k)$, $\forall\alpha\in\mathbb{R}$. This shows that, when applied consistently in both the kinetic term and the interaction term, the transformations $\phi\to\phi_\alpha$ and $\phi^\ddagger\to\phi^\ddagger_\alpha$, eq. \eqref{fields-family}, do not affect the one-loop quantum corrections to the 2-point function. Therefore, we shall restrict in the following our attention to the case $\alpha=0$, namely by considering the propagator $\Delta_{F}$.\smallskip

Going back to the model with Casimir kinetic operator, one finds that the two contributions, eq. \eqref{2-point}, admit the generic expression 
\begin{subequations}\label{2pt-co}
\begin{align}
&\omega_j=\frac{\kappa}{\pi}\int_0^\infty dy\ \Phi_j(y) J(y),\ j=1,2,\\
&\text{with}\ \ \ J(y):=\int_{\mathbb{R}^3}\frac{d^3\vec{k}}{(2\pi)^3}\ \frac{1}{\|\vec{k}\|^2+\kappa^2\mu_c^2(y)}.\label{2pt-spatial}
\end{align}
The functions $\Phi_j$, which can be read from \eqref{2-point}, are given by
\begin{equation}\label{2pt-coy}
\Phi_1(y):=\frac{3g_2-g_1}{1+y^3}+g_1,\ \ \Phi_2(y):=\frac{3g_1-g_2}{1+y^3}+\frac{g_2}{y^3},
\end{equation}
\end{subequations}
where we have used, for latter computational convenience, the decompositions
\begin{equation}
\frac{y^3}{1+y^3}=1-\frac{1}{1+y^3},\ \ \frac{1}{y^3(1+y^3)}=\frac{1}{y^3}-\frac{1}{1+y^3}.
\end{equation}

Let us go back a moment to eq. \eqref{full-mass}. Recall that this equation gives the expression of the (energy dependent) masslike term, $M^2(k^0;m_1,m_2)$, which was shown to be bounded from above and below by the masses $m_1$ and $m_2$. To make the presentation (slightly) more complete, let us assume for a moment that $m_1\neq m_2$. Because of our correspondence principle (CP), we know that the two masses, which both admit the same commutative limit, differ from their commutative counterpart, \textit{i.e.} $\bar{m}_0$ in eq. \eqref{action-limit}, only by term of order $\kappa^{-1}$. Therefore, we now assume that $\vert m_1^2 -m^2_2\vert=:\varepsilon^2\ll m_1^2,m^2_2$ together with $m_1^2\geq m_2^2$. It follows that
\begin{equation}
M^2(k^0;m_2)=m_2^2+\frac{\varepsilon^2}{1+y^3}.
\end{equation}
Under this assumption, eq. \eqref{c-propagator} now reads
\begin{equation}
\mu_c^2(y)=R_c(y)+\frac{(\varepsilon/\kappa)^2y}{1+y^3},\ \ R_c(y)=1-\left(\frac{2\kappa^2-m_2^2}{\kappa^2}\right) y + y^2.\label{decomposition-muc}
\end{equation}
We now return to the computation of the one-loop corrections to the 2-point function.\smallskip

Observe that we have performed the following change of variables
\begin{equation}\label{ychange-variable}
k^0\mapsto y:=e^{-k^0/\kappa},
\end{equation}
to obtain \eqref{2pt-co} from \eqref{2-point}. First, note that both the lower (0) and upper ($\infty$) bounds of integrations in $\int_0^\infty dy$ correspond to the UV (\textit{i.e.} large $\vert k^0\vert$) regime. Then, observe that the $y$ variable is formally related to the quantity $\mathcal{P}_0$ replacing the (ordinary) energy at the level of the $\kappa$-deformed field theory as it is apparent from the expression of the first Casimir of the $\kappa$-Poincar\'{e} algebra; see eq. \eqref{Casimir}.\footnote{To be more precise, $y$ is related to $\mathcal{E}=e^{-P_0/\kappa}\in\mathfrak{T}_\kappa$, eq. \eqref{twist0}.} We have
\begin{equation}\label{deformed-energy}
\mathcal{P}_0(k^0)=\kappa(1-y),
\end{equation}
which reduces obviously to $k^0$ in the commutative ($\kappa\to\infty$) limit.\\
Hence, according to the discussion given in $\S$\ref{sec-kinetic}, the $y$-integrals will be regularised with respect to this quantity rather than $k^0$ in a sense explained below.\\
Let $\Lambda_0$ be a cutoff for $\mathcal{P}_0$ defined by $\vert\mathcal{P}_0\vert\leq\Lambda_0$. From eq. \eqref{deformed-energy}, we easily infer the following bounds for the $y$ variable
\begin{equation}\label{ybounds}
\frac{\kappa}{\kappa+\Lambda_0} \leq y \leq \frac{\kappa+\Lambda_0}{\kappa},
\end{equation}
which will be used to regularised the $y$-integral; see eq. \eqref{2pt-coreg}. Of course, to assume $\vert\mathcal{P}_0\vert\leq\Lambda_0$ automatically implies the variable $k^0$ to be bounded as well. Denoting by $M_\kappa(\Lambda_0)$ the cutoff for the Fourier parameter $k^0$, we find that the two regulators are related via the relation
\begin{equation}
M_\kappa(\Lambda_0):=\kappa\ln\left(1+\frac{\Lambda_0}{\kappa}\right),
\end{equation}
with $M_\kappa(\Lambda_0)\to\Lambda_0$ as $\mathcal{P}_0(k^0)\to k^0$ in the limit $\kappa\to\infty$.\smallskip

The analysis of the 2-point function can be more conveniently carried out upon using a Pauli-Villars regularisation to extract the singular behaviour of the 3-dimensional (spacelike) integral. This is formally achieved by introducing a cutoff $\Lambda$ (\textit{a priori} different from $\Lambda_0$), then substituting $J(y)\to J_\Lambda(y)$ in \eqref{2pt-co} with
\begin{equation}
J_{\Lambda}(y):=\int_{\mathbb{R}^3}\frac{d^3\vec{k}}{(2\pi)^3}\left(\frac{1}{\|\vec{k}\|^2+\kappa^2\mu_c^2(y)}-\frac{1}{\|\vec{k}\|^2+\Lambda^2}\right).
\end{equation}
This latter integrals is easily computed upon using the two relations
\begin{subequations}\label{formulassss}
\begin{align}
\frac{1}{A^aB^b}=\frac{\Gamma(a+b)}{\Gamma(a)\Gamma(b)} & \int_0^1 du \ \frac{u^{a-1}(1-u)^{b-1}}{\left(uA+(1-u)B\right)^{a+b}},\ \ a,b>0, \\
\int \frac{d^np}{(2\pi)^n} \frac{1}{(p^2+M^2)^{m}}&=M^{n-2m}\frac{\Gamma(m-n/2)}{(4\pi)^{n/2} \Gamma(m)},\ \ m>n/2> 0,
\end{align}
\end{subequations}
where $\Gamma(z)$ is the Euler gamma function. Explicitly, we have
\begin{align}
J_{\Lambda}(y)&=\int_{\mathbb{R}^3}\frac{d^3\vec{k}}{(2\pi)^3}\frac{\Lambda^2-\kappa^2\mu_c^2(y)}{\big(\|\vec{k}\|^2+\kappa^2\mu_c^2(y)\big)\big(\|\vec{k}\|^2+\Lambda^2\big)}\\
&=\int_0^1du\int_{\mathbb{R}^3}\frac{d^3\vec{k}}{(2\pi)^3}\frac{\Lambda^2-\kappa^2\mu_c^2(y)}{\big(\|\vec{k}\|^2+\Lambda^2+u(\kappa^2\mu_c^2(y)-\Lambda^2)\big)^2}\nonumber\\
&=\frac{1}{8\pi}\int_0^1 du\ \frac{\Lambda^2-\kappa^2\mu_c^2(y)}{\sqrt{\Lambda^2+u(\kappa^2\mu_c^2(y)-\Lambda^2)}}=\frac{1}{4\pi}\big(\Lambda-\kappa\mu_c(y)\big),\nonumber
\end{align}
exhibiting a linear (UV) divergence in $\Lambda$. It remains to compute
\begin{equation}\label{2pt-coreg}
\omega_j(\Lambda,\Lambda_0)=\frac{\kappa}{4\pi^2}\int_{\Lambda_0} dy\ \Phi_j(y)\big(\Lambda-\kappa\mu_c(y)\big),
\end{equation}
where the $y$-integral is understood to be regularised as $\int_0^\infty\to\int_{\Lambda_0}$ by means of using the bounds of integrations given by \eqref{ybounds}. The integration over the $y$ variable consists in standard integrals that can be found in any handbook of mathematics or table of integrals.\footnote{See e.g. I. S. Gradshteyn and I. M. Ryzhik, \textit{Table of Integrals, Series, and Products} (Boston: Academic Press, 2007).} Nevertheless, for the sake of completeness, we present the full computation of all of these integrals below. We suggest the reader familiar of such calculations, and who would skip the presentation of the computational details, to go directly to the final results and discussions starting just after eq. \eqref{go-to-result1}.\smallskip

We begin with the computation of $\omega_1$. The computation of the first term in \eqref{2pt-coreg}, which involves the cutoff $\Lambda$, amounts to compute
\begin{equation}
\int_{\Lambda_0} dy\ \frac{1}{1+y^3}=\frac{2\pi}{3\sqrt{3}},\ \ \int_{\Lambda_0} dy=1+\frac{\Lambda_0}{\kappa},
\end{equation} 
while the second term involves integrals of the form
\begin{equation}\label{integral1}
\int_{\Lambda_0} dy\ \frac{\mu_c(y)}{1+y^3},\ \ \int_{\Lambda_0} dy\ \mu_c(y).
\end{equation} 
Simple inspection of the decay properties of the integrand of the first integral in \eqref{integral1} shows that the integrand behaves like a constant when $y\to0$ while it behaves like $y^{-2}$ when $y\to\infty$, indicating that the first integral is finite. In contrast, a similar analysis shows that the second integral diverges at most quadratically. The singularities are obtained by expanding $\mu_c$, eq. \eqref{decomposition-muc}, in power of $\varepsilon$. This leads to
\begin{equation}
\int_{\Lambda_0} dy\ \mu_c(y)=\sum_{n=0}^\infty c_n \left(\frac{\varepsilon}{\kappa}\right)^{2n} \int_{\Lambda_0} dy\ \frac{y^n\sqrt{R_c(y)}}{(1+y^3)^nR_c^n(y)},
\end{equation}
where the $c_n$ are the coefficients of the Taylor expansion (around 0) of $x\mapsto\sqrt{1+x}$. We easily find that the integrals involved in the series expansion converge for all $n\geq1$. The diverging part is then given by
\begin{align}
&\int dy\ \sqrt{R_c(y)}=\frac{\big(2\kappa^2y-(2\kappa^2-m_2^2)\big)\sqrt{R_c(y)}}{4\kappa^2}+\frac{\big(4\kappa^2-m_2^2\big)m_2^2}{8\kappa^4}\int \frac{dy}{\sqrt{R_c(y)}},\\
&\text{with}\ \ \ \int \frac{dy}{\sqrt{R_c(y)}}=\ln\left(2\sqrt{R_c(y)}+2y-\frac{2\kappa^2-m^2_2}{\kappa^2}\right),\nonumber
\end{align}
which leads to
\begin{equation}
\int_{\Lambda_0} dy\ \sqrt{R_c(y)}=\frac{\Lambda_0^2}{2\kappa^2}+\frac{m_2^2\Lambda_0}{2\kappa^3}+\frac{(4\kappa^2-m_2^2)m_2^2}{8\kappa^4}\ln\left(1+\frac{\Lambda_0}{\kappa}\right)+\lbrace\text{finite terms}\rbrace.
\end{equation}
Putting these results all together, we finally find
\begin{subequations}\label{result-co1}
\begin{align}
\omega_1(\Lambda,\Lambda_0)=&\ \frac{g_1}{4\pi^2}\left(\Lambda-\frac{\Lambda_0}{2}\right)\Lambda_0+\frac{6\pi g_2+(3\sqrt{3}-2\pi)g_1}{12\pi^2\sqrt{3}}\ \kappa\Lambda-\frac{m_2^2g_1}{8\pi^2\kappa}\ \Lambda_0-\\
&-\frac{(4\kappa^2-m_2^2)m_2^2g_1}{32\pi^2\kappa^2}\ln\left(1+\frac{\Lambda_0}{\kappa}\right)+F_1(\kappa,\varepsilon),\nonumber
\end{align}
where the finite terms are given by
\begin{align}
F_1(\kappa,\varepsilon):=&-\frac{m_2^2g_1}{16\pi^2}\left(1+4\ln\left(\frac{2\kappa}{m_2}\right)\right)-\frac{(3g_2-g_1)\kappa^2}{4\pi^2}\int dy\ \frac{\mu_c(y)}{1+y^3}-\\
&-\frac{g_1\varepsilon^2}{4\pi^2}\sum_{n:=1} c_n \left(\frac{\varepsilon}{\kappa}\right)^{2n-2} \int dy\ \frac{y^n\sqrt{R_c(y)}}{(1+y^3)^nR_c^n(y)}+\mathcal{O}(\kappa^{-1}).\nonumber
\end{align}
\end{subequations}

The computation of the second contribution $\omega_2$ is quite similar, some of the integrals to compute being the same as for $\omega_1$. The integrals which differ from $\omega_1$ are
\begin{equation}
\int_{\Lambda_0} \frac{dy}{y^3}=\frac{\kappa^2+2\kappa\Lambda_0+\Lambda_0^2}{2\kappa^2},
\end{equation}
for the first term in eq. \eqref{2pt-coreg}, while for the second term we have 
\begin{equation}\label{integral2}
\int_{\Lambda_0} dy\ \frac{\mu_c(y)}{y^3}=\sum_{n=0}^\infty c_n \left(\frac{\varepsilon}{\kappa}\right)^{2n} \int_{\Lambda_0} dy\ \frac{y^n\sqrt{R_c(y)}}{y^3(1+y^3)^nR_c^n(y)}.
\end{equation}
Mere analysis of the integrand in \eqref{integral2} shows that the integrals involved in the above series are finite for all $n\geq3$. The diverging parts of the remaining integrals are given by
\begin{subequations}
\begin{align}
&\int dy\ \frac{\sqrt{R_c(y)}}{y^3}=\big((2\kappa^2-m_2^2)y-2\kappa^2\big)\frac{\sqrt{R_c(y)}}{4\kappa^2y^2}+\frac{(4\kappa^2-m_2^2)m_2^2}{8\kappa^4}\int\frac{dy}{y\sqrt{R_c(y)}},\\
&\int \frac{dy}{y^2\sqrt{R_c(y)}}=-\frac{\sqrt{R_c(y)}}{y}+\frac{2\kappa^2-m_2^2}{2\kappa^2}\int\frac{dy}{y\sqrt{R_c(y)}},\\
&\int \frac{dy}{y\sqrt{R^3_c(y)}}=\frac{2\left((2\kappa^2-m_2^2)\kappa^2y+(4\kappa^2-m_2^2)m_2^2-2\kappa^4\right)}{(4\kappa^2-m_2^2)m_2^2\sqrt{R_c(y)}}+\int\frac{dy}{y\sqrt{R_c(y)}},\\
&\text{with}\ \ \ \int \frac{dy}{y\sqrt{R_c(y)}}=-\ln\left(\frac{2\kappa^2-(2\kappa^2-m_2^2)y+2\kappa^2\sqrt{R_c(y)}}{\kappa^2y}\right).\nonumber
\end{align}
\end{subequations}
Straightforward computations yield
\begin{equation}\label{integral4}
\int_{\Lambda_0} dy\ \frac{\mu_c(y)}{y^3}=\frac{\Lambda_0^2}{2\kappa^2}+\frac{m_1^2\Lambda_0}{2\kappa^3}+\frac{(4\kappa^2-m_1^2)m_1^2}{8\kappa^4}\ln\left(1+\frac{\Lambda_0}{\kappa}\right)+\lbrace\text{finite terms}\rbrace.
\end{equation}
Putting these results all together, we finally find
\begin{subequations}\label{result-co2}
\begin{align}
\omega_2(\Lambda,\Lambda_0)=&\ \frac{g_2}{8\pi^2\kappa}\ \Lambda\Lambda_0^2+\frac{g_2}{4\pi^2}\left(\Lambda-\frac{\Lambda_0}{2}\right)\Lambda_0+\frac{12\pi g_1+(3\sqrt{3}-4\pi)g_2}{24\pi^2\sqrt{3}}\ \kappa\Lambda-\\
&-\frac{m_1^2g_2}{8\pi^2\kappa}\ \Lambda_0-\frac{(4\kappa^2-m_1^2)m_1^2g_2}{32\pi^2\kappa^2}\ln\left(1+\frac{\Lambda_0}{\kappa}\right)+F_2(\kappa,\varepsilon),\nonumber
\end{align}
where the finite terms are given by
\begin{align}\label{go-to-result1}
F_2(\kappa,\varepsilon):=&\ -\frac{m_1^2g_2}{4\pi^2}\ln\left(\frac{2\kappa}{m_2}\right)-\frac{(\kappa^2+\varepsilon^2)g_2}{8\pi^2}-\frac{(3g_1-g_2)\kappa^2}{4\pi^2}\int dy\ \frac{\mu_c(y)}{1+y^3}-\\
&\ -\frac{g_2\varepsilon^2}{8\pi^2}\int \frac{ydy}{(1+y^3)\sqrt{R_c(y)}}-\frac{g_2\varepsilon^4}{32\pi^2\kappa^2}\int dy\ \frac{y^2(2+y^3)}{(1+y^3)^2\sqrt{R^3_c(y)}}-\nonumber\\
&\ -\frac{g_2\varepsilon^2}{4\pi^2}\sum_{n=3}^\infty c_n \left(\frac{\varepsilon}{\kappa}\right)^{2n-2} \int dy\ \frac{y^n\sqrt{R_c(y)}}{y^3(1+y^3)^nR_c^n(y)}+\mathcal{O}(\kappa^{-1}).\nonumber
\end{align}
\end{subequations}\smallskip

We now summarise the results we have found so far and discuss them in light of the well known quantum behaviour of the commutative $\vert\phi\vert^4$ field theory. To make the comparison more relevant, let us begin by presenting what are exactly the one-loop quantum corrections $\Omega$ to the corresponding 2-point function within the same regularisation scheme as we have used for the NCFT. We have $\Gamma_1^{(2)}[\phi,\bar{\phi}]=\frac{1}{2}\int \frac{d^4k}{(2\pi)^4}\ \bar{\phi}(k)\Omega\phi(k)$ with
\begin{subequations}
\begin{equation}
\Omega\to\Omega(\Lambda,\Lambda_0):=g\int_{-\Lambda_0}^{\Lambda_0}\frac{dk_0}{2\pi}\int_{\mathbb{R}^3}\frac{d^3\vec{k}}{(2\pi)^3}\left(\frac{1}{\|\vec{k}\|^2+(k_0^2+m_0^2)}-\frac{1}{\|\vec{k}\|^2+\Lambda^2}\right).
\end{equation}
Standard computations yield
\begin{align}\label{integral3}
\Omega(\Lambda,\Lambda_0)&=\frac{g}{8\pi^2}\left(2\Lambda-\sqrt{\Lambda_0^2+m_0^2}\right)\Lambda_0-\frac{m_0^2g}{8\pi^2}\ \text{arcsinh}\left(\frac{\Lambda_0}{m_0}\right)\\
&=\frac{g}{4\pi^2}\left(\Lambda-\frac{\Lambda_0}{2}\right)\Lambda_0-\frac{m_0^2g}{8\pi^2}\ln\left(\frac{2\Lambda_0}{m_0}\right)-\frac{m^2g}{16\pi}.\nonumber
\end{align}
\end{subequations}
where we made use of the series expansion (around 0) of $x\mapsto\sqrt{1+x}$, together with the asymptotic expression of $x\mapsto\text{arcsinh}(x)$, \textit{i.e.} $\text{arcsinh}(x)=\ln(2x)+\mathcal{O}(x^{-2})$, to go from the first line to the second one in \eqref{integral3}. Thus, $\Omega$ diverges quadratically.\smallskip

Going back to the present NCFT, we have shown that the first order radiative corrections to the mass operator which can be read from the expression of the one-loop quadratic part of the effective action
\begin{subequations}\label{result-co}
\begin{equation}
\Gamma_1^{(2)}[\phi,\bar{\phi}]=\frac{1}{2}\int \frac{d^4k}{(2\pi)^4}\ \bar{\phi}(k)\left(\omega_1+\omega_2e^{-3k^0/\kappa}\right)\phi(k),
\end{equation}
are given by
\begin{align}
&\omega_1(\Lambda)=\frac{g_1}{8\pi^2}\ \Lambda^2+\left(\frac{(3g_2-g_1)\kappa}{6\pi\sqrt{3}}+\frac{(2\kappa^2-m_2^2)g_1}{8\pi^2\kappa}\right)\Lambda-\\
&\hspace{1.5cm}-\frac{(4\kappa^2-m_2^2)m_2^2g_1}{32\pi^2\kappa^2}\ln\left(1+\frac{\Lambda}{\kappa}\right)+\lbrace\text{finite terms}\rbrace,\nonumber\\
&\omega_2(\Lambda)=\frac{g_2}{8\pi^2\kappa}\ \Lambda^3+\frac{g_2}{8\pi^2}\ \Lambda^2+\left(\frac{(3g_1-g_2)\kappa}{6\pi\sqrt{3}}+\frac{(\kappa^2-m_1^2)g_2}{8\pi^2\kappa}\right)\Lambda-\\
&\hspace{1.5cm}-\frac{(4\kappa^2-m_1^2)m_1^2g_2}{32\pi^2\kappa^2}\ln\left(1+\frac{\Lambda}{\kappa}\right)+\lbrace\text{finite terms}\rbrace,\nonumber
\end{align}
\end{subequations}
where we have set $\Lambda=\Lambda_0$ for simplicity.\smallskip

First of all, in view of the above results, we conclude that setting $g_2=0$ while keeping $g_1\neq0$ the NCFT behaves the same (at leading order in the cutoff) as its commutative counterpart. On the contrary, turning off $g_1$ while restoring $g_2\neq0$ we find that the NCFT diverges cubically, thus slightly worse than its commutative counterpart. Anyhow, in both cases, we find that the mass degeneracy is lifted by quantum fluctuations since the two masses receive one-loop corrections whose respective (leading order) dependences on $\Lambda$ are different. In particular, assuming $g_1,g_2\neq0$, we find
\begin{equation}
\left\vert\frac{\omega_1-\omega_2}{\omega_1}\right\vert=\frac{g_2}{\kappa g_1}\ \Lambda+\mathcal{O}(1),
\end{equation}
which is far from being small. This seems to indicate that the ansatz $\varepsilon\ll m_1,m_2$ does not survive the quantum fluctuations. However, as we are going to see when considering the case with nonorientable interactions below, the planar contributions arising in the one-loop 2-point function of the nonorientable model reverse in some sense this situation. In this case, we will find that the one-loop corrections to $m_1$ are proportional (at leading order in $\Lambda$) to $\Lambda^3$ while the corrections to $m_2$ are proportional (at leading order) to $\Lambda^2$. Therefore, the relation $\varepsilon\ll m_{1},m_2$ might be preserved if considering the full theory involving all of the interactions in eq. \eqref{interactions}, \textit{i.e.} both orientable and nonorientable.\smallskip

Then, it is interesting to notice that the integral, eq. \eqref{integral4}, appearing in the computation of $\omega_2$, involves terms of the form $m_2^2+\varepsilon^2$ which corresponds to $m_1^2$ by definition. Hence, at least in the approximation $\vert m_1^2-m_2^2\vert\ll m_1^2,m_2^2$, the respective masses of the fields $\phi$ and $\phi^\ddagger$ are in some sense mixed together by quantum fluctuations.\smallskip

Finally, let us comment on the peculiar role played by the deformation parameter $\kappa$ within such model of NCFT. Taking the (formal commutative) limit $\kappa\to\infty$ in eq. \eqref{result-co}, while keeping $\Lambda$ finite,\footnote{Note that similar results would have been obtained, to within unessential rescaling of the coefficients in eq. \eqref{limit-co}, by assuming the ratio $\Lambda/\kappa$ to remain constant in the limit $\kappa\to\infty$.} we find
\begin{subequations}
\begin{equation}\label{limit-co}
\omega_j(\Lambda)\xrightarrow[\kappa\to\infty]{}\frac{g_j}{8\pi^2}\ \Lambda^2+a_j\kappa\Lambda+\lim_{\kappa\to\infty}F_j(\kappa,0),
\end{equation}
where we have set $\varepsilon=0$ in order to simplify the discussion,\footnote{Indeed, to set $\varepsilon=0$ before taking the commutative limit simplifies significantly the computation of $\lim_{\kappa\to\infty}F_j(\kappa,0)$. Note however that this is formally consistent with the fact that $\varepsilon\to0$ when $\kappa\to\infty$.} and the $a_j$'s are some constant coefficients we can read from eq. \eqref{result-co}. Surprisingly, although not strictly speaking a regulator for the the NCFT (in the sense that the integrals involved in $\omega_j$ have to be regularised in both timelike and spacelike variables), we find that $\kappa$ plays a role similar to $\Lambda$ in the commutative limit, \textit{i.e.} when $\kappa\to\infty$. This can be compared, for instance, with what happens in the context of $\mathbb{R}^3_\theta$. In this case, we will find that the NCFT is (UV) finite at fixed (nonzero) $\theta$, while the usual UV behaviour is recovered when taking the commutative limit $\theta\to0$. It follows that, $\theta$ plays the role of a (natural) UV cutoff within such model of NCFT on $\mathbb{R}^3_\theta$; see Chap. \ref{sap-ncftsu2}.\\
Going back to the case under consideration, the reason we kept track of the finite terms in eq. \eqref{result-co1} and \eqref{result-co2} is that these terms $F_j(\kappa,\varepsilon)$ become singular in the limit $\kappa\to\infty$. Indeed, we find
\begin{align}\label{limit-co2}
F_j(\kappa,0)\xrightarrow[\kappa\to\infty]{}b_j\kappa^2-\frac{m^2g_j}{4\pi^2}\ln\left(\frac{2\kappa}{m}\right)-\frac{m^2g_j}{16\pi^2},\ b_j\in\mathbb{R}.
\end{align}
\end{subequations}
Hence, combining eq. \eqref{limit-co} with \eqref{limit-co2}, we recover the same behaviour as for the ordinary (commutative) $\vert\phi\vert^4$-theory. See eq. \eqref{integral3} for a comparison.
\subsubsection{Nonorientable model.}
We now turn to the analysis of the model with nonorientable interactions. Combining eq. \eqref{no-vertex} with \eqref{gamma-2pts}, we find that the one-loop quadratic part of the effective action decomposes into two families of contributions. The first family involves planar contributions for which a delta of conservation depending only on the external momenta can be factorised out from the Wick contracted vertex functions appearing in \eqref{gamma-2pts}. These contributions are of the same nature as the contributions encounter in ordinary quantum field theory and are similar to the ones found above for the orientable model. The other family of contributions involves nonplanar contributions which result from the Wick contraction of two nonadjacent fields in the expression of the interaction term \eqref{no-int2}. We write
\begin{equation}
\Gamma_1^{(2)}[\phi,\bar{\phi}]=\Gamma^{(2)}_\text{N}[\phi,\bar{\phi}]+\Gamma^{(2)}_\text{NP}[\phi,\bar{\phi}],
\end{equation}
where $\Gamma^{(2)}_{N}$ (resp. $\Gamma^{(2)}_\text{NP}$) denotes the planar (resp. nonplanar) component of the quadratic part of the effective action.\smallskip

Let us begin with the planar contributions. Mere analysis of the expressions of $\Gamma_1^{(2)}$ shows that we have to set $\alpha=0$ in eq. \eqref{nov-gen} in order the tree-level structure of the mass operator to be preserved by quantum fluctuations. A situation we assume from now on. We find
\begin{subequations}\label{contributions-cnop}
\begin{equation}
\Gamma^{(2)}_\text{N}[\phi,\bar{\phi}]=\frac{1}{2}\int \frac{d^4k}{(2\pi)^4}\ \bar{\phi}(k)\left(\omega_3+\omega_4e^{-3k^0/\kappa}\right)\phi(k),
\end{equation}
with
\begin{align}
&\omega_3:=\frac{\kappa g_3}{2\pi}\int\frac{dy}{y^4} \frac{d^3\vec{k}}{(2\pi)^3}\left(1+y^3\right)\Delta^{c}_{F}(k),\\
&\omega_4:=\frac{\kappa g_4}{2\pi}\int \frac{dy}{y}\frac{d^3\vec{k}}{(2\pi)^3}\left(1+y^3\right)\Delta^{c}_{F}(k).
\end{align}
\end{subequations}
Setting formally $g_1=3g_2$ in \eqref{2-pointa} and $g_2=3g_1$ in \eqref{2-pointb} we obtain
\begin{equation}
\omega_3=\frac{g_3}{g_2}\omega_2,\ \omega_4=\frac{g_4}{g_1}\omega_1.
\end{equation}
Hence, no additional computations are needed here and we can read the singular UV behaviour of the planar contributions for the nonorientable model from the results already obtained for the orientable model. Mere adaptation of \eqref{result-co} yields
\begin{subequations}
\begin{align}
&\omega_3(\Lambda)=\frac{g_3}{8\pi^2\kappa}\ \Lambda^3+\frac{g_3}{8\pi^2}\ \Lambda^2+\frac{(\kappa^2-m_1^2)g_3}{8\pi^2\kappa}\ \Lambda-\frac{(4\kappa^2-m_1^2)m_1^2g_3}{32\pi^2\kappa^2}\ln\left(\frac{\Lambda_0}{\kappa}\right)+\\
&\hspace{1.5cm}+\lbrace\text{finite terms}\rbrace,\nonumber\\
&\omega_4(\Lambda)=\frac{g_4}{8\pi^2}\ \Lambda^2+\frac{(2\kappa^2-m_1^2)g_4}{8\pi^2\kappa}\ \Lambda-\frac{(4\kappa^2-m_2^2)m_2^2g_4}{32\pi^2\kappa^2}\ln\left(\frac{\Lambda_0}{\kappa}\right)+\\
&\hspace{1.5cm}+\lbrace\text{finite terms}\rbrace,\nonumber
\end{align}
\end{subequations}
and the same conclusions as for the orientable model still apply here.\smallskip

Notice that, unlike the orientable model for which the cubically divergent term ($\sim\Lambda^3$) appears as a correction to the mass of the field $\phi^\ddagger$, within the nonorientable model this term appears as a correction to the mass of the field $\phi$. It follows that, because of the twist factor which splits the mass operator into two components (which are given at the level of the classical action by $m_1^2$ and $m_2^2$ respectively), the cubic divergence cannot be cancelled out by an appropriate combination of the coupling constants except by setting $g_2=g_3=0$. However, as we are going to see next section, $\S$\ref{sec-4point}, when considering the one-loop 4-point function, the quantum fluctuations tend to restore the configuration $g_2,g_3\neq0$. This could have been expected since the relation $g_2=g_3=0$ is not related to a symmetry of the action. On the other hand, as already mentioned when discussing the results for the orientable model, this helps to restore the balance between the two components of the mass operator and in particular ensures the relation $\vert m_1^2-m_2^2\vert\ll m_1^2,m^2_2$ to be preserved at the quantum level.\smallskip

We now turn to the analysis of the nonplanar contributions which constitute the main difference between the present model and the model involving orientable interactions. We find
\begin{subequations}\label{contributionNP}
\begin{align}
&\Gamma^{(2)}_\text{NP}(k_1,k_2)=\big(\Xi(k_1,k_2)+\Xi(k_2,k_1)\big)\delta(k_2^0-k_1^0),\label{gammaNP}\\
&\Xi(k_a,k_b):=\frac{\kappa}{2\pi}\int\frac{dy}{y}\int\frac{d^3\vec{k}}{(2\pi)^3}\ \big(g_3+g_4e^{-3k_a^0/\kappa}y^3\big)\Delta_F^{c}(k)\times\label{fonctionXI}\\
&\hspace{6cm}\times\delta^{(\hspace{-1pt}3\hspace{-1pt})}\left((1-e^{-k_a^0/\kappa})\vec{k}-\vec{k}_a+y\vec{k}_b\right).\nonumber
\end{align}
\end{subequations}
This time, the internal momentum, $\vec{k}$, does not cancel out in the 3-dimensional delta function. Instead, integrating over $\vec{k}$ yields
\begin{subequations}
\begin{equation}\label{integrandNP}
\Xi(k_a,k_b)=\frac{\kappa e^{-3k_a^0/\kappa}}{(2\pi)^4\vert1-e^{-k_a^0/\kappa}\vert}\int \frac{dy}{c(k_a)-\tilde{c}y+c(k_b) y^2} \left( g_4+\frac{g_3e^{3k_a^0/\kappa}-g_4}{1+y^3} \right),
\end{equation}
where we have assumed that $k_a^0\neq0$ and defined
\begin{equation}
c(k):=\|\vec{k}\|^2+\kappa^2(1-e^{-k^0/\kappa})^2,\ \tilde{c}:=2\vec{k}_a\cdot\vec{k}_b+(2\kappa^2-m^2)(1-e^{-k^0/\kappa})^2.
\end{equation}
\end{subequations}
Mere inspection of the integrand in \eqref{integrandNP} shows that $\Xi(k_a,k_b)$ is finite at ``nonexceptional" external momenta. On the other hand, going back to eq. \eqref{fonctionXI}, we see that setting one of the external momenta to zero, say $(k_a,\vec{k}_a)=(0,\vec{0})$, we recover integrals of the same type as the ones involved in the computation of planar contributions. Namely
\begin{equation}
\Xi(0,k_b)=\frac{\kappa}{2\pi}\int\frac{dy}{y}\int\frac{d^3\vec{k}}{(2\pi)^3}\big(g_3+g_4y^3\big)\Delta_F^{c}(k)\delta^{(\hspace{-1pt}3\hspace{-1pt})}(\vec{k}_b),
\end{equation}
which has to be compared, for instance, with \eqref{2-pointa}. First, note that the conservation law between external momenta is preserved, \textit{i.e.} $k_b\to0$ when $k_a\to0$. Next, in view of \eqref{gammaNP}, similar conclusions would have been obtained setting $k_b$ to zero instead of $k_a$. Finally, we conclude that nonplanar contributions, albeit finite at nonzero external momenta, diverge in the IR sector. This last phenomenon reflects the existence of UV/IR mixing when considering models with nonorientable interactions. \smallskip

The UV/IR mixing is often regarded as problematic since it spoils the renormalisation properties of the quantum field theory. Indeed, although (UV) finite a one-loop order, the above contribution \eqref{contributionNP} may induce (IR) singularities at higher-loop order. One way to cure the theory from the UV/IR mixing would be to extract the singular behaviour (in the external momenta) of \eqref{contributionNP} then to add a counterterm accordingly. This would necessitate to compute exactly \eqref{integrandNP}, then study carefully the behaviour around zero of $\Xi(k_a,k_b)$. We leave this analysis to future studies. Nevertheless, insight in the result can be easily obtained under some assumptions.\smallskip

Let us set $g_3=g_4$. Expanding $\Xi(k_a,k_b)$ around $k^0_a\sim0$ in \eqref{integrandNP} yields 
\begin{equation}
\Xi(k_a,k_b)\underset{k^0_a\to0}{\sim}\frac{g_3\kappa}{(2\pi)^4}\int \frac{dy}{k_a^2-\tilde{c}y+k_b^2 y^2} \left(\frac{\kappa}{\vert k_a^0\vert}+\mathcal{O}(1)\right).
\end{equation}
At leading order, the integration over $y$ can easily be performed to get
\begin{equation}
\Xi(k_a,k_b)\underset{k^0_a\to0}{\sim}\frac{g_3\kappa^2}{(2\pi)^4}\frac{1}{\vert k_a^0\vert\sqrt{4k_a^2k_b^2-\tilde{c}^2}}\left(\frac{\pi}{2}-\arctan\left(\frac{-\tilde{c}}{\sqrt{4k_a^2k_b^2-\tilde{c}^2}}\right)\right),
\end{equation}
exhibiting a quadratic singularity.
\subsection{Models with equivariant kinetic operator.}\label{sec-2pt-equivariant}
In this section, we investigate the quantum properties of another NCFT characterised this time by the equivariant kinetic operator, eq. \eqref{eq-propagator}. This latter is related to the Casimir operator $\mathcal{C}_\kappa$ via the relation $K^\text{eq}=\mathcal{C}_\kappa+\mathcal{C}^2_\kappa/(4\kappa^2)$ and possesses the interesting characteristic to be equivariant under the action of $\kappa$-Poincar\'{e}. Actually, the corresponding propagator can be physically interpreted as a (kind of) Pauli-Villars regularised version of the propagator considered in $\S$\ref{sec-2pt-Casimir} whose cutoff would be related to some function of $\kappa$.\footnote{This is apparent from an appropriate redefinition of the mass parameters in the expression of the Lagrangian density; see eq. \eqref{c-Villars}.} It follow that the analysis of the NCFT with Casimir kinetic operator can be (almost) straightforwardly adapted to the present context. Although the one-loop 2-point function of the equivariant models is still (UV) singular, the actual divergence is milder than for the models with Casimir kinetic operator as it could have been expected from the strong decay properties of the equivariant propagator which decreases as $\|\vec{k}\|^{-4}$ when $\|\vec{k}\|\to\infty$.\smallskip

In view of the negligible benefit obtained in $\S$\ref{sec-2pt-Casimir} under the assumption $m_1\neq m_2$, together with the more involved expression of the equivariant propagator compared to that of the Casimir propagator, we now set $m_1=m_2=m$.
\subsubsection{Orientable model.}
The various contributions to the one-loop 2-point function can be read from the (one-loop) quadratic part of the effective action, eq. \eqref{gamma-2pts}, which reduces after some trivial manipulations to
\begin{subequations}
\begin{equation}
\Gamma_1^{(2)}[\phi,\bar{\phi}]=\frac{1}{2}\int \frac{d^4k}{(2\pi)^4}\ \bar{\phi}(k)\left(\omega_1+\omega_2e^{-3k^0/\kappa}\right)\phi(k),
\end{equation}
where $\omega_1$ and  $\omega_2$ are still given by \eqref{2-point}. Upon integrating over $\int d^3\vec{k}$, we find
\begin{equation}\label{2pt-eqo}
\omega_j(\Lambda_0):=\frac{\kappa^3}{4\pi^2\sqrt{\kappa^2-m^2}}\int_{\Lambda_0}dy\ \Phi_j(y)\big(\mu_{+}(y)-\mu_{-}(y)\big),
\end{equation}
\end{subequations}
where the functions $\Phi_j$ are still given by \eqref{2pt-coy}. Recall that
\begin{equation}
\mu^2_{\pm}(y)=1\pm 2y\sqrt{1-\left(\frac{m}{\kappa}\right)^2}+y^2.
\end{equation}
A mere comparison between eq. \eqref{2pt-eqo} and \eqref{2pt-coreg} shows that the same kind of integrals as in $\S$\ref{sec-2pt-Casimir} have to be computed here. Thus, from straightforward adaptation of the material presented in there, we find
\begin{subequations}\label{result-eqo}
\begin{equation}
\omega_j(\Lambda_0)=\frac{g_j\kappa}{2\pi^2}\ \Lambda_0+F_j(\kappa),\ j=1,2,
\end{equation}
exhibiting a linear UV divergence. The finite contributions are given by 
\begin{align}
F_1(\kappa):=&\ \frac{g_1\kappa^2}{4\pi^2}+\frac{m^2g_1\kappa}{8\pi^2\sqrt{\kappa^2-m^2}}\ln\left(\frac{\kappa+\sqrt{\kappa^2-m^2}}{\kappa-\sqrt{\kappa^2-m^2}}\right)+\\
&+\frac{(3g_2-g_1)\kappa^3}{4\pi^2\sqrt{\kappa^2-m^2}}\int dy\ \frac{\mu_{+}(y)-\mu_{-}(y)}{1+y^3},\nonumber\\
F_2(\kappa):=&\ \frac{g_2\kappa^2}{4\pi^2}+\frac{m^2g_2\kappa}{8\pi^2\sqrt{\kappa^2-m^2}}\ln\left(\frac{\kappa+\sqrt{\kappa^2-m^2}}{\kappa-\sqrt{\kappa^2-m^2}}\right)+\\
&+\frac{(3g_1-g_2)\kappa^3}{4\pi^2\sqrt{\kappa^2-m^2}}\int dy\ \frac{\mu_{+}(y)-\mu_{-}(y)}{1+y^3}.\nonumber
\end{align}
\end{subequations}

Again, the expected quantum behaviour for the complex $\vert\phi\vert^4$-model is recovered in the limit $\kappa\to\infty$; see eq. \eqref{integral3} for a comparison. We find
\begin{equation}
(\omega_1+\omega_2e^{-\frac{3k^0}{\kappa}})\xrightarrow[\kappa\to\infty]{}\frac{g_1+g_2}{4\pi^2}\left(2\kappa\Lambda_0+\kappa^2\big(1+\frac{16\pi}{\sqrt{3}}\big)+m^2\ln\left(\frac{2\kappa}{m}\right)-\frac{4\pi m^2}{\sqrt{3}}\right).
\end{equation}
Hence, as for the model with Casimir kinetic operator, $\kappa$ plays once more time the role of a cutoff for the ordinary quantum field theory. Nevertheless, the situation is a bit different from the previous case considered in $\S$\ref{sec-2pt-Casimir} since $\kappa$ appears to be related to some spacelike (Pauli-Villars) regulator for the model with equivariant kinetic operator and it seems more natural to interpret $\kappa$ as a cutoff for the (commutative) $\vert\phi\vert^4$-model within the present context. It is important to keep in mind that $\kappa$ is fixed once and for all when working at the level of the NCFT, however. In particular, $\kappa$ will not be interpreted as a cutoff in the sense of the renormalisation scheme when deriving the beta function in the next section, $\S$\ref{sec-4point}.\smallskip

As it could have been expected from the decay properties of the equivariant propagator, see eq. \eqref{eq-decay}, the quantum behaviour of the NCFT with equivariant kinetic operator is milder than that of its commutative counterpart, although the one-loop 2-point function remains singular in the UV. This seems to indicate that NCFT equipped with equivariant kinetic operator are more suitable for describing realistic physical model of $\kappa$-Poincar\'{e} invariant quantum field theory at least in comparison with the NCFT characterised by the Casimir kinetic operator. Even more interesting is the apparent symmetry of the one-loop corrections for the two components of the mass operator, namely of the respective masses of the fields $\phi$ and $\phi^\ddagger$, already at the level of the orientable model (provided $g_1,g_2\neq0$). This reflects likely the existence of a symmetry of the action functional under the exchange $\phi\leftrightarrow\phi^\ddagger$, as it is apparent from the expression of $\mathcal{S}_\kappa$; see $\S$\ref{sec-action}. Indeed, setting one of the coupling constant (either $g_1$ or $g_2$) to zero would lift the degeneracy between the two masses. As already mentioned in $\S$\ref{sec-2pt-Casimir}, this is supported by the fact that the quantum fluctuations tend to restore the symmetry $\phi\leftrightarrow\phi^\ddagger$ as we are going to see later on when considering the one-loop 4-point function.
\subsubsection{Nonorientable model.}
Again, the quadratic part of the effective action decomposes into a planar component and a nonplanar one. On the one hand, the planar contributions are similar to the contributions computed for the orientable model, namely
\begin{subequations}
\begin{equation}
\Gamma^{(2)}_\text{N}[\phi,\bar{\phi}]=\frac{1}{2}\int \frac{d^4k}{(2\pi)^4}\ \bar{\phi}(k)\left(\omega_3+\omega_4e^{-3k^0/\kappa}\right)\phi(k),
\end{equation}
with
\begin{equation}
\omega_j(\Lambda_0)=\frac{g_j\kappa}{2\pi^2}\ \Lambda_0+\frac{g_j\kappa^2}{4\pi^2}+\frac{m^2g_j\kappa}{8\pi^2\sqrt{\kappa^2-m^2}}\ln\left(\frac{\kappa+\sqrt{\kappa^2-m^2}}{\kappa-\sqrt{\kappa^2-m^2}}\right),\ j=3,4.
\end{equation}
\end{subequations}
On the other hand, the nonplanar contributions take the form
\begin{subequations}
\begin{align}
&\Gamma^{(2)}_\text{NP}(k_1,k_2)=\big(\Xi(k_1,k_2)+\Xi(k_2,k_1)\big)\delta(k_2^0-k_1^0),\\
&\Xi(k_a,k_b):=\frac{\kappa}{2\pi}\int\frac{dy}{y}\int\frac{d^3\vec{k}}{(2\pi)^3}\ \big(g_3+g_4e^{-3k_a^0/\kappa}y^3\big)\Delta_F^\text{eq}(k)\times\\
&\hspace{6cm}\times\delta^{(\hspace{-1pt}3\hspace{-1pt})}\left((1-e^{-k_a^0/\kappa})\vec{k}-\vec{k}_a+y\vec{k}_b\right),\nonumber
\end{align}
\end{subequations}
analogous to eq. \eqref{contributionNP}. Although the computation are slightly more involved than for the model with Casimir kinetic operator, the nonplanar contributions are found to diverge when one of the external momenta is set to zero albeit finite otherwise. We conclude that UV/IR mixing is also present in this case.
\section{One-loop 4-point function and beta function.}\label{sec-4point}
Previous section, we have considered various models of $\kappa$-Poincar\'{e} invariant scalar field theory. Each model was characterised by a specific choice of kinetic operator together with interaction potential. In view of the results obtained in $\S$\ref{sec-2point}, we now restrict our attention on the model which exhibits the best behaviour, namely the model with equivariant kinetic operator. Recall that the other model, with Casimir kinetic operator, diverges cubically, \textit{i.e.} worst than the commutative $\vert\phi\vert^4$ model. Moreover, since the tree-level structure of the 2-point function, as well as the relation $m_1=m_2$, are preserved at one-loop order when considering only the orientable interaction, we now focus on this interaction. Note that, preliminary computations for the model with Casimir kinetic operator and orientable interaction indicate that some of the contributions to the one-loop 4-point function are linearly divergent and some others exhibit UV/IR mixing.\bigskip

The one loop order corrections to the 4-point functions are obtained by expanding the generating functional of the connected correlation function $W$ up to the second order in the coupling constant. Standard computations, which are recalled in Appendix \ref{sap-perturbation}, yield the following expression for the quartic part of the effective action
\begin{subequations}\label{gamma4pts}
\begin{align}
&\Gamma^{(4)}_1[\phi,\bar{\phi}]:=\frac{1}{4!}\int\prod_{\ell=1}^n\left[\frac{d^4k_\ell}{(2\pi)^4}\right]\ \bar{\phi}(k_1)\phi(k_2)\bar{\phi}(k_3)\phi(k_4)\Gamma^{(4)}_1(k_1,k_2,k_3,k_4),\\
&\Gamma^{(4)}_1(k_1,k_2,k_3,k_4)=\frac{1}{(2\pi)^8}\int \frac{d^4k_5}{(2\pi)^4}\frac{d^4k_6}{(2\pi)^4}\ \Delta_F(k_5)\Delta_F(k_6)\times\\
&\hspace{5cm}\times\Big[2\mathcal{V}_{5462}\mathcal{V}_{3615}+2\mathcal{V}_{5462}\mathcal{V}_{3516}+2\mathcal{V}_{5216}\mathcal{V}_{3465}+\nonumber\\
&\hspace{5.175cm}+2\mathcal{V}_{1652}\mathcal{V}_{3465}+2\mathcal{V}_{5612}\mathcal{V}_{6435}+2\mathcal{V}_{5612}\mathcal{V}_{3564}+\nonumber\\
&\hspace{5.175cm}+2\mathcal{V}_{5216}\mathcal{V}_{3564}+2\mathcal{V}_{5612}\mathcal{V}_{3465}+\hspace{2pt}\mathcal{V}_{5612}\mathcal{V}_{6534}\hspace{3pt}+\nonumber\\
&\hspace{5.175cm}+\hspace{2pt}\mathcal{V}_{5216}\mathcal{V}_{6435}\hspace{3pt}+\hspace{2pt}\mathcal{V}_{1652}\mathcal{V}_{3564}\hspace{3pt}+\mathcal{V}_{1256}\mathcal{V}_{3465}\Big].\nonumber
\end{align}
\end{subequations}
As we are going to see, at the level of the one-loop 4-point function, even the orientable interaction leads to nonplanar contributions. \smallskip

From now on, we focus on the model with orientable interactions, eq. \eqref{o-int}. The symmetries of the 4-vertex function associated to this interaction, regarded as a distribution, can be easily read from eq. \eqref{o-delta}. They are given by
\begin{subequations}
\begin{align}
&V_\textit{o}(k_1,k_2,k_3,k_4)\equiv V_\textit{o}(k_4,k_3,k_2,k_1),\\
&V_\textit{o}(k_1,k_2,k_3,k_4)\equiv e^{3(k^0_3-k_4^0)/\kappa}V_\textit{o}(k_3,k_4,k_1,k_2).
\end{align}
\end{subequations}
Combining these above symmetry properties with the fusion rules
\begin{subequations}\label{fusion-rule}
\begin{align}
&V_\textit{o}(k_1,k_2,\mathbf{k_6},\mathbf{k_5})V_\textit{o}(\mathbf{k_5},\mathbf{k_6},k_3,k_4)\equiv V_\textit{o}(\mathbf{k_5},\mathbf{k_6},k_3,k_4)V_\textit{o}(k_1,k_2,k_3,k_4),\\
&V_\textit{o}(k_1,\mathbf{k_5},\mathbf{k_6},k_4)V_\textit{o}(\mathbf{k_5},k_2,k_3,\mathbf{k_6})\equiv V_\textit{o}(\mathbf{k_5},k_2,k_3,\mathbf{k_6})V_\textit{o}(k_1,k_2,k_3,k_4),
\end{align}
\end{subequations}
where the bold characters denote the Wick contracted momentum we sum over in \eqref{gamma4pts}, we find that $\Gamma_1^{(4)}$ decomposes into four families of contributions some being planar, the other being not. As we are going to see, the tree-level structure of the action functional is preserved by radiative corrections provided both $g_1$ and $g_2$ are different from zero.\\
We write 
\begin{equation}
\Gamma^{(4)}_1=\Gamma^P_1+\Gamma^P_2+\Gamma^{N\!P}_3+\Gamma^{N\!P}_4.
\end{equation}

The two families of planar contributions, hereafter denoted by (P1) and (P2), admit respectively the following expressions
\begin{subequations}\label{Pcontrib}
\begin{align}
\Gamma^P_1(k_1,k_2,k_3,k_4):=&\ (2\pi)^{-8}V_{1234}\int \frac{d^4k_5}{(2\pi)^4}\frac{d^4k_6}{(2\pi)^4}\ \Psi_1(k^0_5)\Delta^\text{eq}_F(k_5)\Delta^\text{eq}_F(k_6) V_{5634},\label{planar1}\\
\Gamma^P_2(k_1,k_2,k_3,k_4):=&\ (2\pi)^{-8}V_{1234}\int \frac{d^4k_5}{(2\pi)^4}\frac{d^4k_6}{(2\pi)^4}\ \Psi_2(k^0_5)\Delta^\text{eq}_F(k_5)\Delta^\text{eq}_F(k_6) V_{5236},\label{planar2}
\end{align}
\end{subequations}
where the two functions $\Psi_j$, $j=1,2$, are given by
\begin{subequations}\label{CoeffP}
\begin{align}
&\Psi_1(k^0_5):=a_1+b_1e^{3k_5^0/\kappa}+c_1e^{6k_5^0/\kappa},\\
&\Psi_2(k^0_5):=a_2+b_2e^{3k_5^0/\kappa}+d_2e^{-3k_5^0/\kappa},
\end{align}
\end{subequations}
where the coefficients depend only on the external momenta.\footnote{The coefficients associated with the planar contributions are the following. For $\Psi_1$ we have
\begin{subequations}
\begin{align}
&a_1:=2g_1^2\big(1+e^{3(k^0_1-k^0_2)/\kappa}\big)+g_1g_2\big(1+e^{3(k^0_1-k^0_3)/\kappa}+2e^{3(k^0_1-k^0_4)/\kappa}\big)e^{3k^0_4/\kappa}+g_2^2e^{3(k^0_1+k^0_4)/\kappa},\\
&b_1:=g_1g_2\big(3+e^{3(k^0_1-k^0_2)/\kappa}\big)+2g_2^2e^{3k^0_1/\kappa},\ c_1:=g_2^2.
\end{align}
\end{subequations}
For $\Psi_2$ we have
\begin{subequations}
\begin{align}
&a_2:=2g_1\big(g_1+g_2e^{3k^0_1/\kappa}\big)+g_1g_2\big(e^{3k^0_1/\kappa}+e^{3k_4^0/\kappa}\big),\ d_2:=g_1^2e^{3k^0_1/\kappa}\\
&b_2:=2g_2\big(g_1+g_2e^{3k^0_1/\kappa}\big)+g_2^2e^{3k^0_4/\kappa}+\big(g_1+g_2e^{3k^0_3/\kappa}\big)\big(g_1+g_2e^{3k^0_1/\kappa}\big)e^{-3k^0_2/\kappa}. 
\end{align}
\end{subequations}
}\smallskip

The nonplanar contributions, hereafter denoted by (NP3) and (NP4), are given by
\begin{subequations}\label{NPcontrib}
\begin{align}
\Gamma^{N\!P}_3(k_1,k_2,k_3,k_4):=&\ (2\pi)^{-8}\int \frac{d^4k_5}{(2\pi)^4}\frac{d^4k_6}{(2\pi)^4}\ \Psi_3(k_5^0)\Delta^\text{eq}_F(k_5)\Delta^\text{eq}_F(k_6)V_{5216}V_{3465},\label{nonplanar1}\\
\Gamma^{N\!P}_4(k_1,k_2,k_3,k_4):=&\ (2\pi)^{-8}\int \frac{d^4k_5}{(2\pi)^4}\frac{d^4k_6}{(2\pi)^4}\ \Psi_4(k_5^0)\Delta^\text{eq}_F(k_5)\Delta^\text{eq}_F(k_6)V_{5163}V_{5462},\label{nonplanar2}
\end{align}
\end{subequations}
with
\begin{subequations}\label{CoeffNP}
\begin{align}
&\Psi_3(k^0_5):=a_3+b_3e^{3k_5^0/\kappa}+c_3e^{6k_5^0/\kappa},\\ 
&\Psi_4(k^0_5):=a_4+b_4e^{3k_5^0/\kappa}+c_4e^{6k_5^0/\kappa},
\end{align}
\end{subequations}
where the coefficients depend only on the external momenta.\footnote{The coefficients associated with the nonplanar contributions are the following. For $\Psi_3$ we have
\begin{subequations}
\begin{align}
&a_3:=g_1^2\big(1+e^{3(k_3^0-k_4^0)/\kappa}\big)+g_1g_2e^{3k_3^0/\kappa},\ c_3:=g_1g_2\big(1+e^{-3k_2^0/\kappa}\big)+g_2^2e^{3(k_3^0-k_4^0)/\kappa},\\
&b_3:=g_1^2\big(e^{-3k_1^0/\kappa}+e^{-3k_2^0/\kappa}\big)+g_2^2\big(e^{3k_3^0/\kappa}+e^{3k_4^0/\kappa}\big)+\\
&\hspace{4truecm}+g_1g_2\big(2+e^{3(k_3^0-k_2^0)/\kappa}+e^{3(k_1^0-k_4^0)/\kappa}+e^{3(k_3^0-k_4^0)/\kappa}\big).\nonumber
\end{align}
\end{subequations}
For $\Psi_4$ we have
\begin{equation}
a_4:=g_1\big(g_1+g_2e^{3k_3^0/\kappa}\big),\ b_4:=g^2_1e^{-3k_1^0/\kappa}+2g_1g_2+g_2^2e^{3k_3^0/\kappa},\ c_4:=g_2\big(g_2+g_1e^{-3k_1^0/\kappa}\big).
\end{equation}
}
\subsection{Planar contributions.}
We begin with the study of the planar contributions, eq. \eqref{Pcontrib}. Integrating over $k_6$, we can express this latter in term of $k_5$ and the external momenta. Namely,
\begin{equation}\label{int4pt}
k^0_6=k^0_5+Q_j^0, \ \vec{k}_6= A_j\left(\vec{k}_5+y^{\epsilon_j}\vec{Q}_j\right),\ y=e^{-k_5^0/\kappa},\ j=1,2,
\end{equation}
where $Q_j^0$ and $\vec{Q}_j$ are functions of the external momenta, which are related to the noncommutative counterparts of the usual $s,t,u$ channels. Their expressions can be read from the delta functions \eqref{o-delta} involved in eq. \eqref{Pcontrib}. The coefficients appearing in eq. \eqref{int4pt} are given by $(A_1,\epsilon_1)=(1,1)$ and $(A_2,\epsilon_2)=(e^{-Q_2^0/\kappa},0)$. Note that, anticipating the computation of \eqref{Pcontrib}, we have introduced the variable $y$ corresponding to the change of variables \eqref{ychange-variable} already discussed in $\S$\ref{sec-2point}.\smallskip

Here, we are not interested in the exact expressions for the various contributions. Rather, we are going to show that all of the contributions are finite. To do so, it is convenient to exploit a particular estimate for the equivariant propagator \eqref{eq-propagator}, namely
\begin{equation}\label{borne}
\Delta_F^\text{eq}(k)\leq\frac{e^{-2k^0/\kappa}}{1+e^{-3k^0/\kappa}} \  \frac{8\kappa^2}{\big(\vec{k}^{\hspace{2pt}2}+\kappa^2\mu^2_{-}(k^0)\big)^2},
\end{equation}
which permits us to control the growth of each contributions. Now, upon using the relations \eqref{formulassss} together with \eqref{borne} in \eqref{Pcontrib}, we find after standard computations
\begin{align}\label{eq-planar}
\Gamma_j^P(k_\text{ext})&\leq\ \frac{e^{-2Q^0_j/\kappa}}{A_j(2\pi)^{10}}\int_0^\infty\frac{y^6\Psi_j(y)\ dy}{(1+y^3)(1+y^3e^{-3Q^0_j/\kappa})}\int_0^1\frac{x(1-x)\ dx}{\left(-\alpha_j(y)x^2+\beta_j(y)x+\mu_{-}^2(y)\right)^{5/2}}\\
&\leq\ \frac{e^{-2Q^0_j/\kappa}}{6A_j(2\pi)^{10}}\int_0^\infty\frac{y^6\Psi_j(y)\ dy}{(1+y^3)(1+y^3e^{-3Q^0_j/\kappa})}\ \max\left[\frac{1}{\mu^2_{-}(y)},\frac{A^2_j}{\mu^2_{-}(ye^{-Q^0_j/\kappa})}\right]^{5/2},\nonumber
\end{align}
in which the $\Psi_j$'s are given by eq. \eqref{CoeffP}, and the coefficients appearing in the first line of \eqref{eq-planar} are given by
\begin{equation}
\alpha_j(y):=\frac{\|\vec{Q}_j\|^{2}}{\kappa^2}\ y^{2\epsilon_j},\ \beta_j(y):=\alpha_j(y)+A_j^{-2}\mu_{-}^2(ye^{-Q^0_j/\kappa})-\mu_{-}^2(y).
\end{equation}
Now, setting $j=1$, it is easy to see that the integrand in the second line of \eqref{eq-planar} behaves (at leading order) like $\sim y^3$ when $y\to0$ while it behaves like $\sim y^{-5}$ when $y\to\infty$. This indicates that (P1) is finite. Setting $j=2$, we find that the integrand now behaves like a constant when $y\to0$ while it behaves like $\sim y^{-2}$ when $y\to\infty$. Thus, (P2) is also finite and we conclude that all of the planar contributions are UV finite.
\subsection{Nonplanar contributions.}
We now turn to the study of the nonplanar contributions, eq. \eqref{NPcontrib}. In this case, it is not possible to factorise out a delta function depending only on the external momenta as it was the case for the planar contributions. In particular, it is not possible to use factorisation rules like those given in eq. \eqref{fusion-rule}. Rather, after integrating over $k_6$, one of the two 3-momenta delta function involving $\vec{k}_5$ remains and the integration with respect to this latter variable can be done easily. However, a simple analysis based on the respective positions of the contracted momenta involved in \eqref{nonplanar1} and \eqref{nonplanar2} shows that the two families (NP3) and (NP4) are of different nature, see below.\smallskip

Integrating over $k_6$ in \eqref{nonplanar1}, we obtain
\begin{equation}\label{NP1p2}
k^0_6=k^0_5+(k^0_4-k^0_3),\ \ \vec{k}_6=\vec{c}_1y+\vec{c}_2,\ \ \vec{k}_5=\vec{c}_3y+\vec{c}_4.
\end{equation}
Making use of the bound \eqref{borne} on the propagator, we find
\begin{equation}\label{eqNP1}
\Gamma^{N\!P}_3(k_\text{ext})\leq\kappa^5\int\frac{\delta\left(k_4^0-k_3^0+k_2^0-k_1^0\right)y^9(1+y^3)^{-1}(1+c_0^3y^3)^{-1}\Psi_3(y)\ dy}{\left[(\vec{c}_3y+\vec{c}_4)^2+\kappa^2\mu_{-}^2(y)\right]^2\left[(\vec{c}_1y+\vec{c}_2)^2+\kappa^2\mu_{-}^2(c_0y)\right]^2},
\end{equation}
where $c_0$, $\vec{c}_i$, $i=1,\cdots,4$, are (non vanishing) functions of the external momenta (constant of $y$) whose explicit expressions are unessential for the ensuring analysis. Note that we have dropped the overall $2\pi$ factors. In the limit $y\to0$ we find that the integrand behaves (at leading order) like $\sim y^3$, while it behaves like $\sim y^{-5}$ when $y\to\infty$. Hence, (NP3) is (UV) finite.\smallskip

In the same way, integrating over $k_6$ in \eqref{nonplanar2}, we obtain
\begin{equation}\label{NP2p2}
k^0_6=-k^0_5+(k^0_1+k^0_3),\ \vec{k}_6=\frac{1}{y}\left(\vec{c}_1^{\hspace{3pt}'}y+\vec{c}_2^{\hspace{3pt}'}\right),\ \vec{k}_5=\vec{c}_3^{\hspace{3pt}'}y+\vec{c}_4^{\hspace{3pt}'},
\end{equation}
such that
\begin{equation}\label{eqNP2}
\Gamma^{N\!P}_4(p_\text{ext})\leq\kappa^5\int\frac{\delta\left(p_4^0-p_3^0+p_6^0-p_5^0\right)y^9(1+y^3)^{-1}(y^3+c_0^{'\hspace{2pt}3})^{-1}\Psi_4(y)\ dy}{[(\vec{c}_4^{\hspace{3pt}'}+\vec{c}_3^{\hspace{3pt}'}y)^2+\kappa^2\mu_{-}^2(y)]^2[(\vec{c}_2^{\hspace{3pt}'}+\vec{c}_1^{\hspace{3pt}'}y)^2+\kappa^2y^2\mu_{-}^2(\frac{c_0^{'}}{y})]^2},
\end{equation}
where $c^{'}_0$, $\vec{c}_i^{\hspace{3pt}'}$, $i=1,\cdots,4$, are other (non vanishing) functions of the external momenta (constant of $y$), and we have dropped the overall $2\pi$ factors. Mere comparison of eq. \eqref{eqNP1} and \eqref{eqNP2} shows that (NP4) is also finite. We conclude that the nonplanar contributions are finite at nonexceptional external momenta.\smallskip

It remains to check that the nonplanar contributions have no singularities at some specific values of the external momenta. Since $\Gamma^{(4)}_1$ involves four external momenta, the analysis is a bit less straightforward than for the 2-point functions. However, a careful analysis shows that neither (NP3) nor (NP4) become singular for some values of the external momenta. To proceed, we have to turn off successively (one by one) the external momenta, and repeat the operation for all possible configurations. It is easy to show that turning off only one of the external momenta in eq. \eqref{NPcontrib}, the contributions (NP3) and (NP4) remain nonplanar and are still finite. When turning off two external momenta, we find that there are two configurations of the external momenta for which (NP3) and (NP4) become planar. For all of the other configurations they remain nonplanar and finite.\smallskip 

Setting $(k^0_1,\vec{k}_1)=(k^0_2,\vec{k}_2)$ in \eqref{nonplanar1}, one finds
\begin{equation}
\Gamma^{N\!P}_3(k_1,k_1,k_3,k_4)\leq\frac{4\kappa^{5/2}}{(2\pi)^{10}}V_{1134}\int\ \frac{y^6(1+y^3)^{-2}\Psi_3(y)\ dy}{(\kappa-2\sqrt{\kappa^2-m^2}y+\kappa y^2)^{5/2}},
\end{equation}
which behaves asymptotically like $\int^\infty y^{-5}$ and $\int_0dy$. Similar computations show that $\Gamma^{N\!P}_3(k_1,k_2,k_3,k_3)\propto V_{1233}$ is finite. Thus, no singularity shows up in this case.\smallskip

Setting $(k^0_2,\vec{k}_2)=(k^0_3,\vec{k}_3)$ in \eqref{nonplanar2} and using \eqref{borne} together with \eqref{formulassss}, we find
\begin{align}\label{eq-nonplanar2}
\Gamma^{NP}_2(k_1,k_2,k_2,k_4)&\leq\int dy\ \frac{y^6\Psi_4(y)}{(1+y^3)(y^3+c_0^{'\hspace{2pt}3})} \int_0^1 \frac{(1-x)x\ e^{(k_1^0-k^0_2)/\kappa}V_{1224}\ dx}{\left(-h_1(y)x^2+h_2(y)x+\mu^2_{-}(y)\right)^{5/2}},\\
&\leq\int dy\ \frac{y^6\Psi_4(y)}{(1+y^3)(y^3+c_0^{'\hspace{2pt}3})}\frac{e^{(k_1^0-k^0_2)/\kappa}V_{1224}}{(\kappa-2\sqrt{\kappa^2-m^2}y+\kappa y^2)^{5/2}},\nonumber
\end{align}
where the new coefficients are given by
\begin{equation}
h_1(y):=\frac{(\vec{k}_4+\vec{k}_2e^{k_2^0/\kappa}y)^2}{\kappa^2},\ h_2(y):=h_1(y)+y^2\mu^2_{-}\left(\frac{e^{(k_2^0+k_4^0)/\kappa}}{y}\right)-\mu^2_{-}(y).
\end{equation}
From the last inequality of eq. \eqref{eq-nonplanar2}, we conclude that $\Gamma^{N\!P}_4(k_1,k_2,k_2,k_4)$ is finite since the integrand in \eqref{eq-nonplanar2} behaves like $\sim y^{-5}$ when $y\to\infty$ while it behaves like a constant when $y\to0$. A similar conclusion hold when setting $(k_1^0,\vec{k}_1)=(k_4^0,\vec{k}_4)$. Thus, we conclude that no singularity shows up in this case neither.
\subsection{Beta function.}
In the previous section we have shown that the one-loop 4-point function for the model with equivariant kinetic operator and orientable interactions is finite and without UV/IR mixing. This indicate that the (one-loop) beta function is zero and that the coupling constant is not renormalised. This will be discussed in the conclusion. On the other hand, from the computations of both $\Gamma^{(2)}_1$ and $\Gamma^{(4)}_1$, we can deduce the counterterms entering the definition of the renormalised action. Keeping in mind eq. \eqref{kinetic-map} and \eqref{o-int2}, we can write
\begin{subequations}
\begin{align}
\mathcal{S}_{\kappa,r}[\phi,\bar{\phi}]=&\ \int \frac{d^4k}{(2\pi)^4}\left(\bar{\phi}_r(k)\tilde{\mathcal{K}}_0^\text{eq}(k)\phi_r(k)+\bar{\phi}_r(k)(m_1^2+m_2^2e^{-3k_0/\kappa})\phi_r(k)\right)+\\
&\hspace{1truecm}+\int \prod_{i=1}^4\left[\frac{d^4k_i}{(2\pi)^4}\right]\bar{\phi}_r(k_1)\phi_r(k_2)\bar{\phi}_r(k_3)\phi_r(k_4) \mathcal{V}_\textit{o}{(k_1,k_2,k_3,k_4)},\nonumber
\end{align}
where $\tilde{\mathcal{K}}^\text{eq}_0$ is given by \eqref{kin-hypa} with $M=0$. In particular, neither the wave functions ($\phi_r=\phi$) nor the coupling constant(s) are renormalised. The renormalized mass terms are related to the  counterterms $\delta m_i^2$ and the bare quantities involved in the classical action by
\begin{align}
m_i^2+\delta m_i^2=m^2,
\end{align}
\end{subequations}
where $\delta m_i$ can be read from \eqref{result-eqo}. Note that $m_1$ and $m_2$ differ from each other only by finite renormalisation terms in such a way that the tree-level structure of the action functional is preserved.
\chapter{Quantum field theory on \texorpdfstring{$\mathfrak{su}(2)$}{su(2)} noncommutative spacetime.}\label{sap-ncftsu2}
In this chapter, we consider both real and complex scalar field theories with quartic interactions and massive Laplacian of $\mathbb{R}^3$ as kinetic operators. The models are built from the material introduced in $\S$\ref{sukont}, namely making use of the Kontsevich product associated with $\mathbb{R}^3_\theta$. Because of the unimodularity of the compact Lie group $SU(2)$, a natural notion of involution is provided by the ordinary complex conjugation $f\mapsto\bar{f}$, which we shall use in the following to define reasonable reality condition for the action functional. Both UV and IR behaviours of the corresponding one-loop 2-point functions are analysed. By a simple inspection of the perturbative expansion of the effective action $\Gamma$ (see Appendix \ref{sap-perturbation}), it can be easily realised that the one-loop 2-point correlation function for the $\mathbb{R}$-valued field case receives two types of contributions, hereafter called Type-I and Type-II contributions, depending on whether or not the contracted lines giving rise to the propagator are related to two consecutive exponential factors or not, upon taking into account the cyclicity of the trace $\int d^3x$. In the case the fields are $\mathbb{C}$-valued, the form of the interaction term determines which type of contributions have to be taken into account for the 2-point function. Only Type-I contributions matter when the interaction is given by $\int d^3x\ \bar{\phi}\star_\mathcal{K}\phi\star_\mathcal{K}\bar{\phi}\star_\mathcal{K}\phi$, while both Type-I and Type-II actually contribute when the interaction is given by $\int d^3x\ \bar{\phi}\star_\mathcal{K}\bar{\phi}\star_\mathcal{K}\phi\star_\mathcal{K}\phi$. As already mentioned in $\S$\ref{sec-interaction}, the first (resp. second) type of interactions is known as orientable (nonorientable) interactions. The terminology invariant (resp. noninvariant) interactions is also sometimes used to designate these interactions. Invariance or noninvariance is with respect to the transformations defined by the natural action of the automorphisms of the algebra viewed as a (right-)module on itself, compatible with the canonical Hermitian structure used here, namely $h(a_1,a_2)=\bar{a}_1\star_\mathcal{K} a_2$. Thus, we can write $\phi^g=g\star_\mathcal{K} \phi$, for any $g$ with $\bar{g}\star_\mathcal{K} g=g\star_\mathcal{K} \bar{g}=\bbone$ so that $h(\phi_1^g,\phi_2^g)=h(\phi_1,\phi_2)$.

In Sec. \ref{sec-typeI}, we first consider the real field case and focus on the analysis of Type-I contributions, showing that they are both IR and UV finite. The extension to the case of complex scalar field with interaction $\int d^3x\ \bar{\phi}\star\phi\star\bar{\phi}\star\phi$ is then given. Similar conclusions are obtained for the $\mathbb{C}$-valued NCFT. In Sec. \ref{sec-typeII}, we go back to the real field case and consider Type-II contributions. These are found to be IR finite. The corresponding UV behaviour is then analysed. The case of complex model with noninvariant interaction is also discussed. The results presented here are published in
\cite{moi:2016}.
\section{Type-I contributions.}\label{sec-typeI}
We first consider a $\mathbb{R}$-valued scalar field theory with quartic interaction whose classical action is defined by
\begin{equation}\label{real-clas-action}
\mathcal{S}:=\int d^3x\left(\frac{1}{2}\partial_\mu\phi\star_\mathcal{K}\partial_\mu\phi+\frac{1}{2}m^2\phi\star\phi+\frac{\lambda}{4!}\phi\star_\mathcal{K}\phi\star_\mathcal{K}\phi\star_\mathcal{K}\phi\right)(x),
\end{equation}
where $\star_\mathcal{K}$ is given by \eqref{kontsev-product}. The fields and parameters are assumed to have the usual $\mathbb{R}^3$ mass dimensions, namely $[\phi]=\frac{1}{2}$, $[\lambda]=1$ and $[m]=1$.

Thanks to the property \eqref{starclos} of the star product under integration sign, the kinetic term of \eqref{real-clas-action} reduces to
\begin{equation}
\mathcal{S}^\text{kin}:=\int d^3x\ \big(\partial_\mu\phi\partial_\mu\phi+m^2\phi\phi\big)(x),
\end{equation}
where the $\partial_\mu$'s are the usual derivatives (with respect to $x^\mu$) on $\mathbb{R}^3$.\smallskip

The interaction term can be conveniently recast into the form
\begin{subequations}\label{interaction0}
\begin{equation}\label{interaction}
\mathcal{S}^\text{int}:=\frac{\lambda}{4!}\int d^3x\int\ \left[\prod_{i=1}^4 \frac{d^3k_i}{(2\pi)^3}
\widetilde\phi(k_i)\right](e^{ik_1\cdot x}\star_\mathcal{K} e^{ik_2\cdot x}\star_\mathcal{K} e^{ik_3\cdot x}\star_\mathcal{K} e^{ik_4\cdot x})(x),
\end{equation}
which can be equivalently written as, using eq. \eqref{duflopw},
\begin{equation}\label{interaction1}	
\mathcal{S}^\text{int}=\frac{\lambda}{4!}\int\ \left[\prod_{i=1}^4 \frac{d^3k_i}{(2\pi)^3}
\widetilde\phi(k_i)\right]\mathcal{W}(k_1,k_2)\mathcal{W}(k_3,k_4)\delta(B(k_1,k_2)+B(k_3,k_4)).
\end{equation}
\end{subequations}
We will use alternatively \eqref{interaction} and \eqref{interaction1} in the computation of the contributions to the 2-point correlation function. Notice that the standard conservation law of the momenta $\delta(\sum_{i=1}^4k_i)$ of the commutative $\phi^4$ theory on $\mathbb{R}^3$ in replaced by a nonlinear one as it can be seen from the delta function in \eqref{interaction1}. This complicates strongly the perturbative calculations, which however can be partly overcome by a suitable use of \eqref{interaction} combined with properties of the plane waves and cyclicity of the trace.\smallskip

A typical contribution of Type-I to the one-loop effective action is easily found to be given by
\begin{equation}
\Gamma^{(I)}_2=\int d^3x \left[\prod_{i=1}^4 \frac{d^3k_i}{(2\pi)^3}\right]\widetilde\phi(k_3)\widetilde\phi(k_4)\frac{\delta(k_1+k_2)}{k_1^2+m^2}(e^{ik_1\cdot x}\star_\mathcal{K} e^{ik_2\cdot x}\star_\mathcal{K} e^{ik_3\cdot x}\star_\mathcal{K} e^{ik_4\cdot x})(x)\label{type1}
\end{equation}
where we dropped the overall constant $\sim\lambda$. Combining \eqref{type1} with \eqref{Bproperties}, \eqref{theweight} and
\begin{equation}\label{plane-norm}
(e^{ikx}\star_\mathcal{K} e^{-ikx})(x)=\frac{4}{\theta^2}\frac{\sin^2(\frac{\theta}{2}|k|)}{|k|^2},
\end{equation}
we obtain
\begin{subequations}
\begin{align}\label{amplitude-I}
\Gamma^{(2)}_{1;I}&=\int d^3x\frac{d^3k_3}{(2\pi)^3}\frac{d^3k_4}{(2\pi)^3}\widetilde\phi(k_3)\widetilde\phi(k_4)(e^{ik_3x}\star_\mathcal{K} e^{ik_4x})(x) \omega^{(I)}\\
&=\int d^3x(\phi\star_\mathcal{K}\phi)(x)\omega^{(I)}=\int d^3x\ \phi(x)\phi(x)\omega^{(I)},\nonumber
\end{align}
with
\begin{equation}
\omega_{I}=\frac{4}{\theta^2}\int\frac{d^3k}{(2\pi)^3}\frac{\sin^2(\frac{\theta}{2}|k|)}{k^2(k^2+m^2)} \label{type1omega}.
\end{equation}
\end{subequations}
This integral can be easily compute upon using spherical coordinates, \textit{i.e.}
\begin{equation}\label{type1-final}
\omega_{I}=\frac{4}{\theta^2}\int\frac{d^3k}{(2\pi)^3}\frac{\sin^2(\frac{\theta}{2}|k|)}{k^2(k^2+m^2)}=\frac{1}{\pi^2\theta^{2}}\int^{\infty}_{0}dr\frac{1-\cos(\theta r)}{r^2+m^2}=\frac{1-e^{-\theta m}}{2\pi m\theta^2},
 \end{equation}
from which we conclude that $\omega_{I}$ is finite whenever $\theta\neq0$. Moreover, expanding the exponential around $m\sim 0$, we find $\omega_{I}=(2\pi\theta)^{-1}+\mathcal{O}(m)$ indicating that the massless case is also not singular. Hence, Type-I contributions are UV finite and do not exhibit IR singularity. We conclude that whenever $\theta\ne0$, Type-I contributions cannot generate UV/IR mixing. Note that the closed star product structure of the quadratic part of the effective action survives the one-loop quantum corrections at it is apparent from \eqref{amplitude-I}.\smallskip

From \eqref{type1-final}, we readily obtain the small $\theta$ expansion of $\omega_{I}$, namely
\begin{equation}\label{comutlim1}
\omega_{I}\xrightarrow[\theta\rightarrow 0]{}\Lambda+\mathcal{O}(1),\ \Lambda:=\frac{1}{2\pi\theta}.
\end{equation}
Thus, we recover the expected linear divergence (showing up when $\Lambda\to\infty$) which occurs in the 2-point function for the commutative theory with $\Lambda$ as the UV cutoff. In physical words, the present noncommutativity of $\mathfrak{su}(2)$ type gives rise to a natural UV cutoff for the scalar field theory \eqref{real-clas-action} which regularises both the UV and the IR (massless case).
\paragraph{Complex scalar field theories.}{}
The above one-loop analysis extends easily to Type-I contributions for the 2-point function of the complex scalar field theories with orientable or nonorientable interactions. \\
In the case of invariant interaction, the 2-point function only receives Type-I contributions. The action is
\begin{equation}
\mathcal{S}=\int d^3x\big[\partial_\mu\bar{\phi}\star_\mathcal{K}\partial_\mu\phi+m^2\bar{\phi}\star_\mathcal{K}\Phi+
{\lambda}\bar{\phi}\star_\mathcal{K}\phi\star_\mathcal{K}\bar{\phi}\star_\mathcal{K}\Phi\big]\label{complx-clas-action}.
\end{equation}
The one-loop quadratic part of the effective action is now given by
\begin{equation}
\Gamma^{(2)}_{1;I}=\int d^3x \left[\prod_{i=1}^4 \frac{d^3k_i}{(2\pi)^3}\right]\widetilde{\bar{\phi}}(k_3)\widetilde{\phi}(k_4)\frac{\delta(k_1+k_2)}{k_1^2+m^2}(e^{ik_1x}\star_\mathcal{K} e^{ik_2x}\star_\mathcal{K} e^{ik_3x}\star_\mathcal{K} e^{ik_4x})(x)\label{type1-complx}
\end{equation}
where we dropped again the overall constant $\sim\lambda$. Adapting the previous analysis, we immediately obtain 
\begin{equation}\label{omega-complx}
\Gamma^{(2)}_{1;I}=\int d^3x\ \bar{\phi}\phi\ \omega_{I},
\end{equation}
with $\omega_{I}$ still given by eq. \eqref{type1omega}. We conclude that the Type-I contribution of the complex case is similar to the Type-I contribution of the real case. Hence, the Type-I is finite and the same conclusions as before hold. A similar conclusion obviously holds true for the Type-I contributions involved in the 2-point function related to the scalar theory with noninvariant interaction. However, Type-II contributions mentioned at the beginning of this section are also involved in that case. 
\section{Type-II contributions.}\label{sec-typeII}
Let us go back to the real scalar field theory \eqref{real-clas-action}. A typical Type-II contribution to the one-loop quadratic part of the effective action is given by
\begin{subequations}
\begin{equation}\label{type2}
\Gamma^{(2)}_{1;I\hspace{-1pt}I}=\int\frac{d^3k_2}{(2\pi)^3}\frac{d^3k_4}{(2\pi)^3}\widetilde\phi(k_2)\widetilde\phi(k_4)\omega^{(I \hspace{-1pt} I)}(k_2,k_4),
\end{equation}
with
\begin{align}\label{type2-omega}
\omega_{I\hspace{-1pt}I}(k_2,k_4)&=\int d^3x\frac{d^3k_1}{(2\pi)^3}\frac{d^3k_3}{(2\pi)^3}\frac{\delta(k_1+k_3)}{k_1^2+m^2}(e^{ik_1\cdot x}\star_\mathcal{K} e^{ik_2\cdot x}\star_\mathcal{K} e^{ik_3\cdot x}\star_\mathcal{K} e^{ik_4\cdot x})(x)\\
&=\int d^3x\frac{d^3k}{(2\pi)^3}\frac{1}{k^2+m^2}(e^{ik\cdot x}\star_\mathcal{K} e^{ik_2\cdot x}\star_\mathcal{K} e^{-ik\cdot x}\star_\mathcal{K} e^{ik_4\cdot x})(x),\nonumber
\end{align}
\end{subequations}
where the internal momentum involves two nonadjacent exponential factors.\smallskip

We first consider the infrared regime of \eqref{type2-omega} corresponding to the small external momenta region, \text{i.e.} $k_2\sim0,\ k_4\sim0$. From \eqref{theweight}, we infer
\begin{equation}\label{expo-unit}
\lim_{k_2\to0}(e^{ik_1\cdot x}\star_\mathcal{K} e^{ik_2\cdot x})(x)=e^{ik_1\cdot x}
\end{equation}
which simply reflects the fact that $E_{k=0}(\hat{x})=\bbone$, or equivalently, $Q(1)=\bbone$.\\
Then, we can write
\begin{align}\label{ir-omega2}
\omega_{I\hspace{-1pt}I}(0,k_4)&=\int d^3x\frac{d^3k}{(2\pi)^3}\frac{1}{k^2+m^2}(e^{ik\cdot x}\star_\mathcal{K} e^{-ik\cdot x}\star_\mathcal{K} e^{ik_4\cdot x})(x)\\
&=\delta(k_4)\frac{4}{\theta^2}\int\frac{d^3k}{(2\pi)^3}\frac{\sin^2(\frac{\theta}{2}|k|)}{k^2(k^2+m^2)}=\delta(k_4)\ \omega_{I}.\nonumber
\end{align}
where $\omega_{I}$ is given by \eqref{type1-final} and we have used \eqref{plane-norm} to obtain the second equality. From the discussion about Type-I contributions given in $\S$\ref{sec-typeI}, we conclude that \eqref{ir-omega2} is not IR singular (and also UV finite). A similar result holds true for $\omega_{I\hspace{-1pt}I}(k_2,0)$. The extension to complex scalar field theories is obvious. We conclude that no IR singularity shows up in the 2-point functions for both real and complex (even massless $m=0$) scalar field theories at one-loop so that these NCFT are free from UV/IR mixing.\smallskip

Unfortunately, since exponentials no longer simplify in \eqref{type2-omega}, now we have to deal with infinite expansions stemming from the $B(k_1,k_2)$ function in \eqref{theweight} and/or delta functions with nonlinear arguments which complicate considerably the UV analysis of the Type-II contributions. However, this situation can be slightly simplified by considering a somewhat restricted situation for which the coordinate functions $x_\mu$ satisfying the relation \eqref{su2per} for the Lie algebra $\mathfrak{su}(2)$ are represented as Pauli matrices. Note that a similar representation is used in models related to quantum gravity
\cite{Freidel:2008,Guedes:2013} in which are involved noncommutative structures similar to the ones we considered here. Namely, we introduce the following morphism of algebra $\rho: \mathfrak{su}(2)\to \mathbb{M}_2(\mathbb{C})$
\begin{equation}\label{decadix}
\rho(\hat{x}_\mu) = \theta \sigma_\mu,\ \rho(\bbone) = \bbone.
\end{equation}
From the usual properties of the Pauli matrices, we obtain the following relation
\begin{equation} \label{representation_operateur}
\rho (\hat{x}_i \hat{x}_j) = \rho(\hat{x}_i) \rho(\hat{x}_j) = \theta^2 \delta_{ij} \bbone + i \frac{\theta}{2} \varepsilon_{ij}^{\hspace{5pt} k} \rho (\hat{x}_k),
\end{equation}
which will give rise to a rather simple expression for the exponential factors occurring in \eqref{type2-omega}. Indeed, after some algebraic manipulations, we obtain 
\begin{equation}\label{expo-coupe}
e^{ik^\mu \rho(\hat{x}_\mu)} = \cos\left(\theta |k| \right) \bbone + i \frac{\sin\left(\theta |k| \right)}{\theta |k|} k^\mu \rho(\hat{x}_\mu),
\end{equation}
which finally implies
\begin{equation}\label{commut-expon}
[ e^{ik^\mu\rho(\hat{x}_\mu)} , e^{ip^\nu \rho(\hat{x}_\nu)}] = - i \theta \frac{\sin\left(\theta |k| \right)}{\theta |k|} \frac{\sin\left(\theta |p| \right)}{\theta |p|} \varepsilon_{\sigma \nu}^{\hspace{8pt} \rho} k^\sigma p^\nu \rho(\hat{x}_\rho).
\end{equation}

On the other hand, $\Gamma^{(2)}_{1;I\hspace{-1pt}I}$, eq. \eqref{type2}, can be conveniently rewritten as 
\begin{equation} \label{TypeII_asI}
\Gamma^{(2)}_{1;I\hspace{-1pt}I} = \Gamma^{(2)}_{1;I} + \int \frac{d^3k_2 }{(2\pi)^3} \frac{d^3k_4 }{(2\pi)^3} \widetilde{\phi}(k_2) \widetilde{\phi}(k_4) I(k_2,k_4),
\end{equation}
where in obvious notations
\begin{align}\label{ik2k4}
 I(k_2,k_4) =& \int \frac{d^3k }{(2\pi)^3} \frac{d^3x}{k^2+m^2} \left[ e^{ik^\nu x_\nu} , e^{ik_2^\nu x_\nu} \right]_{\star_\mathcal{K}} \star_\mathcal{K} e^{-ik^\nu x_\nu} \star_\mathcal{K} e^{ik_4^\nu x_\nu}  \\
 =& \left(\frac{2}{\theta}\right)^4 \int \frac{d^3k }{(2\pi)^3} \frac{d^3x}{k^2+m^2} \left(\frac{\sin(\frac{\theta}{2}|k|)}{|k|}\right)^2 \frac{\sin(\frac{\theta}{2}|k_2|)}{|k_2|} \frac{\sin(\frac{\theta}{2}|k_4|)}{|k_4|}\nonumber\\
 &\times \mathcal{Q}^{-1} \left( \left[ e^{ik^\mu \hat{x}_\mu} , e^{ik_2^\nu \hat{x}_\nu} \right] e^{-ik^\sigma \hat{x}_\sigma} e^{ik_4^\rho \hat{x}_\rho} \right)\nonumber .
\end{align}
Hence, making use of \eqref{decadix} together with \eqref{commut-expon}, we arrive after a lengthy computation given in Appendix \ref{appendixb} to the following expression
\begin{align} \label{typeII_spherique}
I(k_2,k_4)&\\
= &\ \frac{J(k_2,k_4)}{\pi^3 \theta^4} \int d\alpha d\beta d r \frac{\sin^2(\frac{\theta}{2}r)}{r^2+m^2} \left[ \frac{1}{2} \sin\left(2 \theta r \right) \sin\gamma + \sin^2\left(\theta r \right) \sin^2\frac{\gamma}{2} \right] \sin\alpha,\nonumber
\end{align}
with
\begin{equation}\label{france-j}
 J(k_2,k_4) = \frac{\sin\left(\theta |k_2| \right) \sin(\frac{\theta}{2}|k_2|)}{|k_2|} u^\mu \delta_\mu^{'}(k_4).
\end{equation}
In the above expressions, we have introduced spherical coordinates for the momentum $k$, namely $k=(r=|k|,\alpha,\beta)$, together with the following notations: $u_\mu$ is the $\mu$-component of a unit vector $u$, $\gamma$ is the angle between the momenta $k$ and $k_2$ (depending only on $\alpha$, $\beta$ and $\alpha_2$, $\beta_2$ entering the spherical coordinates for $k_2$) and $\delta^\prime_\mu$ is defined by $\langle \delta^\prime_\mu, f\rangle=-\frac{\partial f}{\partial k_{4}^\mu}$ for any test function $f$. Upon introducing an estimate on the integrand appearing in eq. \eqref{typeII_spherique}, we can check that $I(k_2,k_4)$ is finite; see eq. \eqref{typeIIbounds}. We conclude that the Type-II contributions are UV finite. The exact derivation can be performed as follow.\smallskip

The radial integration in \eqref{typeII_spherique} can be performed by further using 
\begin{subequations}\label{GS-integral-prime}
\begin{align}
&\int_0^\infty dx\frac{\cos(ax)}{\beta^2+x^2}=\frac{\pi}{2\beta}e^{-a\beta},\ a\geq 0,\ \text{Re}(\beta)>0,\\ 
&\int_0^\infty dx\frac{\sin(ax)}{\beta^2+x^2} = \frac{1}{2\beta} \left( e^{-a \beta} \overline{\text{Ei} (a \beta)} - e^{a \beta} \text{Ei} (- a \beta) \right),\ a>0,\ \beta >0,
\end{align} 
\end{subequations}
where $\text{Ei}$ is the exponential integral function defined by
\begin{equation}
 \text{Ei}(x) = - \underset{e \rightarrow 0^{+}}{\lim} \left[ \int^{-e}_{-x} \frac{e^{-t}}{t}dt + \int^\infty_e \frac{e^{-t}}{t}dt\right],\ x>0
\end{equation}
and we have
\begin{equation}
 \text{Ei}(x) = \textbf{C} + \ln |x| +\sum \limits^\infty_{n=1} \frac{x^n}{n.n!},\ x\neq 0,
\end{equation}
in which \textbf{C} the Euler-Mascheroni constant. We obtain
\begin{subequations}\label{commutative_typeII}
\begin{align}\label{commutative_typeII_1}
\frac{1}{2}\int d r \frac{\sin^2(\frac{\theta}{2}r)}{r^2+m^2}&\sin\left(2 \theta r \right)\\
=&\ \frac{1}{16m} \left[ 2e^{-2 \theta m} \overline{\text{Ei} (2 \theta m)} - 2e^{2 \theta m} \text{Ei} (-2 \theta m) - e^{-3 \theta m} \overline{\text{Ei} (3 \theta m)} \right. \nonumber \\
&\hspace{1.5truecm}+ \left. e^{3 \theta m} \text{Ei} (-3 \theta m) - e^{- \theta m} \overline{\text{Ei} ( \theta m)} + e^{ \theta m} \text{Ei} (- \theta m) \right], \nonumber 
\end{align}
and
\begin{equation}\label{commutative_typeII_2}
\int d r \frac{\sin^2(\frac{\theta}{2}r)}{r^2+m^2} \sin^2\left(\theta r \right) = \frac{\pi}{8m} \left[ 1-\left(1+\sinh(\theta m)\right)e^{-2\theta m} \right].
\end{equation}
\end{subequations}
In the small $\theta$ limit (i.e formal commutative limit $\theta\to0$), we infer
\begin{subequations}\label{decadix-1}
\begin{equation}
\frac{J(k_2,k_4)}{\pi^3 \theta^4} = \frac{1}{\pi^3 \theta^2} |k_2| u^n \delta^{'}_n(k_4) + \mathcal{O}(1), 
\end{equation}
and eq. \eqref{commutative_typeII} becomes
\begin{align}
&\frac{1}{2}\int d r \frac{\sin^2(\frac{\theta}{2}r)}{r^2+m^2}\sin\left(2 \theta r \right)=(6\ln3-8\ln2)\theta+\mathcal{O}(\theta^2),\\
&\int d r \frac{\sin^2(\frac{\theta}{2}r)}{r^2+m^2} \sin^2\left(\theta r \right) =\frac{\pi}{8} \theta + \mathcal{O}(\theta^2).
\end{align}
\end{subequations}
Combining \eqref{decadix-1} with \eqref{typeII_spherique} yields the following small $\theta$ limit:
\begin{equation}
I(k_2,k_4)=\frac{C(\alpha_2,\beta_2)}{\theta} |k_2| u^n \delta^{'}_n(k_4) + \mathcal{O}(1),
\end{equation}
where $C(\alpha_2,\beta_2)$ is finite. Hence, as for the Type-I contributions, the $\theta$ expansion of $I(k_2,k_4)$ is 
\begin{equation}
I\xrightarrow[\theta\rightarrow 0]{}\Lambda+\mathcal{O}(1),\ \Lambda=\frac{1}{2\pi\theta}.
\end{equation}
Thus, combining this result with the decomposition \eqref{TypeII_asI} of $\Gamma_{1;I \hspace{-1pt} I}^2$, we recover once more the expected linear divergence when $\Lambda\to\infty$ ($\theta\to0$) occurring in the 2-point function for the commutative theory. Again, the present $\mathfrak{su}(2)$ noncommutativity generates a natural UV cutoff for the scalar field theory. Notice that this holds true even when $m=0$. This result extends obviously to complex scalar field theories.
\part*{Summary and outlook.\addcontentsline{toc}{part}{Summary and outlook.}}
In the present dissertation, we have considered various candidates for a quantum spacetime to be involved in the description of physical phenomena at some quantum gravity scale. These candidates were characterised by different Lie algebras of coordinate operators, say $\mathfrak{g}=\text{Lie}(\mathcal{G})$, which can be gathered into two different categories. In one case, $\mathfrak{g}$ was semisimple, while in the other case it was solvable. Accordingly, in the former case the corresponding Lie group was unimodular, in the latter it was not. Starting from their respective algebras of coordinate operators, we have constructed various families of star products within different approaches. Then, focusing on $\mathbb{R}^3 _\theta$ (a deformation of  $\mathbb{R}^3$ with $\mathfrak{su}(2)$ noncommutativity) and $\kappa$-Minkowski, we have constructed various models of noncommutative field theory and studied their quantum properties at one-loop order. Part \ref{ch-ncst} was devoted to the construction of star products associated with these spaces. Part \ref{part-ncft} was devoted to the study of the one-loop order quantum properties of various models of noncommutative scalar field theory with quartic interactions.\smallskip

In Chap. \ref{sec-products}, we have considered various quantum spaces whose algebras of coordinates  were semisimple Lie algebras. We have shown that, assuming the existence of a linear, invertible, quantisation map $Q$, it is possible to construct star products, eq. \eqref{sustar}, only by determining the expression for the deformed plane waves $Q(e^{ip\cdot x})$. Identifying these deformed plane waves with projective representations of $\mathcal{G}$, eq. \eqref{projective}, we have highlighted that inequivalent families of star products can in principle be classified by the second cohomology group of $\mathcal{G}$ with value in ${\mathbb{C}\hspace{-2pt}\setminus\hspace{-2pt}\lbrace0\rbrace}$, \textit{i.e.} $H^2\big(\mathcal{G},{\mathbb{C}\hspace{-2pt}\setminus\hspace{-2pt}\lbrace0\rbrace}\big)$. The advantage of this point of view is that many examples of $H^2\big(\mathcal{G},\mathcal{A}\big)$, with $\mathcal{A}$ an Abelian group, are already classified in the mathematical literature. On the other hand, in view of application to noncommutative field theories (\textit{i.e.} with actual numbers in the end), it is necessary to exhibit (at least) one representative of the deformed plane waves, \textit{i.e.} one representative of (one of the) cohomology class of equivalence which are classified by $H^2\big(\mathcal{G},\mathcal{A}\big)$. And, as we have shown, even in the simple(st) case of $\mathcal{G}=SU(2)$, the derivation of the expression of $Q(e^{ip\cdot x})$ is not trivial and necessitates a few assumptions. This is true despite the fact that $H^2\big(\mathcal{G},\mathcal{A}\big)$ may be trivial. To this end, we have proposed a systematic method to derive explicit expressions for the deformed plane waves, which we now briefly recall the main lines.

This approach amounts to represent the abstract coordinate operators (generators of $\mathfrak{g}$) as differential operators acting on some Hilbert space of functions. Although such idea is not new, we have emphasised that in the mathematical-physics literature the preservation of the various involutive structures underlying the construction of the noncommutative spacetime is not always taken seriously; namely, the selfadjoint coordinate operators are not represented as selfadjoint differential operators. In our opinion, this may have dramatic consequences on the construction of the algebra of fields modelling the quantum space, as well as on the study of noncommutative field theories (and on any other physical applications) as it prevent us from defining a reasonable reality condition for the action functional describing the dynamics of such models. Therefore, in our derivation, particular attention has been drawn to the preservation of the various involutive structures all along the various steps leading to the expressions of the star products. We have shown that the admissible expressions for the differential involutive representations can be obtained from a set of four master differential equations, eq. \eqref{diff-rep3} and \eqref{diff-rep4}. Assuming such a representation to be chosen, we have shown that the expression for $Q(e^{ip\cdot x})$ can be determined by enforcing $Q(f)\triangleright f=f\star g$ and $Q(f)\triangleright1=f$, together with polar decomposition of the deformed plane waves, eq. \eqref{polar}.

Focusing on $\mathfrak{g}=\mathfrak{su}(2)$, we have shown that further assuming the representation to be $SO(3)$-equivariant singles out a particular family of involutive representations indexed by two real $SO(3)$-invariant functionals, eq. \eqref{general_rep}. In this case, the deformed plane waves are found to be characterised by two (representation dependent) functions of the momenta whose expressions are given by two Volterra equations, eq. \eqref{volterra}. The product of two plane waves is merely given (up to some details) by the Baker-Campbell-Hausdorff formula for $SU(2)$, reflecting the nonlinear composition law between momenta. Among these representations, we have shown that one of them is equivalent to the Kontsevich product for the Poisson manifold dual to the finite dimensional Lie algebra $\mathfrak{su}(2)$,  namely closed for the trace functional defined by the usual Lebesgue integral $\int d^3x$. Using this product, we have computed in Chap. \ref{sap-ncftsu2} the one-loop 2-point functions for both real and complex, massive and massless, scalar field theories with quartic interactions. We have exhibited two types of contribution, depending on whether or not the Wick contracted lines giving rise to the propagator are related to two adjacent exponential factors or not; see eq. \eqref{interaction0}. We have found that both Type-I and Type-II contributions were ultraviolet (UV) finite with no infrared (IR) singularity even in the massless case. This likely indicates that such theories are free of UV/IR mixing. Moreover, we have found that the deformation parameter $\theta$ plays the role of a natural UV cutoff in this context. These results qualitatively agree with previous similar investigations; see, e.g.,
\cite{JCW:2013b}.\smallskip

One of the result of Chap. \ref{sap-ncftsu2} is the highlighting of the limitations of the star product formulation for studying noncommutative field theories. At least for some choice of spacetime noncommutativity. This is clear from the computation of the Type-II contribution in $\S$\ref{sec-typeII}. The reason is the very complicated structure of the Baker-Campbell-Hausdorff formula for semisimple Lie algebra. An alternative to this approach consists in working directly at the level of the operators. However, this presumes the existence of a matrix basis in which the operators can be conveniently decomposed. This is the case for the Moyal space
\cite{Gracia:1988}, as well as for any quantum space whose underlying group is compact. In this latter case, a basis can be obtained by mere application of the Peter-Weyl theorem which states (among other things) that the matrix coefficients of the irreducible unitary representations of $\mathcal{G}$ form an orthonormal basis of $L^2(\mathcal{G})$, the convolution algebra of  $\mathcal{G}$. Unfortunately no such a basis exists for $\kappa$-Minkowski (in this case, we have to deal with objects such that direct integral of representations). Hopefully, because the $\kappa$-Minkowski algebra of coordinates is a solvable Lie algebra, the corresponding Baker-Campbell-Hausdorff formula enjoys nice properties as illustrated in $\S$\ref{sec-convolution}. This fact holds for any other quantum space of solvable Lie algebra noncommutativity. In the case $\mathfrak{g}$ is nilpotent, the Baker-Campbell-Hausdorff formula even admits finite expansion. Unlike $\mathfrak{su}(2)$, for this type of noncommutativity the star product formulation proves efficient. This has been illustrated in the present dissertation by the computation of the one-loop 2-point and 4-point functions for noncommutative field theories built on $\kappa$-Minkowski.\smallskip

Natural extensions of this work would be to consider other Lie groups than $SU(2)$, taking benefit from the natural cohomological setting underlying the construction of the deformed plane waves; see $\S$\ref{sec-dpw}. Indeed, group cohomology with value in an Abelian group is the proper tool to investigate the socalled central extension of a group. This would enable us to study how noncommutative field theory on a given quantum space is modified when built from inequivalent star products. For instance we could applied this framework to $SL(2,R)$ of which the $ax+b$ group is a subgroup. About the results obtained in Chap. \ref{sap-ncftsu2}, it would be interesting to study the full renormalisation properties of noncommutative field theories on $\mathbb{R}^3_\theta$. Intermediate steps would be to compute the one-loop 4-point functions for $\phi^4$ theory, as well as consider other polynomial interactions such as $\phi^6$. Indeed, it is known that, in the commutative case, the 3-dimensional $\phi^4$ model is super renormalisable. Therefore, exploring the $\phi^6$ model would provide a way to test if the quantum behaviour is really improved on $\mathfrak{su}(2)$ noncommutative space. In particular, if the deformation parameter still provides a cutoff. However, for the reason already mentioned, the right approach to adopt to undertake such study should be to use the natural matrix basis stemming from the Peter-Weyl decomposition of $SU(2)$. This approach already proves useful in the context of gauge theory on $\mathbb{R}^3_\theta$, see, e.g.,
\cite{JCW:2016}.\smallskip

The other main concern of the present study was noncommutative field theory on $\kappa$-Minkowski background. In Chap. \ref{sec-Minkowski}, we have derive a star product associated with $\kappa$-Minkowski using an approach slightly different from the above-mentioned one. In this case, we have taken advantage of tools from abstract harmonic analysis and group C*-algebras, identifying quantisation maps with *-representations of the convolution algebra of $\mathcal{G}_{d+1}$, the nonunimodular locally compact Lie group obtained by exponentiating the $\kappa$-Minkowski algebra of coordinate operators. This approach was motivated by the original Weyl quantisation scheme used to construct the Groenewold-Moyal product in quantum mechanics. This has been achieved by replacing the Heisenberg group with the $ax+b$ group which is isomorphic to $\mathcal{G}_{d+1}$ in 2-dimensions ($d=1$). This enables us to construct quite easily a well controlled star product, indicating that the Weyl quantisation scheme provides a natural and powerful framework to describe $\kappa$-deformations of the Minkowski spacetime. Actually, this approach might be applied profitably to the construction of star products for any other quantum space whose algebra of coordinates is a solvable or nilpotent Lie algebra (note that this is the case for the Heisenberg algebra entering the construction of the Groenewold-Moyal product). Moreover, within this framework, star product, involution and measure of integration are canonically provided from their corresponding notions at the level of the convolution algebra of the group. Note that these structures actually underlies the three most well known examples of quantum spacetime, \textit{i.e.} Moyal space, $\mathbb{R}^3_\theta$ and $\kappa$-Minkowski.\smallskip

In Chap. \ref{ch-ncft}, we have discussed the properties a reasonable action functional describing the dynamics of noncommutative scalar fields on $\kappa$-Minkowski should have. In view of the very important role played by the Poincar\'{e} algebra in ordinary quantum field theory, together with the fact that $\kappa$-Minkowski support a natural action of  the $\kappa$-Poincar\'{e} algebra (a deformation of the ordinary Poincar\'{e} Lie algebra) which plays the role of the algebra of symmetry of the quantum space, it is physically relevant to require the $\kappa$-Poincar\'{e} invariance of any physically reasonable action functional. This conditioned the use of the Lebesgue integral in the construction of the action functional. Despite its very simple expression, it turns out that the Lebesgue integral is not a trace for the star product associated with $\kappa$-Minkowski. Instead, we have shown that the Lebesgue integral defines a KMS weight on the algebra of fields modelling $\kappa$-Minkowski. This indicates that the $\kappa$-Poincar\'{e} invariance of the action functional trades the cyclicity of the Lebesgue integral for a KMS condition. As discussed in $\S$\ref{sec-KMS} this KMS condition
\begin{equation}\label{finaleq}
\zeta\big((\sigma_t\triangleright f)\star g\big)=\zeta\big(g\star (\sigma_{t-i}\triangleright f)\big),
\end{equation}
represent an abstract version of the KMS condition introduced a long time ago as a tool to characterise equilibrium temperature states of quantum systems in field theory and statistical physics. However, in this case the KMS condition holds at the level of the correlation function $\langle\Sigma_t(A)B\rangle_\beta$ computed from some thermal vacuum where $A$ and $B$ are now functionals of the fields and $\Sigma_t$ is the Heisenberg evolution operator, hence elements pertaining to the algebra of observables of the theory. But whenever a KMS condition holds true on the algebra of observables of a quantum system or a quantum field theory, the flow generated by the modular group, \textit{i.e.} the Tomita flow, may be used to define a global (observer independent) time which can be interpreted as the ``physical time." This reflects the deep correspondence between KMS and dynamics, and underlies the interesting proposal about the thermal origin of time introduced in
\cite{Connes:1994} namely, the ``emergence of time" from noncommutativity. Therefore, it would be tempting to interpret $\sigma_t$, eq. \eqref{defautomorphism}, as generating a ``physical time" for the present system, akin to the thermal time mentioned above. However, no conclusion can yet be drawn. In fact, eq. \eqref{finaleq}, linked to the modular group, only holds at the level of $\mathcal{M}_\kappa$, the algebra of (classical) fields modelling the $\kappa$-Minkowski space. To show that a natural global time can be defined requires to determine if eq. \eqref{finaleq} forces a KMS condition to hold true at the level of the algebra of observables. This could be achieved by actually showing the existence of some KMS state on this latter algebra built from the path integral machinery. Also, it might be useful to exploit objects from C*-dynamical systems to reach this goal. Indeed, C*-dynamical systems naturally arise whenever the Lie group underlying the construction of the C*-algebra of fields has a structure of semidirect product as it is the case for $\kappa$-Minkowski. In view of the possibility to associate to $\kappa$-Poincar\'{e} invariant noncommutative field theories a natural global (cosmological) time, a physically appealing property, the implications of the KMS condition \eqref{finaleq} shared by all these theories obviously deserves further study. Finally, notice that if the KMS condition, eq. \eqref{finaleq}, is transferred at the level of the algebra of observables, any $\kappa$-Poincar\'{e} invariant scalar field theory could be interpreted as a thermal field theory whose thermal bath temperature would be given by some function of $\kappa$. Thus, providing us with a model (possibly) describing the Higgs dynamics at Planck scale.\smallskip

Another interesting feature arising from the construction of the action functional is the central role played by the (canonical) involution $\ddagger$ within such theory. Recall that this involution naturally arises from the construction of the (group) C*-algebra of fields modelling $\kappa$-Minkowski. We have shown that this involution ensures the implementation of a reasonable reality condition for the action functional, namely through the definition of the Hilbert product $\langle f,g\rangle_\star:=\int d^4x (f\star g^\ddagger)(x)$. Note by the way that this Hilbert product is related to the positive linear functional (KMS weight) $\zeta$ via $\langle f,g\rangle_\star=\zeta(f\star g^\ddagger)$. Therefore, $\ddagger$ replaces (and differs from) the ordinary complex conjugation of function. Thus, it would be interesting to explore more deeply how this could modify the usual charge conjugation at the level of $\kappa$-deformed theories, hence its impact on the CPT theorem. Note that deviation from (or violation of) CPT theorem (in a quantum gravity prospect) is an active research area; see, e.g., 
\cite{Mavromatos:2005} for a review.\smallskip

Thanks to the above Hilbert product we were able to construct the expressions for various candidates of $\kappa$-Poincar\'{e} invariant action functional. We have restricted our analysis to two kinetic operators which are related to the square of some Dirac operator, namely the first Casimir operator of the Poincar\'{e} algebra and the $\mathcal{U}_\kappa(\text{iso}(4))$-equivariant Dirac operator constructed in
\cite{dAndrea:2006}. The decay properties of the corresponding propagators have been analysed. On the other hand, we have restricted our attention to polynomial quartic interactions such that the full action functional tends to the ordinary complex $\vert\phi\vert^4$ model in the commutative (low energy) limit, $\kappa\to\infty$. It turned out that the unique interaction of the commutative case is replaced at the level of $\kappa$-Minkowski by four inequivalent interaction potentials. We shows that these interactions can actually be gathered into two families of interaction depending on the relative position of the fields $\phi$ and $\phi^\ddagger$ entering in them. One family was identified with orientable interactions, the other with nonorientable interactions. Finally, we have emphasised that thanks to the relatively simple expression of the star product associated with $\kappa$-Minkowski it was possible to represent any noncommutative field theory as an ordinary field theory (\textit{i.e.} involving pointwise product among functions) with nonlocal interactions.\smallskip

This latter fact enabled us to carry out the first complete study of the one-loop (quantum) properties of the 2-point and 4-point functions for various models of interacting $\kappa$-Poincar\'{e} invariant field theory. In $\S$\ref{sec-2point}, we have computed the one-loop 2-point functions for four different models of noncommutative field theory, resulting from various combinations of the two above kinetic operators with the various interactions. In the case the kinetic operator was provided by the Casimir kinetic operator, we find that it is necessary to consider the full set of interactions (\textit{i.e.} both orientable and nonorientable)  to ensure that the two mass terms are renormalised the same way. We have found that in this case the theory diverges cubically, thus slightly worst than in the commutative case. On the contrary, we show that the models built from the equivariant kinetic operator diverges only linearly, thus slightly milder than in the commutative case. In this case, the two masses are renormalised in the same way already when considering only one type of interactions (\textit{i.e.} either orientable or nonorientable) indicating likely a symmetry of the action functional by exchange $\phi\leftrightarrow\phi^\ddagger$. For both kinetic operators, we have found that the models were plagued with UV/IR mixing whenever the interaction considered was nonorientable; see Table \ref{tableau2}. In section $\S$\ref{sec-4point}, we have computed the one-loop 4-point functions for the model with equivariant kinetic operator and orientable interactions. Thanks to an estimate for the propagator, we have shown that the one-loop 4-point function is UV finite. We have concluded that the beta function is zero and that only the mass terms have to be renormalised.

\begin{table}[h!]
\centering
\includegraphics[scale=0.46]{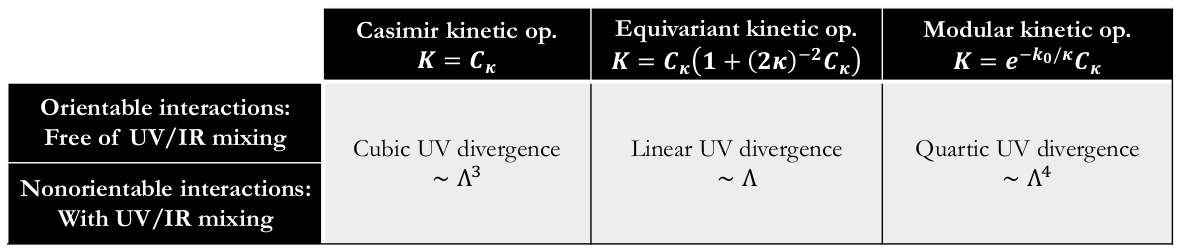}
{\caption{\label{tableau2} \small \noindent One-loop quantum properties of the 2-point functions for various models of $\kappa$-Poincar{\'e} invariant scalar field theory with quartic interactions. All of the kinetic operators are functions of the first Casimir operator of the $\kappa$-Poincar{\'e} algebra, \textit{i.e.} $\mathcal{C}_\kappa=4\kappa^2\sinh^2\left(\frac{P_0}{2\kappa}\right)+e^{P_0/\kappa}P_iP^i $.}}
\end{table}

A immediate extension of this work would be to renormalise at all order the above theory (\textit{i.e.} with equivariant kinetic operator and orientable interactions). This would require to investigate more closely the expressions of the various amplitudes, and define a reasonable power counting to characterise all of the (possibly) singular contributions. Already at the level of the one-loop theory it remains to define a ``physically reasonable" subtraction scheme to complete the one-loop renormalisation. Which subtraction point to choose is not obvious due to the peculiar role played by the parameter $\kappa$ within such models. Indeed, although $\kappa$ cannot be interpreted as a cutoff ($\kappa$ is fixed), it still regularises in some way the theory. One guideline however is provided by the desire to recover the ordinary $\phi^4$ model in the commutative limit. On the one hand, the fact that the beta function is exactly zero in our model indicates that the coupling constant is constant at the scale at which the effects of the $\kappa$-deformations become relevant (\textit{i.e.} $\kappa$). On the other hand, in the commutative case, we know that the coupling constant is increasing with increasing momenta. Therefore, for the two pictures to coincide, the flow of the (commutative) coupling constant must be bounded from above when the energy scale tends to $\kappa$. This indicates that, within this picture, it must exist a crossover (inflection point) in order the two models (low energy and high energy) to be compatible. In particular, the scale independent (noncommutative) coupling constant must recover its usual growth properties in the commutative limit ($\kappa\to\infty$). Of course, the inflection point should be given by some function of $\kappa$. Hence, we conclude that characterising this inflection point would enable us to provide an estimate for the value of $\kappa$, namely at which energy the crossover takes place. To conclude, under some physically reasonable subtraction scheme, our model of $\kappa$-Poincar\'{e} invariant scalar field theory would admit the ordinary scalar field theory which describes the Higgs dynamics in the standard model of particle physics. Said the other way around, our model of $\kappa$-Poincar\'{e} invariant scalar field theory would provide a high energy (quantum gravity) extension for studying the Higgs dynamics at Planck scale, resolving by the way the socalled triviality of the $\phi^4$ theory. Accordingly, it would be interesting to study the fate of Higgs mechanism at Planck scale within this picture.\smallskip

Finally, we conclude by mentioning that the star product considered in the description of $\kappa$-Minkowski could be used in the construction of other (not necessarily $\kappa$-Poincar\'{e} invariant) noncommutative field theories as well as gauge versions of them. In this latter case, due to the twisted trace property of the Lebesgue integral used in the construction of the action functional, the differential calculus should presumably be adapted to this situation, namely by defining a (reasonable) twisted differential calculus. In particular it has been proved that there is no 4-dimensional bicovariant differential calculi that are also Lorentz covariant
\cite{Sitarz:1995}; see also
\cite{Mercati:2016}. Nevertheless, the study of gauge theories on $\kappa$-Minkowski constitutes one of the important issues to investigate. In particular, a better understanding of the structures of gauge theory compatible with the $\kappa$-Poincar\'{e} invariance of the action functional would provide more insight in the symmetries entering the construction of any $\kappa$-Poincar\'{e} invariant action functional.

\appendix
\part*{Appendices.\addcontentsline{toc}{part}{Appendices.}}
\chapter{Basics on abstract harmonics analysis.}\label{sap-harmonic}
In this appendix, we gather a few fundamental results of harmonic analysis on locally compact group which prove to be useful in the characterisation of quantum spaces of Lie algebra type noncommutativity, as illustrated in chapter \ref{ch-ncst}. Starting from a not necessarily unimodular, locally compact, group, say $\mathcal{G}$, we present the main steps leading to the construction of the convolution algebra $\big(L^1(\mathcal{G}),\hat{\circ}\big)$, fixing by the way some notations. Then, we construct the reduced group C*-algebra $C_r^*[\mathcal{G}]$. This is achieved by defining a representation of $L^1(\mathcal{G})$ from the unitary left regular representation of $\mathcal{G}$. Finally, assuming $\mathcal{G}$ to be amenable (which is the case for the $ax+b$ group underlying the construction of $\kappa$-Minkowski) we construct the group C*-algebra $C^*[\mathcal{G}]$, which we identify with the C*-algebra of fields modelling the quantum space.\\
Proofs of the material presented here, as well as complementary information, can be found in any textbook on harmonic analysis; see, e.g., ref.
\cite{Loomis:1953,Folland:1995}.\bigskip

Let $\mathcal{G}$ be a locally compact Hausdorff group. We can show that, $\mathcal{G}$ carries a unique, up to a strictly positive multiplicative scalar, left invariant Borel measure, called the Haar measure.\footnote{See, e.g., theorems 2.10 and 2.20 in ref.
\cite{Folland:1995}.} Accordingly, there exists a unique, to within a positive multiplicative constant, left invariant integral on the algebra $C_c(\mathcal{G})$ of compactly supported continuous $\mathbb{C}$-valued functions on $\mathcal{G}$. This integral is defined by the positive linear functional
\begin{equation}\label{left functional-ap}
I:C_c(\mathcal{G})\to\mathbb{C},\ I(f):=\int_\mathcal{G}f(x)d\mu(x),
\end{equation}
the left invariance of which can be symbolically written as $I(L_yf)=I(f)$, where $L_yf(x):=f(y^{-1}x)$ is the left translated of $f$. It follows that
\begin{equation}
\| f \|_1:=\int_\mathcal{G}\vert f(x) \vert d\mu(x),
\end{equation}
defines a norm on $C_c(\mathcal{G})$, the completion with respect to which is $L^1(\mathcal{G})$. Hence, the Haar integral $I$ can be extended to all of $L^1(\mathcal{G})$ owing to the density of $C_c(\mathcal{G})$ in $L^1(\mathcal{G})$.\smallskip

In general, the Haar measure needs not be also right  invariant, \textit{i.e.} $I(R_yf)\neq I(f)$, where $R_yf(x):=f(xy)$ is the right  translated of $f$. However, for $y\in\mathcal{G}$ fixed, $I(R_yf)$ is left invariant, and because of the uniqueness of the Haar measure there exists a strictly positive scalar, say $\Delta_{\mathcal{G}}(y^{-1})$, such that
\begin{equation}\label{modular1-ap}
I(R_yf)=\Delta_{\mathcal{G}}(y^{-1})I(f),
\end{equation}
which can be symbolically written  as $d\mu(xy)=\Delta_{\mathcal{G}}(y)d\mu(x)$. The above continuous homomorphism $\Delta_{\mathcal{G}}:\mathcal{G}\to\mathbb{R}_{+}^*$ is called the modular function of $\mathcal{G}$. By definition, a group for which $\Delta_{\mathcal{G}}$ is trivial, namely for which the Haar measure is bi-invariant, is said to be unimodular. Otherwise, it is said to be nonunimodular. For example, this is the case for the $ax+b$ group which underlies the construction of the algebra modelling the $\kappa$-Minkowski space; see Sec. \ref{sec-Minkowski}. In fact, whenever the group under consideration is a semidirect product $\mathcal{G}=A\ltimes_\tau B$, $A<\mathcal{G}$, $B\triangleleft\mathcal{G}$, $A\cap B=\lbrace1\rbrace$, the corresponding modular function can be related to the group homomorphism $\tau:A\times B\to B$. This can be easily realised by observing that, in this case, $\tau$ is given by the adjoint action of $A$ on $B$.\smallskip

A right  invariant measure on $\mathcal{G}$ can be constructed as follow. First, we define the map ${f}^\flat(x):=f(x^{-1})$, $\forall x\in\mathcal{G}$. Straightforward computations show that
\begin{subequations}
\begin{align}
&L_y{f}^\flat(x)={f}^\flat(y^{-1}x)=f(x^{-1}y)=(R_yf)^\flat(x),\label{involution-ap}\\
&R_y{f}^\flat(x)={f}^\flat(xy)=f(y^{-1}x^{-1})=(L_yf)^\flat(x).
\end{align}
\end{subequations}
Next, we define the positive linear functional 
\begin{equation}
J:C_c(\mathcal{G})\to\mathbb{C},\ J(f):=I({f}^\flat).
\end{equation}
Thus, combining the identity \eqref{involution-ap} together with the left invariance of $I$ in the definition of $J$, we easily find that
\begin{equation}
J(R_yf)=I(L_yf^\flat)=I(f^\flat)=J(f),
\end{equation}
namely $J$ is invariant under right translations. Moreover, we infer from the above result that the map $f\mapsto f^\flat$ is an isometric isomorphism from $L^1(\mathcal{G},\mu)$ to $L^1(\mathcal{G},\nu)$.\\
It is convenient to rewrite $J$ by symbolically defining $d\nu(x)=d\mu(x^{-1})$, such that
\begin{equation}
J(f)=\int_\mathcal{G} f(x) d\nu(x).
\end{equation}
Now, upon using the fact that there exists a unique, up to strictly positive constant, right  invariant measure on $\mathcal{G}$, together with eq. \eqref{modular1-ap}, we can show that left  and right  invariant Haar measures are related via the relation\footnote{See, e.g., proposition 2.31 in
\cite{Folland:1995}.} 
\begin{equation}\label{modular2-ap}
d\nu(x)=\Delta_{\mathcal{G}}(x^{-1})d\mu(x),
\end{equation}
exhibiting the equivalence between left  and right  invariant Haar measures. This implies $d\nu(x^{-1})=\Delta_{\mathcal{G}}(x)d\nu(x)$ and $d\mu(x^{-1})=\Delta_{\mathcal{G}}(x^{-1})d\mu(x)$. From now on, we restrict our attention to the left invariant Haar measure and simply write $L^1(\mathcal{G})$ to denote $L^1(\mathcal{G},\mu)$. We shall return to the right  invariant case at the end of this appendix.\smallskip

We are now in position to construct the socalled convolution algebra $\big(L^1(\mathcal{G}),\hat{\circ}\big)$ of $\mathcal{G}$. The convolution product $\hat{\circ}$ between two functions $f,g\in L^1(\mathcal{G})$ is defined by
\begin{equation}\label{convolution-ap}
(f\hat{\circ} g)(x):=\int_\mathcal{G} f(xy)g(y^{-1})d\mu(y)=\int_\mathcal{G} f(xy^{-1})g(y)d\nu(y),\ x\in\mathcal{G},
\end{equation}  
with $f\hat{\circ} g\in L^1(\mathcal{G})$ from H\"{o}lder's inequality. It is further convenient to define the following involution on $L^1(\mathcal{G})$
\begin{equation}\label{involution2-ap}
f^*(x):=\Delta_{\mathcal{G}}(x^{-1})\overline{f^\flat(x)},\ (f\hat{\circ}g)^*=g^*\hat{\circ}f^*,\ \forall f,g\in L^1(\mathcal{G}).
\end{equation}
With the convolution product \eqref{convolution-ap} and the involution \eqref{involution2-ap}, the algebra $\big(L^1(\mathcal{G}),\hat{\circ}\big)$ forms a *-Banach algebra, called the convolution algebra of $\mathcal{G}$. Owing to the density of $L^1(\mathcal{G})$ in $L^2(\mathcal{G})$, the completion of $L^1(\mathcal{G})$ with respect to the norm
\begin{equation}
\| f \|_2:=\sqrt{\int_\mathcal{G} \vert f(x) \vert^2 d\mu(x)},\ \| f\hat{\circ}g \|_2\leq\| f \|_2\| g \|_2,\ f,g\in L^1(\mathcal{G}),
\end{equation}
enables us to conveniently extend the above construction to $\big(L^2(\mathcal{G}),\hat{\circ}\big)$.\smallskip

The last step, needed for constructing the group C*-algebra of $\mathcal{G}$, consists in defining representations of $L^1(\mathcal{G})$, say $\pi:L^1(\mathcal{G})\to\mathcal{B}(\mathcal{H}_\pi)$, which are induced from representations of $\mathcal{G}$, say $\pi_U:\mathcal{G}\to\mathcal{B}(\mathcal{H}_\pi)$, where $\mathcal{H}_\pi$ is some suitable Hilbert space and $\mathcal{B}(\mathcal{H}_\pi)$ is the C*-algebra of bounded operators on $\mathcal{H}_\pi$. The crucial point is that, whenever $\pi_U$ defines a strongly continuous unitary representation of $\mathcal{G}$, then the (induced) representation $\pi$ defined, for any $f,g\in L^1(\mathcal{G})$, by
\begin{equation}\label{induced-rep}
\pi(f):=\int_\mathcal{G}f(x)\pi_U(x)d\mu(x),\ \pi(f\hat{\circ}g)=\pi(f)\pi(g),\ \pi(f^*)=\pi(f)^\dag,
\end{equation}
is a nondegenerate bounded *-representation of the convolution algebra.\footnote{See, e.g., theorem 3.9 in
\cite{Folland:1995}.} Furthermore, the map $\pi_U\to\pi$ defines a one-to-one correspondence between the strongly continuous unitary representations of $\mathcal{G}$ and the nondegenerate bounded *-representation of $L^1(\mathcal{G})$.\footnote{See, e.g., theorem 3.11 \textit{ibid}.}\smallskip

Let us assume that this representation is induced by the left regular representation of $\mathcal{G}$ on $L^2(\mathcal{G},\mu)$, \textit{i.e} $\pi_{U}^{\ell}:y\mapsto L_y$, $\pi_{U}^{\ell}:\mathcal{G}\to\mathcal{B}\big(L^2(\mathcal{G},\mu)\big)$, then
\begin{equation}
\pi^\ell(f)g=f\hat{\circ}g.
\end{equation}
The norm closure of $\pi^\ell\big(L^1(\mathcal{G})\big)$ in $\mathcal{B}\big(L^2(\mathcal{G},\mu)\big)$ is called the reduced group C*-algebra of $\mathcal{G}$ and is denoted by $C_r^*[\mathcal{G}]$. Even though $C_r^*[\mathcal{G}]$ is not isomorphic to the group C*-algebra $C^*[\mathcal{G}]$ of $\mathcal{G}$ in general, whenever $\mathcal{G}$ is amenable, we can show that the above left regular representation is faithful on $C^*[\mathcal{G}]$ and $C_r^*[\mathcal{G}]\cong C^*[\mathcal{G}]$.\smallskip

This concludes the construction of the group C*-algebra of $\mathcal{G}$, that we identify with the C*-algebra of fields modelling the quantum space, at least for $\mathcal{G}$ amenable, which is the case whenever $\mathfrak{g}=\text{Lie}(\mathcal{G})$ is solvable.\footnote{For a comprehensive presentation of amenable groups see, e.g., sec. 7.3 in
\cite{Pedersen:1979}, Appendix A in
\cite{Williams:2007}, and reference therein.} Note that, the Lie group underlying $\kappa$-Minkowski belongs to this class of groups, see in sec. \ref{sec-convolution}.\smallskip 

\chapter{Basics on \texorpdfstring{$\kappa$}{k}-Poincar\'{e} algebra.}\label{sap-poincare}
In this appendix, we recall some basic algebraic properties of the $\kappa$-Poincar\'{e} algebra $\mathfrak{P}_\kappa$, adopting the formal point of view of bicrossproduct Hopf algebra, as introduced in
\cite{Majid:1994}. Within this picture, we have $\mathfrak{P}_\kappa=\mathcal{U}\big(\mathfrak{so}(1,3)\big)\triangleright\!\!\!\blacktriangleleft \mathfrak{T}_\kappa$, where $\mathcal{U}\big(\mathfrak{so}(1,3)\big)$, resp. $\mathfrak{T}_\kappa$, denotes the universal enveloping algebra of the Lorentz algebra, resp. the algebra of translations. In other words, considered separately, none of the algebraic structures of $\mathcal{U}\big(\mathfrak{so}(1,3)\big)$ nor $\mathfrak{T}_\kappa$ are deformed. Rather, on the one hand, $\mathcal{U}\big(\mathfrak{so}(1,3)\big)$ now acts in a deformed way on $\mathfrak{T}_\kappa$ from the right, \textit{i.e.} $\mathcal{U}\big(\mathfrak{so}(1,3)\big)\triangleright\!\!\!< \mathfrak{T}_\kappa$. This is reflected in eq. \eqref{hopf10}. On the other hand, $\mathfrak{T}_\kappa$ coacts back in a nontrivial way on $\mathcal{U}\big(\mathfrak{so}(1,3)\big)$ from the left, \textit{i.e.} $\mathcal{U}\big(\mathfrak{so}(1,3)\big)>\!\blacktriangleleft \mathfrak{T}_\kappa$. This is reflected in eq. \eqref{hopf2}. For a recent comprehensive review on the development of the $\kappa$-deformed Poincar\'{e} algebra see, e.g.,
\cite{Lukierski:2017}, and references therein.\bigskip

First, recall that the ordinary Poincar\'{e} algebra $\mathfrak{P}=\mathcal{U}\big(\mathfrak{so}(1,3)\big)\ltimes \mathfrak{T}$, describing the symmetries of the 4-dimensional Minkowski spacetime, is characterised by
\begin{align}\label{ap-hopf-alg1}
&[M_i,M_j]=i\varepsilon_{ijk}M_k,\ [N_i,N_j]=-i\varepsilon_{ijk}M_k,\ [M_i,N_j]=i\epsilon_{ijk}N_k,\ [P_\mu,P_\nu]=0,\\
&[M_i,P_j]=i\varepsilon_{ijk}P_k,\ [M_i,P_0]=0,\ [N_i,P_0]=iP_i,\  [N_i,P_j]=i\delta_{ij}P_0,\nonumber
\end{align}
where $P_\mu$ are the generators of the translations, $M_i$ those of rotations and $N_i$ those of boosts.\footnote{Here, Greek (resp. Latin) indices label as usual spacetimelike (resp. spacelike) coordinates. Summation over repeated indices is assumed.} Hence, the Poincar\'{e} algebra defined a Lie algebra.\smallskip

As any Lie algebra, $\mathfrak{P}$ can be endowed with a (trivial) Hopf algebraic structure. The corresponding coalgebraic sector is characterised by
\begin{equation}\label{hopf0}
\Delta h = h\otimes 1 + 1\otimes h,\ \epsilon(h)=0,\ S(h)=-h,
\end{equation}
where $\Delta:\mathfrak{P}\to\mathfrak{P}\otimes\mathfrak{P}$, $\epsilon:\mathfrak{P}\to\mathbb{C}$ and $S:\mathfrak{P}\to\mathfrak{P}$ are respectively the coproduct, the counit and the antipode, and $h$ is any of the generators of $\mathfrak{P}$.\smallskip

One of the idea behind quantum groups is then to deformed, in some smooth way, the above (co-)algebraic structures, in such a way that the deformed structures still form a Hopf algebra; for a comprehensive introduction on quantum groups see, e.g.,
\cite{Majid:1995}. The $\kappa$-Poincar\'{e} algebra is such an example, we now turn to the presentation.\smallskip

A convenient presentation of $\mathfrak{P}_\kappa$ is obtained from the 11 elements $(P_i, N_i,M_i, \mathcal{E},\mathcal{E}^{-1})$, where
\begin{equation}\label{twist0}
\mathcal{E}:=e^{-P_0/\kappa},
\end{equation}
satisfying the Lie algebra relations
\begin{subequations}\label{ap-hopf-alg2}
\begin{align}
&[M_i,M_j]= i\epsilon_{ijk}M_k,\ [N_i,N_j]=-i\epsilon_{ijk}M_k,\ [M_i,N_j]=i\epsilon_{ijk}N_k,\\
&[M_i,P_j]= i\epsilon_{ijk}P_k,\ [P_i,P_i]=[P_i,\mathcal{E}]=[M_i,\mathcal{E}]=0,\\
&[N_i,\mathcal{E}]=iP_i\mathcal{E},\ [N_i,P_j]=i(2\kappa)^{-1}\Big(\kappa^2(1-\mathcal{E}^{2})+\vec{P}^2\Big)\delta_{ij}-i\kappa^{-1}P_iP_j.\label{hopf10}
\end{align}
\end{subequations}
The full picture of $\mathfrak{P}_\kappa$ is obtained by specifying its coalgebraic structure, namely
\begin{subequations}
\begin{align}
&\Delta P_0=P_0\otimes1+1\otimes P_0,\ \Delta \mathcal{E}=\mathcal{E}\otimes\mathcal{E},\ \Delta P_i=P_i\otimes1+\mathcal{E}\otimes P_i,\label{hopf1}\\
&\Delta M_i=M_i\otimes1+1\otimes M_i,\ \Delta N_i=N_i\otimes1+\mathcal{E}\otimes N_i-\kappa^{-1}\epsilon_{ijk}P_j\otimes M_k,\label{hopf2}
\end{align}
\end{subequations}
together with
\begin{subequations}
\begin{align}
&\epsilon(P_0)=\epsilon(P_i)=\epsilon(M_i)=\epsilon(N_i)=0,\  \epsilon(\mathcal{E})=1,\label{hopf3}\\
&S(P_0)=-P_0,\ S(\mathcal{E})=\mathcal{E}^{-1},\  S(P_i)=-\mathcal{E}^{-1}P_i,\label{hopf4}\\
&S(M_i)=-M_i,\ S(N_i)=-\mathcal{E}^{-1}\big(N_i-\kappa^{-1}\epsilon_{ijk}P_jM_k\big).\label{hopf5}
\end{align}
\end{subequations}

Now, recall that the algebra of $\kappa$-Minkowski coordinates $\mathfrak{m}_\kappa$ has been originally defined as the dual, say $\mathfrak{T}^*_\kappa$, of the Hopf subalgebra $\mathfrak{T}_\kappa$ generated by $P_\mu$ on which $\mathfrak{T}_\kappa$ acts covariantly as quantum vector field, and vice versa. Hence, thanks to the bicrossproduct structure of $\mathfrak{P}_\kappa$, the (right) action of $\mathcal{U}\big(\mathfrak{so}(1,3)\big)$ on $\mathfrak{T}_\kappa$ dualises to a (left) action on $\mathfrak{T}^*_\kappa$ and the whole $\kappa$-Poincar\'{e} acts on $\mathfrak{T}^*_\kappa$; see, e.g.,
\cite{Majid:1994,Majid:1990}. Therefore, the (Hopf) algebraic structure of $\mathfrak{T}^*_\kappa$ is fully determined by requiring the duality pairing $\langle P_\mu,\hat{x}_\nu\rangle=i\delta_{\mu\nu}$ to be compatible with the Hopf algebraic structures of $\mathfrak{T}_\kappa$ and $\mathfrak{T}^*_\kappa$. Namely, for any $t,s\in \mathfrak{T}_\kappa$, $x,y\in \mathfrak{T}_\kappa^*$, we require
\begin{subequations}
\begin{align}
&\langle t, xy \rangle \equiv \langle \Delta t, x \otimes y \rangle = \langle t_{(1)}, x \rangle \langle t_{(2)},  y \rangle,\label{hopf11}\\
&\langle ts, x \rangle \equiv \langle t\otimes s, \Delta x \rangle = \langle t, x_{(1)} \rangle \langle s,  x_{(2)} \rangle,\label{hopf12}
\end{align}
where used has been made of Sweedler's notations, together with
\begin{equation}
\langle 1,x \rangle = \epsilon(x),\ \langle t,1 \rangle = \epsilon(t),\ \langle S(t),x \rangle = \langle t,S(x) \rangle.
\end{equation}
\end{subequations}
Combining \eqref{hopf12} with the commutativity of the generators of translation, we easily find that the coalgebraic sector of $\mathfrak{T}^*_\kappa$ is undeformed, namely $\hat{x}_\mu$ are primitive
\begin{equation}
\Delta x_\mu = 1\otimes\hat{x}_\mu+\hat{x}_\mu\otimes1.
\end{equation}
While combining \eqref{hopf11} with the deformed coalgebraic structure of $\mathfrak{T}_\kappa$, we obtain 
\begin{equation}
[\hat{x}_0,\hat{x}_i]=\frac{i}{\kappa}\hat{x}_i,\ [\hat{x}_i,\hat{x}_j]=0,
\end{equation}
exhibiting the noncommutative Lie algebraic structure of $\kappa$-Minkowski.\smallskip

Finally,  the structure of $\mathcal{M}_\kappa$ as left module over the Hopf algebra $\mathfrak{P}_\kappa$ can be expressed, for any $f\in\mathcal{M}_\kappa$, in terms of the bicrossproduct basis
\cite{Majid:1994}, by
\begin{subequations}\label{module-action}
\begin{align}
&(\mathcal{E}\triangleright f)(x)=f(x_0+\frac{i}{\kappa},\vec{x}),\ P_\mu\triangleright f=-i\partial_\mu f,\label{left module1}\\
&M_i\triangleright f=\epsilon_{ijk}L_{x_j}P_k\triangleright f,\ N_i\triangleright f=\Big((2\kappa)^{-1}L_{x_i}\big(\kappa^2(1-\mathcal{E}^2)+\vec{P}^{2}\big)-L_{x_0}P_i\Big)\triangleright f,\label{left module3}
\end{align}
\end{subequations}
where $L_a$ denotes the left multiplication operator, i.e. $L_af:=af$.

In particular, it must be stressed that the $P_i$'s act as twisted derivations on $\mathcal{M}_\kappa$ while $P_0$ satisfies the usual Leibniz rule, simply reflecting the coproduct structure \eqref{hopf1} of $\mathfrak{P}_\kappa$. We have, for any $f,g\in\mathcal{M}_\kappa$,
\begin{subequations}\label{derivtwist}
\begin{align}
&P_0\triangleright(f\star g)=(P_0\triangleright f)\star g+f\star(P_0\triangleright  g ),\label{deriv-twist2}\\
&P_i\triangleright(f\star g)=(P_i\triangleright f)\star g+(\mathcal{E}\triangleright f)\star (P_i\triangleright g).\label{deriv-twist1}
\end{align}
On the other hand, $\mathcal{E}$ acts as group element on $\mathcal{M}_\kappa$, namely
\begin{equation}\label{relation-calE}
\mathcal{E}\triangleright(f\star g)=(\mathcal{E}\triangleright f)\star(\mathcal{E}\triangleright g).
\end{equation}
\end{subequations}

In view of further applications to NCFT, we conclude this appendix by collecting (important) additional material on the compatibility between the involutive structures of the various above mentioned (Hopf) algebras. The $\kappa$-Poincar\'{e} becomes a $^*$-Hopf algebra through the requirement $P_\mu^\dag=P_\mu$, $\mathcal{E}^\dag=\mathcal{E}$. Then, by promoting the above duality to a duality between $^*$-algebras, we obtain
\begin{equation}\label{pairing-involution}
(t\triangleright f)^\dag=S(t)^\dag\triangleright f^\ddagger,
\end{equation}
which holds true for any $t\in T$ and any $f\in\mathcal{M}_\kappa$. This, combined with \eqref{hopf4}, implies
\begin{equation}\label{dag-hopfoperat}
(P_0\triangleright f)^\dag=-P_0\triangleright f^\ddagger,\ (P_i\triangleright f)^\dag=-\mathcal{E}^{-1}P_i\triangleright f^\ddagger,\ (\mathcal{E}\triangleright f)^\dag=\mathcal{E}^{-1}\triangleright f^\ddagger.
\end{equation}
\chapter{Basics on path integral quantisation and perturbation theory.}\label{sap-perturbation}
The study of the physical properties of noncommutative field theories can be carried out by identifying the noncommutative space background with its algebra of fields endowed with noncommutative (star) product, then constructing an action functional on that algebra. This is made possible by the existence of closed (integral) form (in contrast to the formal power series expansion in the deformation parameter) for the star product. This is the case, for instance, when applying the Weyl quantisation scheme %
or, more generally, when the quantisation map associated to the star product coincides with a well-defined integral transform. It follows that the star product of two functions can be written as an integral transform with nontrivial (i.e. nonlocal) kernel. That latter fact usually enables us to represent the NCFT (involving star products) by an ordinary (commutative) nonlocal field theory.\bigskip

When the kinetic part of the ``noncommutative" action functional remains symmetric and positive\footnote{More precisely, when the kernel of the kinetic operator is symmetric and positive.} and only the interaction becomes nonlocal, standard tools and computational techniques from path integral quantisation and perturbation theory can be used to study the quantum properties of the NCFT. This is precisely the case for the $\kappa$-Poincar\'{e} invariant scalar field theories considered in the present dissertation as it is apparent from eq. \eqref{kinetic-map}, \eqref{o-int2} and \eqref{no-int2}. The main steps leading to the computation of the one-loop order corrections to both the 2-point and 4-point functions are recalled in this appendix. Note, by the way, that the material presented here is very general since no assumption on the explicit expressions for the kinetic operator and the vertex function is made (we only assumed the kernel of the kinetic operator to be symmetric and positive in order to compute the Gaussian integral associated to the free field theory).\bigskip

Let $\mathcal{Z}$ denote the generating functional of correlation functions (or partition function) describing the NCFT under consideration. $\mathcal{Z}$ is a function of the sources $J$ and $\bar{J}$ conjugate to the fields $\bar{\phi}$ and $\phi$ respectively, and is defined by
\begin{equation}
\mathcal{Z}(\bar{J},J):=\int d\bar{\phi}d\phi \ e^{-S_\kappa(\bar{\phi},\phi)+\int d^4x\left(\bar{J}(x)\phi(x) +J(x)\bar{\phi}(x)\right)},
\end{equation}
where we formally integrate over all the scalar fields $\phi$ and $\bar{\phi}$ which are regarded as independent variables. It is further assumed that $S_\kappa$ admits the decomposition $S_\kappa^\text{kin}+S_\kappa^\text{int}$, similar to \eqref{actiondecomposition}. We have
\begin{equation}\label{functional}
\mathcal{Z}(\bar{J},J)=e^{-S^\text{int}_\kappa\left(\frac{\delta}{\delta J},\frac{\delta}{\delta \bar{J}}\right)}\mathcal{Z}_G(\bar{J},J),
\end{equation}
where the Gaussian partition function is defined by
\begin{equation}\label{functional-gauss}
\mathcal{Z}_G(\bar{J},J):=\int d\bar{\phi}d\phi \ e^{-S^\text{kin}_\kappa(\bar{\phi},\phi)+\int d^4x\left(\bar{J}(x)\phi(x) +J(x)\bar{\phi}(x)\right)},
\end{equation}
and the functional derivatives with respect to $J$ and $\bar{J}$ appearing in \eqref{functional} are defined as usual by
\begin{equation}
\frac{\delta J_a(x_2)}{\delta J_b(x_1)}=\delta_{ab}\delta^{(\hspace{-1pt}4\hspace{-1pt})}(x_2-x_1),
\end{equation}
where the indices $a,b$ labelled the nature (i.e. either $J$ or $\bar{J}$) of the various sources.\\The integration in \eqref{functional-gauss} can be performed by shifting $\phi$ (resp. $\bar{\phi}$) by $\Delta_FJ$ (resp. $\Delta_F\bar{J}$) and leads to
\begin{equation}
\mathcal{Z}_G(\bar{J},J)= N e^{\int d^4x_1d^4x_2\bar{J}(x_1)\Delta_F(x_1,x_2)J(x_2)},
\end{equation}
where $N=\left(\det K\right)^{-1}$ and $\Delta_F$ is the inverse of the kinetic operator $K$ (possibly supplemented by a mass term) defined (in a distributional sense) by
\begin{equation}
\int d^4y\ \Delta_F(x_1,y)K(y,x_2)=\delta^{(4)}(x_1-x_2).
\end{equation}\bigskip

The radiative corrections to the various correlation (or n-point) functions can be read from the expansion in power of the coupling constant $g$ of the effective action $\Gamma$. That latter is defined as the Legendre transform of the generating functional of connected correlation functions, $W:=\ln\mathcal{Z}$, namely
\begin{equation}\label{ap-Legendre}
\Gamma(\bar{\phi},\phi):=\int\frac{d^4k}{(2\pi)^4}\left(\bar{J}(k)\phi(k) +J(k)\bar{\phi}(k)\right)-W(\bar{J},J),
\end{equation}
together with
\begin{equation}\label{ap-conjugatefields}
\bar{\phi}(k)=\frac{\delta W(\bar{J},J)}{\delta J(k)},\ \ \phi(k)=\frac{\delta W(\bar{J},J)}{\delta \bar{J}(k)},
\end{equation}
where we switch from position to momentum space for computational convenience.\\Note that the effective action admits the following development
\begin{equation}
\Gamma(\bar{\phi},\phi)=\sum_{n,m}\frac{\hbar^m}{n!}\int\prod_{\ell=1}^n\left[\frac{d^4k_\ell}{(2\pi)^4}\right] \bar{\phi}(k_1)\cdots\phi(k_n)\Gamma^{(n)}_m(k_1,\cdots,k_n),
\end{equation}
and coincides with the classical action $S_\kappa$ at the lowest, zeroth, order in $\hbar$, namely
\begin{align}
&S^\text{kin}_\kappa(\bar{\phi},\phi)=\frac{1}{2}\int\prod_{\ell=1}^2\left[\frac{d^4k_\ell}{(2\pi)^4}\right] \bar{\phi}(k_1)\phi(k_2)\Gamma^{(2)}_0(k_1,k_2),\\
&S^\text{int}_\kappa(\bar{\phi},\phi)=\frac{1}{4!}\int\prod_{\ell=1}^4\left[\frac{d^4k_\ell}{(2\pi)^4}\right] \bar{\phi}(k_1)\phi(k_2)\bar{\phi}(k_3)\phi(k_4)\Gamma^{(4)}_0(k_1,k_2,k_3,k_4).
\end{align}
The one-loop order corrections to the 2-point (resp. the 4-point) function are obtained by expanding the quadratic part $\Gamma^{(2)}$ (resp. quartic part $\Gamma^{(4)}$) of the effective action up to the first (resp. second) order in $g$. In practice, this is achieved by expanding $W$ up to the desired order in the coupling constant, then expressing the sources $J$ and $\bar{J}$ in term of the fields $\bar{\phi}$ and $\phi$ by inverting the relations \eqref{ap-conjugatefields} and finally plugging the result in \eqref{ap-Legendre}. Explicitly, we start by expressing $W$ as
\begin{subequations}\label{ap-W}
\begin{align}
&W(\bar{J},J)=\ln(N)+W_G(\bar{J},J)+\ln\left(1+\mathbb{D}\right),\label{ap-Wa}\\
&W_G(\bar{J},J)=\int\frac{d^4k}{(2\pi)^4}\bar{J}(k)\Delta_F(k)J(k),\label{ap-Wb}
\end{align}
\end{subequations}
where
\begin{equation}\label{ap-functional}
\mathbb{D}(\bar{J},J):=e^{-W_G(\bar{J},J)}\left(e^{-S^\text{int}_\kappa\left(\frac{\delta}{\delta J},\frac{\delta}{\delta \bar{J}}\right)}-1\right)e^{W_G(\bar{J},J)},
\end{equation}
can be formally written as a series in the coupling constant, namely $\mathbb{D}=\sum_n g^n \mathbb{D}_n$, each coefficient $\mathbb{D}_n$ being obtained by expanding in power of $g$ the exponential involving the interaction term in \eqref{ap-functional}\footnote{The first terms of the expansion which are needed in our derivation are $$\mathbb{D}_0:=0,\ \ \mathbb{D}_1:=\frac{-1}{g}\ e^{-W_G}\left(S^\text{int}_\kappa\triangleright e^{W_G}\right),\ \ \mathbb{D}_2:=\frac{1}{g^2}\ e^{-W_G}\left((S^\text{int}_\kappa)^2\triangleright e^{W_G}\right).$$}. Then, the formal expansion of the logarithm in \eqref{ap-Wa} yields
\begin{equation}\label{ap-Dexpansion}
\ln\left(1+\mathbb{D}\right)=g\mathbb{D}_1+g^2\left(\mathbb{D}_2-\frac{1}{2}\mathbb{D}_1^2\right)+\mathcal{O}(g^3),
\end{equation}
giving rise to a suitable expansion of $W$. The inversion of \eqref{ap-conjugatefields} yields at the first order in $g$
\begin{equation}
\bar{J}(k)=K(k)\bar{\phi}(k)+\mathcal{O}(g),\ \ J(k)=K(k)\phi(k)+\mathcal{O}(g),
\end{equation}
which plugged in \eqref{ap-Legendre} combined with \eqref{ap-W} and \eqref{ap-Dexpansion} yield the following expression for the one-loop order corrections to the 2-point function
\begin{align}\label{ap-gamma2pts}
\Gamma^{(2)}_1(k_1,k_2)=\frac{g}{(2\pi)^4}\int\frac{d^4k_3}{(2\pi)^4}\ \Delta_F(k_3)\Big[\mathcal{V}_{3312}+\mathcal{V}_{1233}+\mathcal{V}_{1332}+\mathcal{V}_{3213}\Big].
\end{align}
where we have introduced the compact notation $\mathcal{V}_{abcd}:=\mathcal{V}(k_a,k_b,k_c,k_d)$. Similar computations yield the following expression for the one-loop 4-point function
\begin{align}\label{ap-gamma4pts}
\Gamma^{(4)}_1(k_1,k_2,k_3,k_4)=\frac{g^2}{(2\pi)^8}\int& \frac{d^4k_5}{(2\pi)^4}\frac{d^4k_6}{(2\pi)^4}\ \Delta_F(k_5)\Delta_F(k_6)\times\\
\times\Big[&2\mathcal{V}_{5462}\mathcal{V}_{3615}+2\mathcal{V}_{5462}\mathcal{V}_{3516}+2\mathcal{V}_{5216}\mathcal{V}_{3465}+\nonumber\\
+&2\mathcal{V}_{1652}\mathcal{V}_{3465}+2\mathcal{V}_{5612}\mathcal{V}_{6435}+2\mathcal{V}_{5612}\mathcal{V}_{3564}+\nonumber\\
+&2\mathcal{V}_{5216}\mathcal{V}_{3564}+2\mathcal{V}_{5612}\mathcal{V}_{3465}+\hspace{2pt}\mathcal{V}_{5612}\mathcal{V}_{6534}\hspace{3pt}+\nonumber\\
+&\hspace{2pt}\mathcal{V}_{5216}\mathcal{V}_{6435}\hspace{3pt}+\hspace{2pt}\mathcal{V}_{1652}\mathcal{V}_{3564}\hspace{3pt}+\mathcal{V}_{1256}\mathcal{V}_{3465}\Big].\nonumber
\end{align}
\chapter{Supplement to Chapter \ref{sap-ncftsu2}.}\label{appendixb}
In this appendix we collect some additional computational details on the derivation of the Type-II contribution entering the computation of the one-loop 2-point function for NCFT on $\mathbb{R}^3_\theta$ in Chap. \ref{sap-ncftsu2}. Below are the steps leading from \eqref{ik2k4} to \eqref{typeII_spherique} in $\S$\ref{sec-typeII}.\bigskip

We start from eq. \eqref{ik2k4}, namely
\begin{align}\label{ik2k4-1}
 I(k_2,k_4) =& \int \frac{d^3k }{(2\pi)^3} \frac{d^3x}{k^2+m^2} \left[ e^{ik^\nu x_\nu} , e^{ik_2^\nu x_\nu} \right]_{\star_\mathcal{K}} \star_\mathcal{K} e^{-ik^\nu x_\nu} \star_\mathcal{K} e^{ik_4^\nu x_\nu}  \\
 =& \left(\frac{2}{\theta}\right)^4 \int \frac{d^3k }{(2\pi)^3} \frac{d^3x}{k^2+m^2} \left(\frac{\sin(\frac{\theta}{2}|k|)}{|k|}\right)^2 \frac{\sin(\frac{\theta}{2}|k_2|)}{|k_2|} \frac{\sin(\frac{\theta}{2}|k_4|)}{|k_4|}\nonumber\\
 &\times \mathcal{Q}^{-1} \left( \left[ e^{ik^\mu \hat{x}_\mu} , e^{ik_2^\nu \hat{x}_\nu} \right] e^{-ik^\sigma \hat{x}_\sigma} e^{ik_4^\rho \hat{x}_\rho} \right)\nonumber .
\end{align}
Then we use \eqref{decadix}-\eqref{expo-coupe} to write
\begin{equation} 
 \left[ e^{ik\hat{x}} , e^{ik_2 \hat{x}} \right] e^{-ik \hat{x}} e^{ik_4 \hat{x}} \overset{\rho}{\longmapsto} - i \theta \frac{\sin\left(\theta |k| \right)}{\theta |k|} \frac{\sin\left(\theta |k_2| \right)}{\theta |k_2|} \varepsilon_{\mu \nu}^{\hspace{8pt} \rho} k^\mu k_2^\nu \rho(\hat{x}_\rho) e^{-ik \rho(\hat{x})} e^{ik_4 \rho(\hat{x})} \label{representationtype2}
\end{equation}
where :
\begin{align}\label{interm-2}
\rho(\hat{x}_\rho) e^{-ik \rho(\hat{x})}&= \cos\left(\theta |k| \right) \rho(\hat{x}_\rho) - i \frac{\sin\left(\theta |k| \right)}{\theta |k|} k^\sigma \rho(\hat{x}_\rho)\rho(\hat{x}_\sigma)  \\
 &= -\theta^2 \frac{\sin\left(\theta |k| \right)}{\theta |k|} k_\rho \mathbb{I}_2 + \left[ \cos\left(\theta |k| \right) \delta^\nu_\rho + \frac{\theta}{2} \frac{\sin\left(\theta |k| \right)}{\theta |k|} \varepsilon_{\rho \mu}^{\hspace{8pt} \nu} k^\mu \right] \rho(\hat{x}_\nu) .\nonumber
\end{align}
After some algebraic manipulations, the right-hand-side of \eqref{representationtype2} can be cast into the simpler form
\begin{align}\label{interm-1}
- i\theta \frac{\sin\left(\theta |k| \right)}{\theta |k|} \frac{\sin\left(\theta |k_2| \right)}{\theta |k_2|} \Big(\cos\left(\theta |k| \right)& \varepsilon_{\mu \nu}^{\hspace{11pt} \rho} k^\mu k_2^\nu\\
&+ \frac{\sin\left(\theta |k| \right)}{2|k|} \left(|k|^2 k_2^\rho - k_\sigma k_2^\sigma  k^\rho \right) \Big) \rho(\hat{x}_\rho) e^{ik_4 \rho(\hat{x})}.\nonumber
\end{align}
Then combining \eqref{representationtype2}-\eqref{interm-1} with the action of \eqref{decadix} on \eqref{ik2k4-1} yields
\begin{equation} \label{finale-1}
 I (k_2,k_4) = - \frac{i}{\theta} \left(\frac{2}{\theta}\right)^4 \int \frac{d^3k }{(2\pi)^3} \frac{d^3x}{k^2+m^2} A^\rho(k,k_2) \mathcal{Q}^{-1} \left[ \rho(\hat{x}_\rho) e^{ik_4 \rho(\hat{x})} \right] \frac{\sin(\frac{\theta}{2}|k_4|)}{|k_4|},
\end{equation}
where
\begin{align}\label{finale-1bis}
A^\rho(k,k_2) =&\ \frac{\sin^2(\frac{\theta}{2}|k|)}{|k|^2} \bigg[\cos(\theta |k| ) \frac{\sin(\theta |k| )}{|k|} \varepsilon_{\mu \nu}^{\hspace{11pt} \rho} k^\mu k_2^\nu\\
&\hspace{1.5truecm}+ \frac{\sin^2(\theta |k| )}{2|k|^2} (|k|^2 k_2^\rho - k_\sigma k_2^\sigma  k^\rho )\bigg] \frac{\sin(\theta |k_2| ) \sin(\frac{\theta}{2}|k_2|)}{|k_2|^2}\nonumber.
\end{align}
Then we write \eqref{ik2k4-1} as 
\begin{equation}
I (k_2,k_4) = \frac{1}{\theta } \left(\frac{2}{\theta}\right)^3 \int \frac{d^3k }{(2\pi)^3} \frac{d^3x}{k^2+m^2} A^\rho(k,k_2) \left(\frac{\partial}{\partial k_4^\rho} e^{ik_4 x} \right).\label{qualityf}
\end{equation}
Integrating over $x$, the last term between parenthesis in \eqref{qualityf} gives the derivative of the Dirac distribution $\delta_\mu^{'}(k_4)$ defined, for any test function $\psi$, by $\langle\delta_\mu^{'} , \psi\rangle = - \left. \frac{\partial \psi}{\partial k_4^\mu} \right\vert_{k_4=0}$.\\
Finally, we introduce $u^\mu$ the unit vector in the direction $\mu$ and $\gamma$ the angle between the two momenta $k$ and $k_2$ and we have
\begin{align}\label{arho}
 A^\rho(k,k_2) =&\ \left. \frac{\sin^2(\frac{\theta}{2}|k|)}{|k|^2} \right[\cos(\theta |k| ) \sin(\theta |k| ) \sin\gamma \\
 &\hspace{1.5truecm} +\left. \sin^2(\theta |k| ) \frac{1-\cos\gamma}{2} \right] \frac{\sin(\theta |k_2| ) \sin(\frac{\theta}{2}|k_2|)}{|k_2|} u^\rho,\nonumber
\end{align}
which combined with \eqref{qualityf} produces
\begin{equation} 
I (k_2,k_4) = \frac{J(k_2,k_4)}{\pi^3 \theta^4} \int d\alpha d\beta d r \frac{\sin^2(\frac{\theta}{2}r)}{r^2+m^2} \left[ \frac{1}{2} \sin\left(2 \theta r \right) \sin\gamma + \sin^2\left(\theta r \right) \sin^2\frac{\gamma}{2} \right] \sin\alpha \label{typeII_spherique-bis}
\end{equation}
with
\begin{equation}\label{france-j-bis}
 J(k_2,k_4) = \frac{\sin\left(\theta |k_2| \right) \sin(\frac{\theta}{2}|k_2|)}{|k_2|} u^\rho \delta_\rho^{'}(k_4)
\end{equation}
which coincide with \eqref{typeII_spherique} and \eqref{france-j} respectively.
In (\ref{typeII_spherique-bis}), we have decomposed $d^3k$ into the spherical coordinates $k = (r=|k|,\alpha,\beta)$ for which the angle $\gamma$ depends only in the angles $\alpha$, $\beta$ that define $k$ and the angles $\alpha_2$, $\beta_2$ that define $k_2$. Thus, the two integrations over $\alpha$ and $\beta$ are finite.\smallskip

One can easily show that the integration over $k$ is finite. Indeed, one has
\begin{equation}\label{typeIIbounds}
\int d\alpha d\beta dr \frac{\sin^2(\frac{\theta}{2}r)}{r^2+m^2} \left[ \frac{1}{2} \sin\left(2 \theta r \right) \sin\gamma + \sin^2\left(\theta r \right) \sin^2\frac{\gamma}{2} \right] \sin\alpha \leqslant \int_0^\infty \frac{dr}{r^2+m^2}.
\end{equation}
Hence, we conclude that the Type-II contribution \eqref{typeII_spherique-bis} is UV finite.
\chapter{R{\'e}sum{\'e} en fran{\c{c}}ais.}\label{app-resumefr}
{------------------------------------------------------------------------------------------------------------------}\\
\noindent Ce mémoire regroupe les résultats que j'ai obtenus dans le cadre de mes études doctorales effectuées sous la direction de Jean-Christophe Wallet au sein du \textit{Laboratoire de Physique Théorique} de l'Université de Paris-Sud 11 entre Octobre 2015 et Septembre 2018.\smallskip

\noindent Y sont présentés divers aspects de théorie non-commutative des champs, c'est-à-dire de théorie des champs compatible avec l'existence d'une échelle de longueur minimale. En particulier, y est détaillée la construction de deux familles d'espaces non-commutatifs, ainsi que l'étude des propriétés quantiques de plusieurs modèles de théorie des champs scalaires qui leurs sont associés. Le résultat majeur de cette thèse est le calcul des corrections radiatives, à une boucle, des fonctions 2-points et 4-points associées à divers modèles de théorie quantique des champs scalaires invariante sous l’action de $\kappa$-Poincaré.\smallskip

\noindent Cet appendice constitue un résumé de ces travaux.\\
{------------------------------------------------------------------------------------------------------------------}\vspace{1cm}

Il est généralement admis que la description de l’espace-temps à la base de la théorie de la relativité générale et du modèle standard de la physique des particules ne fournit pas le cadre adéquat à une description unifiée des processus gravitationnels et quantiques. De nombreuses approches à une théorie quantique de la gravité -- telles que la théorie des cordes et la gravité quantique à boucles -- s’accordent même sur le fait, qu’au-delà d’une certaine échelle d’énergie (proche de l’échelle de Planck), une meilleure description consisterait à supposer une structure quantique à l’espace-temps \cite{Snyder:1947,Mead:1964,Maggiore:1993a,Maggiore:1993b,Doplicher:1994,Doplicher:1995,Amati:1989,Ashtekar:1992,Born:1938}. Dans ce contexte, la géométrie non-commutative \cite{Connes:1990,Landi:1997,Madore:1999,Gracia:2001} nous fournit un cadre mathématique approprié pour entreprendre l’étude des propriétés algébriques de tels espaces, ainsi que l'étude des conséquences d’une telle hypothèse sur la description des phénomènes physiques. En particulier, la théorie non-commutative des champs (TNCC) s’occupe de l’étude des possibles nouvelles propriétés (aussi bien classiques que quantiques) des théories de champs construites sur ces espaces non-commutatifs. L'intérêt d'étudier les TNCC repose sur leur grande généralité. D'une part, relativement peu d'hypothèses sont faites lors de la construction de ces modèles. D'autre part, les TNCC apparaissent dans de nombreux contextes en physique, tels que celui de la gravité quantique \cite{Freidel:2006,Cianfrani:2016}, dans des modèles de cordes \cite{Witten:1986,Seiberg:1999} et de ``branes" \cite{Alekseev:1999}, ainsi qu'en ``group field theory" \cite{Guedes:2013}.\smallskip

Avant de procéder à l’exposition des résultats obtenus dans cette thèse, il est utile de préciser ce que nous entendons par espace-temps quantique. Il est bien connu que la géométrie algébrique (commutative) permet, entre autre, de transcrire dans un langage algébrique le contenu géométrique (et topologique) d’un espace séparable localement compact. En particulier, le théorème de Gelfand-Naimark nous assure que les points d’un espace topologique $X$ peuvent être identifiés aux caractères sur la C*-algèbre (commutative) des fonctions continues sur $X$ s’annulant à l’infini. Réciproquement, toute C*-algèbre commutative abstraite peut être vue comme une telle algèbre de fonctions. Étendant cette correspondance au cadre non-commutatif, nous définissons alors un espace quantique comme étant une algèbre non-commutative de fonctions muni d’un produit déformé, appelé $\star$-produit. Une manière de caractériser un espace quantique revient donc à caractériser le $\star$-produit qui lui est associé. Pour y parvenir, une méthode commode consiste à généraliser le schéma de quantification de Weyl (qui conduit au produit de Groenewold-Moyal dans le cadre de la formulation statistique de la mécanique quantique) à l’algèbre de coordonnées non-commutative, $\mathfrak{g}=\text{Lie}(\mathcal{G})$, caractérisant l’espace non-commutatif étudié. Dans ce cas, le $\star$-produit est défini via l’introduction d’un opérateur de quantification $Q :=\pi\circ\mathcal{F}$, par la relation
\begin{equation}\label{frstar}
Q(f\star g) :=Q(f)Q(g),
\end{equation}
où $Q$ est défini (à une transformée de Fourier $\mathcal{F}$ près) comme une *-representation $\pi$ de l'algèbre de convolution $(L^1(\mathcal{G}),\hat{\circ})$ de $\mathcal{G}$. Cette approche est basée sur l'utilisation de résultats d'analyse harmonique dont les rudiments sont rappelés dans l'appendice \ref{sap-harmonic}. Nous précisons qu’aucune utilisation des triplets spectraux n’est faite.\smallskip

Dans cette thèse, nous proposons une méthode pour construire des $\star$-produits associés à des espaces quantiques dont l’algèbre de coordonnées, $\mathfrak{g}=\text{Lie}(\mathcal{G})$,  est de type algèbre de Lie. Deux familles distinctes d’espaces quantiques sont considérées. Dans un cas, $\mathfrak{g}$ est choisie semi-simple. Dans l’autre, $\mathfrak{g}$ est choisie résoluble. Il s’ensuit que dans le premier cas, le groupe de Lie, $\mathcal{G}$, correspondant est unimodulaire, alors qu’il ne l’est pas dans le second cas. Ceci fait l’objet de la première partie, Part \ref{ch-ncst}. Dans une seconde partie, Part \ref{part-ncft}, nous utilisons les $\star$-produits de la première partie pour construire différents modèles de TNCC scalaires et en étudier les propriétés quantiques. Pour y parvenir, nous remarquons que, grâce à l’expression intégrale des $\star$-produits utilisés, les TNCC peuvent être vues comme des théories de champs ordinaires mais avec un operateur cinétique non-trivial et un potentiel d’interaction non-local. Ceci nous permet d'utilisé les techniques perturbatives usuelles de théorie quantique des champs -- dans le formalisme de l'intégrale de chemin -- pour calculer les corrections radiatives des fonctions 2-points et 4-points.\smallskip

Dans le chapitre \ref{sec-products}, nous considérons une famille d’espaces quantiques dont l’algèbre de coordonnées est semi-simple. Soit $Q$ un opérateur de quantification linéaire et inversible. Nous montrons qu’il est possible de caractériser le $\star$-produit, eq. \eqref{frstar}, en déterminant uniquement l’expression des ondes planes déformées
\begin{equation}
E_p(\hat{x}) :=Q(e^{ipx}),
\end{equation}
apparaissant dans l'expression
\begin{equation}
Q(f)(\hat{x}):=\int \frac{d^np}{(2\pi)^n}\ \mathcal{F}f(p) E_p(\hat{x}).
\end{equation}
En identifiant les $Q(e^{ipx})$ à des représentations projectives de $\mathcal{G}$, \textit{i.e.} $E:\mathcal{G}\to\mathcal{L}(\mathcal{H})$, telles que $E(g)\equiv E_p(\hat{x})$, satisfaisant
\begin{equation}
E(g_1)E(g_2)=\Omega(g_1,g_2)E(g_1g_2),
\end{equation}
avec $\mathcal{H}$ un espace de Hilbert convenablement choisi et $\Omega:\mathcal{G}\times\mathcal{G}\to\mathbb{C}\hspace{-2pt}\setminus\hspace{-2pt}\{0\}$ satisfaisant à une condition de 2-cocycle, nous montrons que des familles de $\star$-produits inéquivalents peuvent être classifiées par le second groupe de cohomologie $H^2(\mathcal{G},\mathbb{C}\hspace{-2pt}\setminus\hspace{-2pt}\{0\})$ de $\mathcal{G}$ à valeur dans $\mathbb{C}\hspace{-1pt}\setminus\hspace{-2pt}\{0\}$. L’intérêt de cette remarque est l’existence de tables mathématiques dans lesquelles de nombreux exemples de $H^2(\mathcal{G},\mathcal{A})$, avec $\mathcal{A}$ un groupe abélien, ont déjà été classifiés. En d'autres termes, la littérature mathématique contient déjà de nombreux résultats que nous pouvons réinvestir dans l'étude des propriétés algébriques des $\star$-produits, ainsi que des espaces quantiques associés.  Cependant, en vue d’une utilisation pratique du $\star$-produit dans le contexte des TNCC, il est nécessaire de déterminer une expression explicite des ondes planes déformées. Ceci revient à choisir un représentant d’une des classes d’équivalence appartenant à $H^2(\mathcal{G},\mathcal{A})$. Comme nous le montrons, cette procédure n’est en général pas triviale et nécessite de faire certaines hypothèses, quand bien même le second group de cohomologie de $\mathcal{G}$ est trivial.\smallskip

Pour y parvenir, nous commençons par représenter l’algèbre abstraite de coordonnées comme une algèbre d’opérateurs différentiels. Bien que cette idée ne soit pas nouvelle, il est à noter que nombre d’études présentes dans la littérature de physique-mathématique ne donnent que peu d’importance à la préservation des structures involutives. En particulier, les opérateurs auto-adjoints (observables) ne sont pas représentés comme des opérateurs différentiels auto-adjoints, ce qui peut avoir de lourdes conséquences sur la construction de l’algèbre des champs modélisant l’espace-temps non-commutatif, ainsi que sur la définition d’une condition raisonnable de réalité lors de la construction de l’action fonctionnelle dans le cas de l’étude de TNCC. Ici, nous insistons donc particulièrement sur la préservation de ces structures involutives. Nous montrons que les expressions admissibles pour la *-représentation différentielle sont obtenues à partir de la résolution d’un ensemble de quatre équations différentielles, eq. \eqref{diff-rep3} et \eqref{diff-rep4}. Enfin, en supposant qu’une telle représentation soit choisie, nous montrons qu’une expression pour les ondes planes déformées peut être obtenue en imposant $Q(f)\rhd f=f\star g$ et $Q(f)\rhd 1=f$, et en utilisant la décomposition polaire d’opérateur, eq. \eqref{polar}. Nous appliquons ensuite cette procédure au cas $\mathcal{G}=SU(2)$. En supposant maintenant que la *-représentation différentielle est $SO(3)$-equivariante, nous montrons que celle-ci ne dépend que deux fonctionnelles réelles et $SO(3)$-invariante, eq. \eqref{general_rep}. Dans ce cas, les ondes planes déformées sont caractérisées par deux fonctions des impulsions dont les expressions sont contraintes par deux équations de Volterra, eq. \eqref{volterra}. De plus, le produit de deux ondes planes déformées résulte (presque uniquement) de la formule de Baker-Campbell-Hausdorff pour $SU(2)$. Parmi cette famille de *-représentations différentielles, nous nous restreignons à celle permettant de dériver un $\star$-produit équivalent au produit de Kontsevich. Nous utilisons ensuite ce produit pour calculer, au chapitre \ref{sap-ncftsu2}, les corrections radiative, à une boucle, de la fonction 2-points pour différents modèles de théorie de champs scalaires avec interactions quartiques. Nous montrons que ces théories sont caractérisées par deux types (inéquivalents) de contributions, l’un planaire, l’autre non-planaire. Dans les deux cas, les corrections sont trouvées finies dans l’ultraviolet, et aucune singularité infrarouge n’est présente (même dans le cas non-massif). Il en découle que ces théories ne présentent pas de mélange infrarouge/ultraviolet. De plus, le paramètre de déformation $\theta$ est trouvé jouer le rôle de coupure ultraviolette et infrarouge.\smallskip

Le deuxième exemple d'espace non-commutatif étudié est connu sous le nom de $\kappa$-Minkowski \cite{Majid:1994}. Dans ce cas, l’algèbre de coordonnées est résoluble et le groupe de Lie associé est donné par le groupe non-unimodulaire, localement-compact, $\mathcal{G}_{d+1}=\mathbb{R}\ltimes\mathbb{R}^d$. Ce groupe est isomorphe au groupe affine de la droite réelle dans le cas bidimensionnel, $d=1$. Le chapitre \ref{sec-Minkowski} est dédié à la construction d’un $\star$-produit associé à $\kappa$-Minkowski. Là encore, l’approche adoptée suit dans les grandes lignes le schéma de quantification de Weyl qui s’avère une fois de plus très commode pour construire un $\star$-produit de manière contrôlée. De plus, l’expression ainsi obtenue, eq. \eqref{star-4d}, est relativement simple comparativement aux expressions d’autres $\star$-produits obtenus pour $\kappa$-Minkowski dans la littérature. À la différence du chapitre \ref{sec-products}, la représentation $\pi$ (qui apparait dans la définition de l’opérateur de quantification) est associée à une représentation unitaire  et non plus projective de $\mathcal{G}_{d+1}$. L'intérêt de $\kappa$-Minkowski réside dans le fait qu'il a été montré que son groupe d'isométries, le groupe quantique de $\kappa$-Poincaré \cite{Lukierski:1991}, correspond au groupe de symétries d'un espace quantique émergeant dans une certaine limite de la gravité quantique \cite{Cianfrani:2016}. Ceci est renforcé par le fait que $\kappa$-Poincaré correspond à une déformation (à haute énergie), dans le langage des algèbres de Hopf, du groupe de Poincaré ; le paramètre de déformation $\kappa$ ayant la dimension d'une masse. Dans le chapitre \ref{ch-ncft}, nous discutons les propriétés qu’une action fonctionnelle, supposée décrire la dynamique de champs définis sur $\kappa$-Minkowski, doit satisfaire. Compte tenu du rôle central joué par le groupe de Poincaré en théorie quantique des champs ordinaire, ainsi que du fait que $\kappa$-Minkowski supporte une action naturelle de l’algèbre de $\kappa$-Poincaré, il semble physiquement raisonnable de requérir cette action invariante sous l’action de $\kappa$-Poincaré. Cette hypothèse conditionne l’emploi de la mesure de Lebesgue dans la construction de l’action fonctionnelle. Cependant, l’intégrale de Lebesgue ne définit pas une trace pour le $\star$-produit associé à $\kappa$-Minkowski. À la place, elle définit une trace ``twistée", eq. \eqref{twistedtrace}.  Cette perte de cyclicité -- à laquelle s’ajoute la complexité des $\star$-produits alors utilisés -- est probablement une des raisons de l’absence d’études des propriétés quantiques des TNCC construites sur $\kappa$-Minkowski. Au contraire, nous donnons ici une interprétation positive de ce fait, puisque nous soulignons que cette perte de cyclicité traduit simplement que la fonctionnelle $\zeta :f\mapsto\zeta(f):=\int d^4x f(x)$ défini un poids KMS sur l’algèbre des champs modélisant $\kappa$-Minkowski. 
En d’autres termes, imposer l’invariance sous $\kappa$-Poincaré implique de remplace la cyclicité de l’intégrale de Lebesgue par une condition KMS, c’est-à-dire
\begin{equation}
\zeta\big((\sigma_t\triangleright f)\star g\big)=\zeta\big(g\star (\sigma_{t-i}\triangleright f)\big),
\end{equation} 
où $\lbrace\sigma_t:=e^{\frac{3t}{\kappa}\partial_0}\rbrace_{t\in\mathbb{R}}$ définit un groupe de *-automorphisme à un paramètre. Il est utile de rappeler que la condition KMS a été originellement introduite dans le cadre de la physique statistique pour caractériser l’état de systèmes quantiques à l’équilibre thermique. Dans ce cas, la condition KMS est vérifiée au niveau des fonctions de corrélation, \textit{i.e.} $\langle \Sigma_t(A)B\rangle_\beta$, calculées pour un certain vide thermique, où $A$ et $B$ sont des fonctionnelles des champs, et $\Sigma_t$ correspond à l’opérateur d’évolution de Heisenberg. Ainsi, une telle condition KMS est vérifiée au niveau de l’algèbre des observables. Il s’ensuit que le ``flow" généré par le groupe modulaire $\lbrace\sigma_t\rbrace_{t\in\mathbb{R}}$, \textit{i.e.} le ``flow" de Tomita, peut être utilisé pour définir un temps physique global. Ceci reflète le lien profond entre condition KMS et dynamique, et est à la base de l’hypothèse de l’origine thermique du temps discutée dans \cite{Connes:1994}, c’est-à-dire l’émergence du temps à partir de la non-commutativité de l’espace. Malheureusement, notre condition KMS est vérifiée au niveau de l’algèbre des champs et non celle des observables. Pour arriver à une telle conclusion, il reste donc à montrer que cette condition KMS induit une condition KMS au niveau de l’algèbre des observables.\smallskip

Cette nouvelle interprétation de la perte de cyclicité nous permet d’aborder plus sereinement la construction de l’action fonctionnelle, et l’étude des TNCC sur $\kappa$-Minkowski. Nous restreignons notre analyse au cas où l’opérateur cinétique est donné par le carré d’un opérateur de Dirac. Trois cas sont considérés : dans le premier cas, l’opérateur cinétique correspond au (premier) Casimir de l’algèbre de $\kappa$-Poincaré ; dans le second cas, il correspond au carré d’un opérateur de Dirac $\mathcal{U}_\kappa(iso(4))$-équivariant ; dans le dernier cas, il correspond au carré d’un opérateur de Dirac modulaire. Les propriétés de décroissance des propagateurs correspondants sont analysées. Nous restreignons également notre attention au cas où le potentiel d’interaction réduit au potentiel $\vert\phi\vert^4$ habituel dans la limite commutative (basse énergie) $\kappa\to\infty$. Nous montrons qu’il existe essentiellement deux familles d’interactions inéquivalentes.  Finalement, l’expression intégrale du $\star$-produit nous permet de représenter les TNCC comme des théories des champs ordinaires mais non-locales. Ceci nous permet de dépasser l'ensemble des difficultés techniques qui empéchaient jusqu'alors l'études des proriétés quantiques des TNCC construites sur $\kappa$-Minkowski, et nous permet de fournir la première étude complète de ces propriétés, à l'ordre une boucle, pour plusieurs modèles de théorie des champs scalaires invariante sous $\kappa$-Poincaré, en calculant les corrections radiatives pour les fonctions 2-points et 4-points associées à ces modèles. Dans le cas où l'opérateur cinétique est donné par le premier Casimir de $\kappa$-Poincaré, ou par le carré de l'opérateur de Dirac modulaire, la fonction 2-point est trouvée diverger plus que dans le cas commutatif. Seul dans le cas où l'opérateur cinétique correspond au carré de l'opérateur de Dirac équivariant la fonction 2-point est trouvée diverger moins que dans le cas commutatif, à savoir linéairement. Dans tous les cas, du mélange infrarouge/ultraviolet est trouvé pour un type d'interaction (non-orientable). Ces résultats sont regroupés dans le tableau \ref{tableau2}. Dans le cas où le cinétique est fonction du Dirac équivariant, et que le potentiel d'ntraction est orientable, nous trouvons que la fonction 4-point est finie à une boucle. Ceci indique que la fonction beta associée est nulle, et seul le terme de masse doit être renormalisé. Une extension immédiate de ce travail consisterait à étudier les propriétés de renormalisation de ces modèles à tous les ordres, ainsi que la question du mélange infrarouge/ultraviolet. Il serait également très intéressant d’étudier de manière précise la limite commutative de ces modèles. En particulier, nous savons que dans le cas commutatif la constante de couplage augmente en même temps que l’échelle d’énergie. Or, nous trouvons ici que la constante de couplage est constante. Ainsi, si nous voulons que les deux régimes coexistent le ``flow de la constante de couplage doit être borné dans la limite des très haute énergie. Ceci implique que la courbe théorique correspondante doit posséder un point d’inflexion qu’il serait important de caractériser. Un tel modèle fournirait un candidat pour un modèle de Higgs à l’échelle de Planck. 
\part*{Bibliography.\addcontentsline{toc}{part}{Bibliography.}}
\bibliographystyle{unsrt}
\bibliography{memoire_vfarxiv}

\end{document}